# Lights Out, Stress In: Assessing Stress Amidst Power and Energy Challenges in Bangladesh

Faisal Quaiyyum[1, *]

Khondaker Golam Moazzem[2]

**Abstract**

This study examines the psychological impact of energy crises on households, utilising the Perceived Stress Scale-10 (PSS-10) to measure the stress induced by disruptions in electricity, gas, and fuel supply and pricing. Through a multivariate analysis incorporating Ordinary Least Squares (OLS) regression, Simultaneous-Quantile Regressions (SQR), Random Forest (RF) and Ordered Probit models, the research identifies the key socio-demographic and environmental factors influencing household stress. Our findings reveal that urban residency, low-income households, older individuals, and those with low environmental awareness are particularly vulnerable to stress during energy crises. Regional disparities and attitudes towards nuclear and renewable energy also significantly shape stress responses. The study emphasises the need for psychologically-informed energy policy, advocating for the inclusion of stress metrics in energy planning to enhance resilience and address the multi-dimensional nature of energy insecurity. This research contributes a novel, human-centric perspective to energy policy, urging policymakers to integrate psychosocial resilience alongside traditional technical and economic considerations in the design of energy interventions.

---

[1, *] Corresponding Author, Research Associate, Centre for Policy Dialogue, Dhaka, Bangladesh, Email Address: faisal@cpd.org.bd, quaiyyumfaisal@gmail.com

[2] Research Director, Centre for Policy Dialogue, Dhaka, Bangladesh.

**Acknowledgement:**

We would like to thank Professor Dr. Atonu Rabbani, Professor Mustafizur Rahman, Dr. Fahmida Khatun, and Professor Dr. M. Shamsul Alam for their valuable feedback. We also appreciate the insights provided by the participants of the 61st Young Scholars' Study Session and our discussant, Mr. Mashfiq Ahasan Hridoy. We are also grateful to Mr. Avra Bhattacharjee, Joint Director of Dialogue and Outreach at CPD. Additionally, we extend our heartfelt thanks to Mr. H.M. Al Imran Khan, Publication Associate, and Mr. Md. Shaiful Hassan, Programme Associate (DTP), CPD, for their contributions in formatting, refining, and finalizing the manuscript. Usual disclaimers apply.

# 1.0 Introduction

In an era where the relentless march of technological progress and urbanisation has led to an insatiable appetite for energy, the plight of Bangladesh is grappled with a burgeoning energy and power crisis. Bangladesh's energy crisis stems from a blend of rising demand due to population growth and economic development, coupled with a heavy reliance on imported fuel and weak foreign currency reserve. Financial challenges in procuring these imports, slow progress in renewable energy generation, and delays in new power plant projects add to the crisis. Moazzem, et al. (2023) show that resource scarcity, factors such as the use of low-quality fuel oils, corruption at distribution points, inadequate maintenance, reduced gas supply pressure, and cuts in government subsidies for electricity and gas, mandated by the International Monetary Fund, further complicate the situation and induce frequent power outages. These elements are poised to drive up the costs of electricity, fuel oil, and gas. Additionally, poor management and administrative practices contribute to the deepening crisis. In response, the Government of Bangladesh (GoB) and other relevant stakeholders have adhered to formulating, following and discussing policies such as the Prosperity Plan 2022-2041, the Energy Efficiency and Conservation Master Plan up to 2030, and the Integrated Energy and Power Master Plan (IEPMP) 2023, etc.

However, what is lacking in all of these policies is the demand dimension of household-level consumers or business-level consumers. The policies are mostly designed to address and cater the supply-side adversities. Therefore, understanding the perceptions of the population is paramount and crucial for policymakers, researchers, and energy providers to develop strategies for energy conservation, demand management, and promotion of sustainable energy sources that is more appropriate and suitable with perceived behaviour (Spence, et al., 2010a). Thus, this study harnesses the power of psychological stress assessment to unveil the hidden currents of public opinion in Bangladesh and to discuss the reasons why integrating the consumers side are crucial, potentially introducing a new narrative of the nation's energy policy from the ground up. Last but not least, the scope of the methodology designed in this study is not confined within the border of Bangladesh.

Determining policy priorities requires balancing crucial components, including government responsibilities, political dynamics, inter-agency coordination, and public sentiment (Quaddus & Chowdhury, 1990; Jacoby & Schneider, 2001; Halpin, Fraussen, & Nownes, 2018). Psychological studies play a crucial role in understanding public perceptions, offering insights into societal views on various issues (Smith, 1971). Stern & Gardner (1981) argue that these studies delve into the emotional and psychological effects of events like energy shortages, identifying key areas for policy focus. This research extends the dialogue on energy policy to take into account the perceptions, measured by psychological effects of the households during energy crises while devising power and energy strategy in both operational and policy level. This deeper understanding aids stakeholders in addressing the most impactful stress factors, making perception studies invaluable for informed policymaking.

The primary goal of this study is to measure the stress levels of Bangladeshi households, directly resulting from the power and energy crisis and the factors that are associated with deciding the stress levels of the households attributed to the crisis. More explicitly, it aims to explore various factors influencing these stress levels, including environmental and political awareness, socio-economic status, and regional differences. Vlek (2000) emphasises that physiological responses to related stressors in power and energy sectors is crucial for individual well-being, workplace health, and community relationships. In our study, we have employed perceived stress scale – 10 (PSS-10) to assess the stress faced by the households due to incidents associated with power and energy crisis in Bangladesh. We have employed multivariate OLS, ordered probit and random forest model to assess the determining factors and contribution of each factor. Additionally, we have employed graphical analysis using GIS to display the geographical disparities associated with the stress level attributed to the power and energy crisis. By examining these determinants, the study seeks to provide a comprehensive view of the factors which contribute to the households' relevant stress. Additionally, the study will assess the reliability of using stress levels as an indicator of the crisis's impact through quantitative measures.

The remainder of this paper is structured as follows: Section 2 highlights a review of the literature; Section 3 details the materials and methodology employed; Section 4 presents the results and discussion; and Section 5 concludes the paper with a summary of insights and implications.

## 2.0 Literature Review

The utilisation of psychological studies and quantitative psychology for gauging perceptions or opinions is a well-established concept, with prior literature demonstrating various instances of its application. This section primarily focuses on a conceptual literature review, as our proposed analytical framework and concepts, while unique in their applications in energy economics and policy-making, draw inspiration from existing works. Since the methodological framework we employ in this study introduces a new approach to behavioural analysis of household consumers in power and energy studies, we first set the background for integrating psychological studies, establishing their significance, capacity, and potentiality of contributions in this field.

Stern and Gardner (1981) underscore the potential of psychological insights in shaping energy and power sector policies, highlighting how consumer behaviour and perception are critical in formulating effective and sustainable strategies. This concept is further supported by findings from subsequent studies (Smith, 1971; Swim, Geiger, & Zawadzki, 2014). This study takes an attempt to bridge this gap, proposing an analytical framework that integrates psychological insights into energy policymaking. We are going to assess how the socio-economic, political and environmental opinions of the households are associated with stress level of the households associated with the power and energy crisis.

Across the global literatures, various psychological methods have been employed over the years to assess perceptions of the general population about energy and power sector. In our discussion on the scope of psychological research in energy policy, we now turn to Whitmarsh's (2011) study, which adeptly employs psychological techniques to measure scepticism and uncertainty about climate change among the UK public. Utilising a multi-dimensional approach, Whitmarsh's methodology focused on developing and refining a measure of scepticism, building upon her previous qualitative work. Her objectives centred on examining the variations in public scepticism and uncertainty about climate change from various dimensions such as reasons, perceptions, remedies, etc., especially in relation to individual and societal factors like demographics, lifestyle, knowledge, and values (Whitmarsh, 2011). Spence et al. (2010b) explore public attitudes towards climate change and different forms of energy production, as well as investigate the evolution of these views over time. To achieve this, they employ surveys, supplemented with various psychological tools, to quantify and analyse public perceptions, underscoring the dynamic nature of public opinion amidst the rapidly shifting environmental and energy landscapes (Spence, et al, 2010b). In a separate study conducted by Carrus, et al. (2021), a meta-analysis was utilised to investigate the influence of various psychological factors, including attitudes, intentions, values, awareness, and emotions, in shaping behaviours related to energy conservation. This methodology facilitated a thorough examination of the robustness of the relationships between different psychological variables and individuals' intentions and actions towards saving energy (Carrus, et al., 2021). Psychological theories such as the 'Theory of Planned Behaviour' have been effectively used to design interventions that align consumer behaviour with their environmental attitudes in enhancing the uptake of green electricity in Switzerland (Litvine & Wüstenhagen, 2011). Ma, Xu, & Zhang (2024) examine consumer decision-making by accounting for bounded rationality and subjective preferences, focusing on how consumers perceive risks and benefits in energy-related choices. Applying Gray correlation analysis and prospect theory, their study evaluates consumer response potential through key indicators, including dispatchable power capacity, available response time, and response reliability (Ma, Xu, & Zhang, 2024). Barsanti, Yilmaz, & Binder (2024) use a quantitative survey on laundry and dishwashing habits to identify behavioural patterns, their determinants, and variations in load-shifting potential through hierarchical clustering, multinomial logistic regression, and analysis of variance. On a different but relevant discourse, behavioural analysis has also been used in renewable energy technologies to explore decision-making dynamics and investment behaviour. Salm (2018), for instance, highlights how differences in risk preferences between incumbent utilities and institutional investors shape investment patterns through financial incentives and risk exposure. Moreover, a study authored by Tiefenbeck, et al. (2013) indicate that environmental campaigns can sometimes lead to unintended behavioural spill overs, such as moral licencing, where an improvement in one environmental behaviour causes a setback in another. These diverse applications underscore the importance of psychological tools and theories in designing more effective policies and interventions for promoting sustainable energy practices and investments.

We have discussed the details of the PSS-10 later in the methodology section. In the literature review section, we are going to discuss the flexible applicability of PSS-10 in various field of study across the world. Vlek (2000) highlights the significance of measuring stress associated with power and energy since the power and energy sectors are pivotal components of modern society, and understanding the psychological and physiological responses to stressors linked to them is essential for safeguarding individual well-being, workplace health, and community relationships (Vlek, 2000). Townsend and Medvedev (2022) demonstrate the remarkable adaptability of the PSS by highlighting its extensive utilisation across diverse general and clinical contexts, amassing over 20,000 citations on Google Scholar. Furthermore, the PSS has exhibited its global reach, with translations available in 28 different languages, including languages used in developing countries, by 2022 (Townsend & Medvedev, 2022). Several examples of utilising PSS measures to examine stress levels include the research conducted by Lushchak et al. (2023), where they applied the PSS-10 to measure the extent of stress experienced by the Ukrainian populace in reaction to events associated with the Russian invasion of Ukraine (Lushchak, et al., 2023). Additionally, the PSS has been employed to assess stress levels in cancer and breast cancer patients, including its translation and validation in different languages for specific clinical studies and the exploration of alternative factor models to understand perceptions of stress in these populations (Mounjid, et al., 2022; Golden-Kreutz, et al., 2004). Moreover, the PSS-10 has been employed to assess stress levels in expectant mothers across culturally diverse settings and to compare stress levels between first-time mothers and those with previous childbirth experiences, demonstrating its utility in prenatal stress research (Katus, et al., 2022). The highlighted and many other studies have opened up the door of employing PSS-10 in our case, upon following the pre-requisites discussed in the next part of this section.

Islam (2020) demonstrate that the Bengali version of the Perceived Stress Scale (PSS-B) is a valid and reliable tool for assessing perceived stress among nonclinical individuals in Bangladesh, thereby expanding the utility of the PSS-10 in nonclinical settings within the country (Islam, 2020). Moreover, a study by Mozumder (2022), it was demonstrated that the PSS-10 maintains its reliability and validity when applied in the context of Bangladesh, affirming the scale's robust psychometric properties and affirming its suitability as a valid and dependable tool for evaluating stress appraisal within Bengali-speaking and Bangladeshi populations (Mozumder, 2022). It validates and expands the scope of using PSS-10 in Bangladesh within a non-clinical setting, making it a better fit for a nationwide household survey.

# 3.0 Methodology

## 3.1 Analytical Framework

Perceived Stress Scale, originally developed by Cohen, Kamarck and Mermelstein (1983), the PSS was initially designed as a 14-item self-report questionnaire to measure the extent to which individuals perceive their lives as stressful. The PSS-10 assesses individuals' perceived stress levels by evaluating the extent to which they find their lives unpredictable, uncontrollable, and overwhelming. The PSS-10 is a shorter adaptation of the original PSS, consisting of 10 items, and has gained popularity for its brevity and efficiency in measuring perceived stress (Cohen, Kamarck, & Mermelstein, 1983). PSS-10 necessitates participants to consider the previous month as the reference period from the survey date. Since the Bengali version of the PSS-10 questionnaire has been validated and proven reliable for the Bengali-speaking population in Bangladesh, particularly in nonclinical settings, it serves as a culturally relevant tool for stress assessment. Its robust psychometric properties in the context of the Bangladeshi population further establish it as an academically appropriate measure for evaluating perceived stress in this demographic.

The Bengali version of the PSS-10 questionnaire, which was used as a reference for developing a modified version to incorporate the context of the power and energy crisis, draws its foundation from two primary sources: one developed by Keya (2006) and the other by Laboratory for the Study of Stress at Carnegie Mellon University[3]. This adaptation aims to retain the psychological essence of each question while tailoring the scenarios or contexts in order to match the experiences of the participants associated with power and energy crisis of Bangladesh (Islam, 2020; Keya, 2006). In the subsequent phase, the questionnaire underwent a thorough review process involving experts from the fields of psychology, behavioural economics, and power and energy sector.

**Figure 1: 6 Broad Scenarios of Power and Energy Crisis of Bangladesh.**

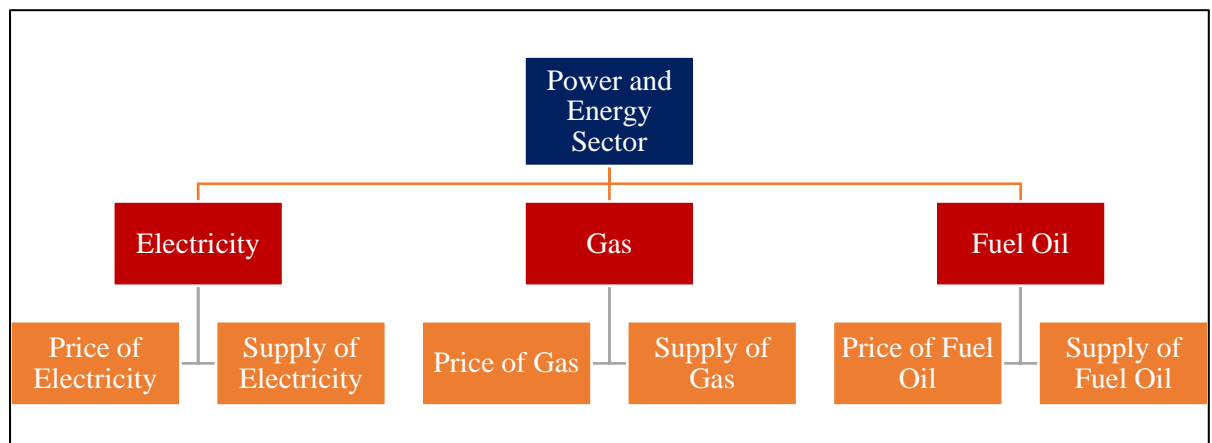

**Source:** Authors' Illustration.

---

[3] https://www.cmu.edu/dietrich/psychology/stress-immunity-disease-lab/scales/pdf/perceived_stress_scale_bengali_translation.pdf

For the purpose of our analytical framework, this study has segmented the power and energy sector in Bangladesh into three primary categories: electricity, gas, and fuel oil. Within each of these categories, two dimensions have been identified: the price aspect and the supply aspect. In essence, this approach results in the identification of six distinct categories that encompass the overall energy and power crisis in Bangladesh. Subsequently, the Bengali version of the PSS-10 questionnaire was tailored to address the specific contexts of each power and energy crisis scenario in Bangladesh (see Figure 1). As a result, our questionnaire comprises a total of 60 PSS questions that assess household stress levels arising from incidents associated with these six categorised power and energy crises. Each of these broad scenarios is represented by a set of PSS-10 questions, totalling 10 questions per scenario.

This study classifies the crises within the power and energy sector into six key scenarios. These scenarios were developed through literature reviews including newspapers, articles, reports, and in consultations with power and energy sector experts. Broadly, the 'Supply of Electricity' scenario encompasses load-shedding, low voltage, corruptions in bill payments, and unexpected outages, etc. The 'Price of Electricity' scenario covers increases in prices, inaccurate metre readings, billing mistakes, pricing uncertainties and so on. For 'Supply of Gas', issues include line leakages, outages, low pressure, and Liquefied Petroleum Gas (LPG) cylinder scarcity, etc. The 'Price of Gas' scenario reflects rising LPG connection costs, cylinder prices, national grid gas prices, and maintenance expenses, among others.. The 'Supply of Fuel Oil' scenario captures problems with fuel oil quality, corruption at filling stations, and availability issues. Finally, the 'Price of Fuel Oil' scenario deals with billing errors, price hikes, and pricing uncertainties, etc.

To assess regional stress level disparities across divisions for each scenario, including the overall situation, we will use Geographic Information System (GIS) techniques. Next, we will apply Multivariate OLS and Random Forest models to scenario-specific stress scores to examine the impact of socioeconomic factors, as well as political and environmental opinions, on stress levels. Lastly, an Ordered Probit model will analyse how these socioeconomic factors influence households' transitions between different stress ranges.

**3.2 Variables**

*3.2.1 Psychological Variables*

Ten questions on the PSS-10 are rated on a 5-point Likert scale. In the responses, 0 indicates 'never' and 4 indicates 'very often'. Responses to question number 4, 5, 7, and 8 are inverted before adding up all of the item scores to create a total stress score. Stress levels for each scenario are categorised into low (0-13), moderate (13-26), and high (27-40) based on the scores obtained (Carnegie Mellon University, 2010). For this study, we created a set of questions for each scenario, averaging them for the overall stress calculation. The variables employed for denoting various stress variables are tabulated in Appendix B.1.

*3.2.2 Socio-economic Variables*

Sex, age, level of education of household heads are considered in this paper. These include the size of the household, the number of students, the income bracket, and whether or not personal or business vehicles are owned. The primary sources of income are separated into the agricultural, industrial, and service sectors, although household income is split into five categories, ranging from less than BDT 10,000 to more than BDT 80,000 per month. The study also considers the availability or ownership of personal or office vehicles to offer insights into household asset ownership.

*3.2.3 Environmental and Political Values*

The variables used to assess the environmental and political values of the households are presented in Appendix B.1 and these variables are used with a view to understanding public perceptions and attitudes towards climate change, energy policy, and environmental responsibility. These variables are measured on a Likert scale. The scores range from 1 to 5. Here, 1 indicates 'strongly disagree', 2 indicates 'roughly disagree', 3 indicates 'neutral', 4 indicates ' roughly agree', and 5 indicates 'strongly agree'. The variables are denoted as 'e-variables' in this paper. Respondents were asked to express their level of agreement with statements on environmental and systematic incidences on climate, energy and power sector. The set addresses the impact of environmental pollution on climate change, and the roles of international organisations and national governments in addressing the climate crisis. Opinions regarding Bangladesh's preparedness for nuclear energy production as well as views on the necessity of nuclear and renewable energy in resolving the energy crisis were taken into consideration. Additional statements evaluated opinions regarding how the crisis was portrayed in the media, whether nuclear power plants in nearby neighbourhoods were accepted by the people in general, and whether households were willing to use less energy in order to protect the environment. These factors reveal information about both personal environmental awareness and general political beliefs.

*3.2.4 Heat Index*

The heat index is defined as how hot it actually feels to the human body when air temperature and relative humidity are combined. This measure is also referred as the apparent temperature (US Dept of Commerce, 2023). This measure is particularly important for assessing thermal comfort and potential health risks, as it captures the physiological experience of heat more accurately than temperature alone. Therefore, the heat index offers a more accurate indicator of ambient heat experienced by individuals, particularly in hot and muggy regions. This is crucial to the current study because it enables us to determine the precise amount of stress that the power and energy issues during the survey period were responsible for. Under various climatic conditions, the same level of energy crisis would not cause the same amount of stress. To put simply, a lower perceived temperature might lessen stress associated the crises, whereas high temperatures could greatly increase stress levels under the same crisis situations. According to the US Dept. of Commerce (2023), the heat index can be calculated as:

$$\begin{aligned}heat = \ &-42.379 + 2.04901523 \times temp + 10.14333127 \times humid \\ &- 0.22475541 \times temp \times humid - 0.00683783 \times temp^2 \\ &- 0.05481717 \times humid^2 + 0.00122874 \times temp^2 \times humid \\ &+ 0.00085282 \times temp \times humid^2 - 0.00000199 \times temp^2 \times humid^2\end{aligned}$$

In Section 4.4 of this paper, it is demonstrated in detail that the heat index cannot be included in our regressions due to significant confounding effects, including reverse causality between this variable and the dependent variables. This reverse causality leads to endogeneity bias, which would distort the results of the analysis. For this reason, we have not included the heat index in our regression models in the following sections. However, the discussion of the scope of including the heat index in our paper is particularly important because of its direct effects on the psychological health of individuals.

*3.2.5 Consumption Behaviour of Electricity, Gas and Fuel Oil*

In a measure to analyse the level of home stresses with respect to power and energy crises, knowing the amount of gas, electricity, and fuel oil consumed within a household is very essential. As the level of dependence increases with energy, consumption levels directly influence the extent to which a household is exposed to price fluctuations and supply interruptions (Zhang, 2024; Guan, et al., 2023). For example, the levels of energy consumption in a household increase the likelihood of them feeling extremely vulnerable, irritated, and angry if there are energy shortages or price hikes (Bardazzi & Pazienza, 2023). Our hypothesis is that the level of dependence that a household has on an affordable and dependable source of energy increases with its energy consumption, and this can result in increased stresses upon any interruptions or price increases.

To ensure a more accurate and fairer comparison across households of different sizes, we will use per capita measures for power and energy consumption. Larger households will naturally consume more energy simply because they have more members, but per capita consumption normalises this by normalising total consumption with household size. While per capita measures offer significant advantages, there are also some limitations. For example, they do not fully capture the variations in energy efficiencies within household usage of power and energy. While a larger household may use energy more efficiently, a smaller household may have a high per capita consumption due to reliance on energy-intensive appliances. Thus, the per capita measure does not reflect energy efficiency within a household in all the cases. However, considering the context of Bangladesh where the variations of efficient machinery usages are outperformed by the degree of consumption, it is logical to assume that the households move towards more power and energy consumption with growing income, which, in consequence, will lead a higher per capita consumption. Since we are using an income variable too, we are expecting the effects of efficiency to be controlled appropriately. Despite these arguments, the per capita measure remains a reliable tool for comparing energy consumption across households since it offers a standardized and scalable approach that accounts for household size.

## 3.3 Sampling Techniques and Sample Characteristics

*3.3.1 Sampling Techniques*

We surveyed 1000 households with access to both gas and electricity in 36 sub-districts spread across 8 divisions of Bangladesh. The respondents were heads of households who were at least 18years old.

The division-wise sample size was determined based on household distributions extracted from the Population Census of 2022. A correspondingly larger sample size was allotted to larger divisions based on the number of households. The assigned sample for each division was then divided equally among the sub-districts that were chosen at random, rounding any fractions. In divisions with larger overall sample allocations, such as Dhaka, a proportionally greater number of sub-districts were selected to reflect the higher population size. This approach ensures that sub-districts within the larger divisions also receive a larger share of the sample, in contrast to divisions with smaller allocations, like Mymensingh, where fewer sub-districts and smaller sample sizes were allocated (see Appendix A).

To ensure heterogeneity and representativeness of the dataset, at least four sub-districts were selected per division using a computer-generated random process. It targets areas with moderate to low poverty levels according to the HIES 2016 Poverty Map (BBS & WFP, 2020). Household selection within sub-districts followed a systematic sampling method. If the initially selected household lacked both gas and electricity connections, the subsequent household possessing both utilities was surveyed. Subsequently, the next household was chosen while maintaining the same interval and the process is repeated.

*3.3.2 Neighbourhood Characteristics (Urban and Sub-urban distribution)*

Sixty-three per cent of the total sample resides in sub-urban neighbourhoods, with the remainder spread across urban neighbourhoods. Notably, there are no rural neighbourhoods included in the sample, as they were excluded due to a precondition requiring confirmed availability of gas and electricity in the households.

### 3.3.3 Sex[4] of the Household Head

Ninety-six per cent of the participants in the study are identified as male household heads. Consequently, incorporating the factor of the sex of household head into our analysis may pose challenges regarding any conclusive statement from the point of view of sex of the household head due to the significantly smaller proportion of female-headed households.

### 3.3.4 Age Distribution of The Household Heads

The mean age of household heads is 42.5 years, ranging from 18 to 65 years. These justify the inclusion of age as a relevant factor in our analysis.

**Figure 2: Distribution of the Years of Education across the Household Heads**

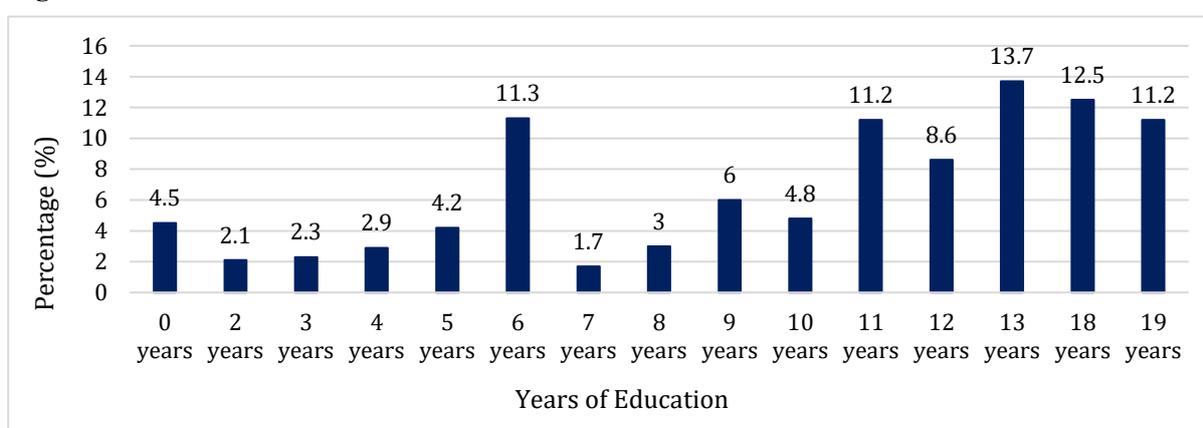

**Source:** Authors' Calculation,

### 3.3.5 Distribution of the Years of Education Across the Households Heads

Our calculation illustrated in Figure 2 shows the distribution of years of education across the household heads. The mean education years for all households is 10.91. Analysis shows female household heads have an average education of 8.76 years versus 11.01 for males, indicating a gender education gap in Bangladesh, with more females having no education. Urban heads average 11.37 years of education compared to 10.65 in sub-urban areas, aligning with the literature on higher urban education (USDA, 2017).

### 3.3.6 Household Size

The sample shows an average household size of 4.71, slightly above the HIES-2022 figure (BBS, 2023) with sizes ranging from 2 to 24. Urban households average 4.3 members, compared to 4.9 in sub-urban areas. Male-headed households have more members (4.7) than female-headed ones (4.3), in line with Bangladesh's typical household patterns (Saad, et al., 2022).

---

[4] In our study, sex is defined as a set of biological attributes associated with physical and physiological features, with a binary sex categorisation (male/female) typically assigned at birth.

*3.3.7 Number of Students in a Household*

On average, the households have 1.47 students, ranging from 0 to 13, with male-headed homes having slightly more (1.47) compared to female-headed ones (1.36). This pattern, not clearly indicating a preference in male-headed households, coincides with existing studies (Bose-Duker, Henry, & Strobl, 2021). Notably, sub-urban households report more students than urban ones, likely due to their generally larger family sizes.

**Figure 3: Income Distribution of Households**

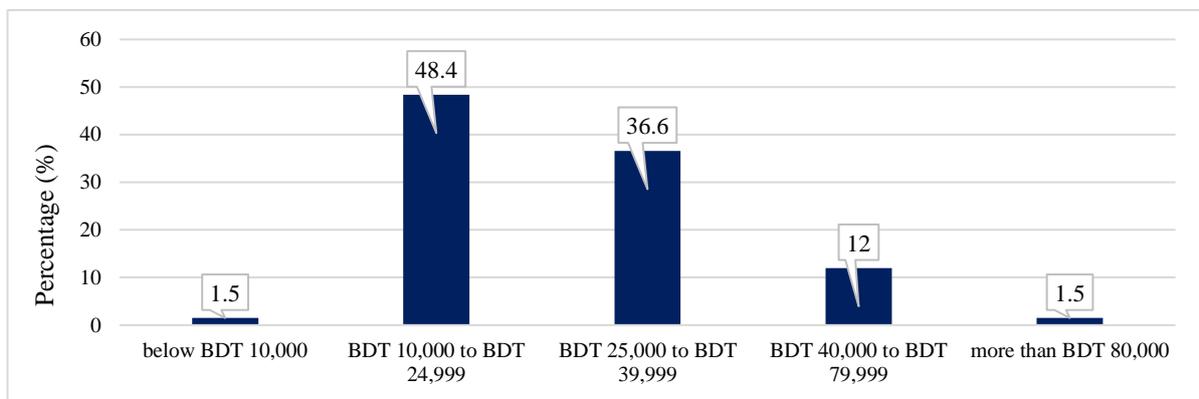

**Source**: Authors' Calculation.

*3.3.8 Primary Income Source*

Our analysis reveals that 72 per cent of the sampled households primarily derive their income from the service sector, while the figures for the industry and agriculture sectors stand at 19 per cent and 9 per cent, respectively. Our analysis shows that sub-urban households are more reliant on agriculture for income, while urban households tend to be more engaged in the industry sector.

*3.3.9 Income Distributions of the Households*

Our calculation, illustrated in Figure 3, shows that at least 80 per cent of households earn between BDT 10,000 to BDT 39,999 per month, reflecting a prevalent middle-income status consistent with national surveys like HIES 2022 (BBS, 2023). Additionally, higher-income households are more clustered in urban areas, a trend seen in prior research (Tripathi, 2020). Moreover, female-headed households tend to have lower incomes than their male-headed counterparts, signifying an economic gap, consistent with conventional literatures.

*3.3.10 Availability of Personal and/or Office Vehicles in The Households*

Our analysis reveals that merely 30 per cent of households possess personal and/or office vehicles.

Overall, the sample characteristics, indicated by years of education of the household head, household size, primary income source distribution, income distribution of the households, and availability of personal or office vehicles, validates the national representativeness of the sample.

### 3.4 Survey and Data Collection Methodology

The survey was conducted from 31st October to 23rd November 2023, during a significant heatwave with temperatures consistently ranging between 21°C and 34°C (AccuWeather, 2023). The temperate and humidity data has been collected from World Weather (World Weather, 2025). Surveyors, selected for their university education in psychology, received two days of training to ensure uniform data collection methods. Household heads provided informed consent and participated voluntarily without incentives, with confidentiality assured. Each household was given a unique identifier derived from two contact numbers. The duration of each survey ranged from 45 to 60 minutes.

### 3.5 Calculation and Econometric Model

*3.5.1 Overall PSS-10:*

The overall PSS-10 score is calculated by averaging the scores from the six scenario-specific PSS-10 assessments. The use of averaging is justified for our purpose as it captures both direct and indirect effects[5] of fuel oil on households, including those without direct usage, reflecting concerns through public transportation, generator fuel, and other means. The resulting average score is then rounded to align with the established ranges commonly used in academic psychology.

*3.5.2 Multivariate OLS*

A multivariate OLS model is employed for our purpose of investigating how various socio-economic and regional factors affect stress level and the specification of the econometric model is given below:

$$pss_{ij} = \alpha_0 + \beta'X_i + \gamma'Z_i + \delta'cons_i + \mu_{ij}$$

Here,

- i = 1, 2, 3......1000th household
- $pss_{ij}$ = PSS-10 for 'j' scenario for 'i'-th household

---

[5] Some households might not have any direct use of fuel oil.

- j = 1, 2, 3, 4, 5, 6, 7 represents stress related to electricity supply (*pss_elcs*), electricity prices (*pss_elcp*), gas supply (*pss_gassp*), gas prices (*pss_gaspr*), fuel supply (*pss_fuelsp*), fuel prices (*pss_fuelpr*) and a general perceived stress measure (*pss*) for 'i'-th household respectively
- $\beta'X_i$ = Vector of linear combination of socioeconomic factors 'i'-th household
- $\gamma'Z_i$ = Vector of linear combination of environmental and political consciousness variable for 'i'-th household
- $\delta'cons_i$ = Power and energy consumption behaviour for 'i'-th household for respective situation. In other words, when we will assess the perceived stress attributed to the price and supply of electricity issues, we will use electricity consumption behaviour as a control variable
- $\mu_{ij}$ = Residual term of the model for 'i'-th household in 'j' scenario

Our initial analysis indicates that some explanatory variables, particularly those related to political and environmental opinions, do not adhere to a normal distribution although these Likert variables are internally consistent (see Appendix C and Appendix D). Consequently, incorporating these variables into a linear model and estimating coefficients could yield spurious and misleading results.

*3.5.3 Quantile Regression*

Multivariate OLS assumes that the effects of predictors are the same across the entire distribution of the dependent variable, which might not be the case in this context since the stress level of households can be skewed towards a tail (higher or lower). Quantile regression, however, is more suitable for this study because it allows us to examine how the explanatory variables affect stress levels at different points of the distribution (e.g., at the lower, median, and upper quantiles), providing a more detailed and realistic understanding of how these factors influence various stress intensities.

Given that stress responses to the energy crisis may not be uniform across the population (e.g., individuals in lower stress levels may be less affected by the crisis than those in higher stress levels), quantile regression is more robust and flexible. This method is particularly advantageous when the dependent variable may be heterogeneously distributed with non-normal errors and potential outliers, as is often the case with survey data on subjective perceptions of stress (Koenker, 2005). Quantile regression provides insight into the conditional distribution of stress levels, offering a detailed understanding of how different factors influence households at different stress levels (e.g., low vs. high stress levels). Moreover, given that the explanatory variables, such as political opinions and environmental consciousness, do not follow a normal distribution and are likely skewed, OLS estimates might be biased and inefficient (Cade & Noon, 2003). Quantile regression is robust to such non-normal distributions, providing reliable estimates without the assumption of normality and improving the precision of the model (Angrist, Chernozhukov, & Fernández-Val, 2006).

In this study, we are interested in using bootstrapped simultaneous-quantile regression to investigate the relationship of various factors with the PSS-10 scores at different quantiles. This allows us to assess how the relationships between socio-economic factors, energy crisis scenarios, and environmental factors vary at different levels of stress. The quantile regression model can be mathematically represented as follows:

$$Q_\tau = \alpha_0(\tau) + \beta'(\tau)X_i + \gamma'(\tau)Z_i + \delta'(\tau)heat_{ik} + \mu_{ij}(\tau)$$

- The variables are defined as the equation of multivariate OLS.
- $\beta'(\tau), \gamma'(\tau)$ and $\delta'(\tau)$ are respectively the vectors of quantile-specific regression coefficients for the explanatory variables.
- $\mu_{ij}(\tau)$ is the residual term at the $\tau$-th quantile of the stress level for the *i*-th household in the *j*-th scenario.

In this study, we estimate quantile regression models at the 15th, 30th, 50th, 70th, and 85th percentiles to capture how the effects of explanatory variables on perceived stress vary across these respective points of the stress distribution from each different range. The number of bootstrap replications to be used in this study to obtain an estimate of the variance-covariance matrix of the standard errors is 400.

*3.5.4 Random Forest*

In order to address and solve the non-normal distribution of the explanatory variables, this study employs the Random Forest ensemble learning technique to predict stress levels (PSS scores) across seven different scenarios. Random Forests are ensemble models that consist of multiple decision trees, and for regression tasks like predicting PSS scores, they aggregate the predictions of individual trees to make the final prediction. For each scenario 'j', we build a separate Random Forest model to predict the corresponding PSS score ($PSS_j$) using a set of explanatory variables ($V_j$). For each scenario 'j', a Random Forest model is constructed to predict the corresponding Perceived Stress Scale (PSS), denoted as $PSS_j$. The model is formulated as:

$$pss_j = RF_j(V_j) + \varepsilon_j$$

Here,

- $pss_j$ = PSS-10 for 'j' scenario
- $V_j$ represents the set of independent variables, which is divided into two parts, mentioned as $X_i$ and $Z_i$ previously
- $RF_j$ is the Random Forest model trained for scenario 'j'
- $\varepsilon_j$ represents the residual error term

To assess the generalisation performance of the Random Forest models, in our study, a simple train-test split approach is used instead of k-fold cross-validation. Specifically, the dataset is split into training and testing sets, with 80 per cent of the data used for training and 20 per cent used for testing.

After building the Random Forest models, we perform feature selection to identify the most relevant independent variables $V_j$ for predicting PSS scores in each scenario. In the analysis, feature importance is visualised for top 10 features for each scenario, helping to understand which independent variables most significantly impact stress levels. For each scenario, we train a Random Forest model ($RF_j$) using the selected features $V_j$. The performance of each Random Forest model is evaluated using regression metrics such as the Mean Squared Error (MSE) and R-square ($R^2$). These metrics provide insights into the model's accuracy and the proportion of variance in the dependent variable that is predictable from the independent variables.

### 3.5.5 Ordered Probit

Another model that will be employed in this study to investigate the factors that is associated with the households shifting from one stress range to another in each specific and overall scenario is an ordered probit model. The specification of the model is:

$$Pr(pss\_rng_{ij} = k) = \Phi_i(\alpha_{jk} - \rho'X - \lambda'Z - \delta'cons_i)$$

and the theoretical latent variable assumed to be embedded here:

$$pss_{ij} = \alpha_0 + \beta'X_i + \gamma'Z_i + \theta'cons_i + \mu_{ij}$$

Here:

- $pss\_rng_{ij}$ = the ordered dependent variable (PSS-10 range) for 'i'-th household and 'j'-scenario PSS-10 score
- k = the stress range (e.g., Low, Moderate, High), where k = 1, 2, 3 represent 'Low Stress', 'Moderate Stress', and 'High Stress' respectively
- $\Phi_j$ = the cumulative distribution function of the standard normal distribution in the context of 'j'-scenario PSS-10 score
- $\alpha_{jk}$ = the threshold parameter for the k-th category in the context of 'j'-scenario PSS-10 score
- $\rho'X$ = Vector of the linear combination of socio-economic independent variables
- $\lambda'Z$ = Vector of the linear combination of environmentally and politically conscious variables
- $\theta'cons_i$ = Power and energy consumption intensity of 'i'-th household

For the defined stress ranges,

$$Pr(pss\_rng_{ij} = 1) = \Phi_i(\alpha_{j1} - \delta'X - \lambda'Z - \delta'cons_i), when, 0 \leq pss_{ij} \leq 13$$

$$Pr(pss\_rng_{ij} = 2) = \Phi_i(\alpha_{j2} - \delta'X - \lambda'Z - \delta'cons_i), when, 14 \leq pss_{ij} \leq 26$$

$$Pr(pss\_rng_{ij} = 3) = \Phi_i(\alpha_{j3} - \delta'X - \lambda'Z - \delta'cons_i), when, 27 \leq pss_{ij} \leq 40$$

In addition, we intend to conduct a post-estimation analysis to calculate marginal effects from the ordered probit models. These marginal effects will provide valuable insights into the influence of various socio-economic and regional factors on households' shifts from one stress range to another within each specific and overall scenario.

*3.5.6 Geographic Information System*

To illustrate regional disparities in stress levels due to energy and power crises across Bangladesh's eight divisions, we used Python, using geospatial shapefiles from Bangladesh Bureau of Statistics[6].

**3.6 Statistical Tests**

The study will implement Breusch-Pagan test to investigate heteroscedasticity. Should the error variances demonstrate heteroscedasticity, the issue will be addressed by applying robust variance method (Gujarati, 2003). Additionally, the study will employ a pairwise correlation coefficient matrix to detect multicollinearity (Gujarati, 2003) and the Shapiro-Wilk W-test to assess normality (Ramachandran & Tsokos, 2021). To ascertain the validity and reliability of the PSS-10 in measuring six scenario-specific scores and an overall score, Cronbach's alpha is calculated (Mozumder, 2022). The similar measure is employed to investigate the internal consistency of environmental and political variables.

**4.0 Result and Discussion**

**4.1 Reliability of the Perceived Stress Scores**

Cronbach's alpha calculations show a reliability coefficient above 0.8 for both specific and overall stress scores related to power and energy crises, indicating strong internal consistency and reliability of our modified PSS-10 questionnaire for Bangladesh's context (see Appendix E).

**4.2 Qualitative Analysis of Perceived Stress Scores for Various Scenarios**

Our calculation outlined in Table 1 shows, on an average, households mostly fall within the moderate stress range across scenarios, with scenarios of electricity and gas prices causing the highest average stress levels nationwide. Stress linked to the cost of electricity, gas, and fuel oil notably exceeds that related to supply scenarios.

---

[6] Online URL: https://github.com/yasserius/bangladesh_geojson_shapefile?tab=readme-ov-file

**Table 1: Summary Statistics of Perceived Stress Score under Various Scenarios**

| PSS Scenarios | Mean | 25% | Median | 75% | Minimum | Maximum |
|---|---|---|---|---|---|---|
| Supply of Electricity | 18.76 | 17 | 20 | 22 | 0 | 40 |
| Price of Electricity | 20.93 | 19 | 22 | 24 | 0 | 40 |
| Supply of Gas | 18.04 | 16 | 19 | 22 | 0 | 40 |
| Price of Gas | 20.41 | 18 | 21 | 23 | 0 | 40 |
| Supply of Fuel Oil | 16.63 | 16 | 17 | 21 | 0 | 31 |
| Price of Fuel Oil | 18.06 | 16 | 18 | 22 | 0 | 40 |
| Overall | 18.81 | 17.5 | 19.67 | 21.67 | 0 | 33.2 |

**Source:** Authors' Calculation.

Table 1 shows that the mean stress levels across scenarios range from 16.63 (Supply of Fuel Oil) to 20.93 (Price of Electricity), with the highest stress linked to electricity pricing. The 25th percentile values (16-19) indicate relatively low stress for many households, while the median (19-22) suggests moderate stress levels. The 75th percentile (21-24) reflects higher stress for a quarter of households. Stress levels vary widely, with some households reporting no stress (minimum 0) and others experiencing maximum stress (40), particularly in the Price of Electricity and Price of Gas scenarios.

*4.2.1 Stress Analysis Across Urban and Sub-urban Neighbourhoods*

**Figure 4: Average Perceived Stress Scale Under Various Scenarios (Across Urban and Sub-urban Households)**

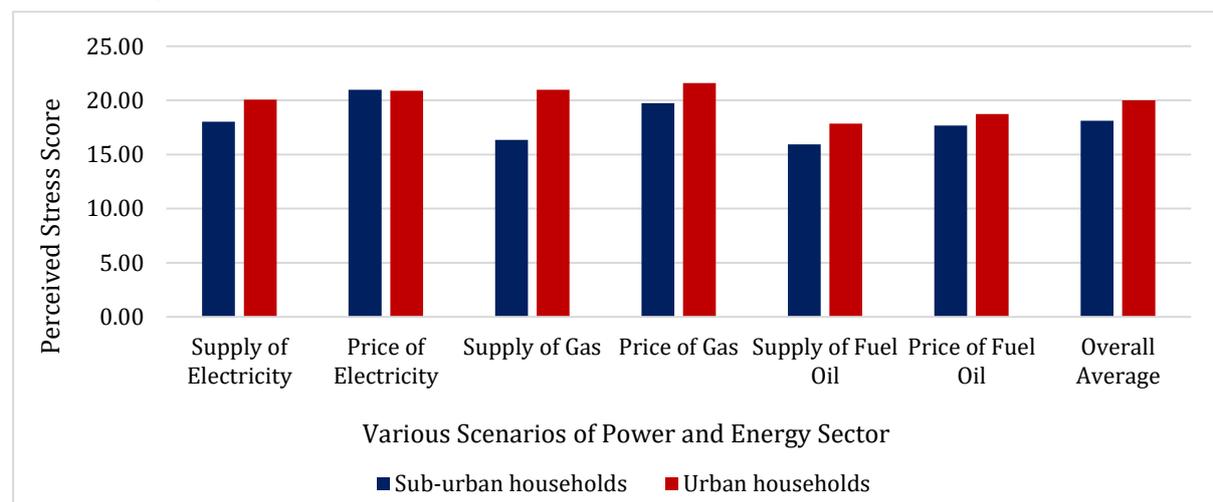

**Source:** Authors' Calculation.

Our calculation illustrated in Figure 4 reveals that, on an average, urban households experience higher stress levels than sub-urban ones across all power and energy scenarios in Bangladesh, with a notable disparity in gas supply and pricing scenarios, underscoring the need for government action on gas availability and affordability in urban areas in accordance with their demand. While stress levels concerning electricity pricing show minimal urban and sub-urban differences, indicating nationwide concern, the stress associated with both aspects fuel oil is comparatively lower than the other energies, highlighting different dimensions of stress in the energy sector and pointing to targeted areas for policy intervention.

*4.2.2 Stress Analysis Across Divisions*

To investigate potential regional disparities in perceived stress scores, the study has incorporated regional dynamics by considering various divisions and analysing the situation accordingly. Our calculation, summarised in Figure 5, showcases Dhaka with lower stress levels from electricity supply issues compared to Rangpur, Chattogram, and Khulna, where stress is markedly higher, aligning with BPDB's reports of frequent load-shedding, especially in Rangpur (Moazzem, et al., 2023).

**Figure 5: Average Perceived Stress Score Across Divisions: Six Different Scenarios**

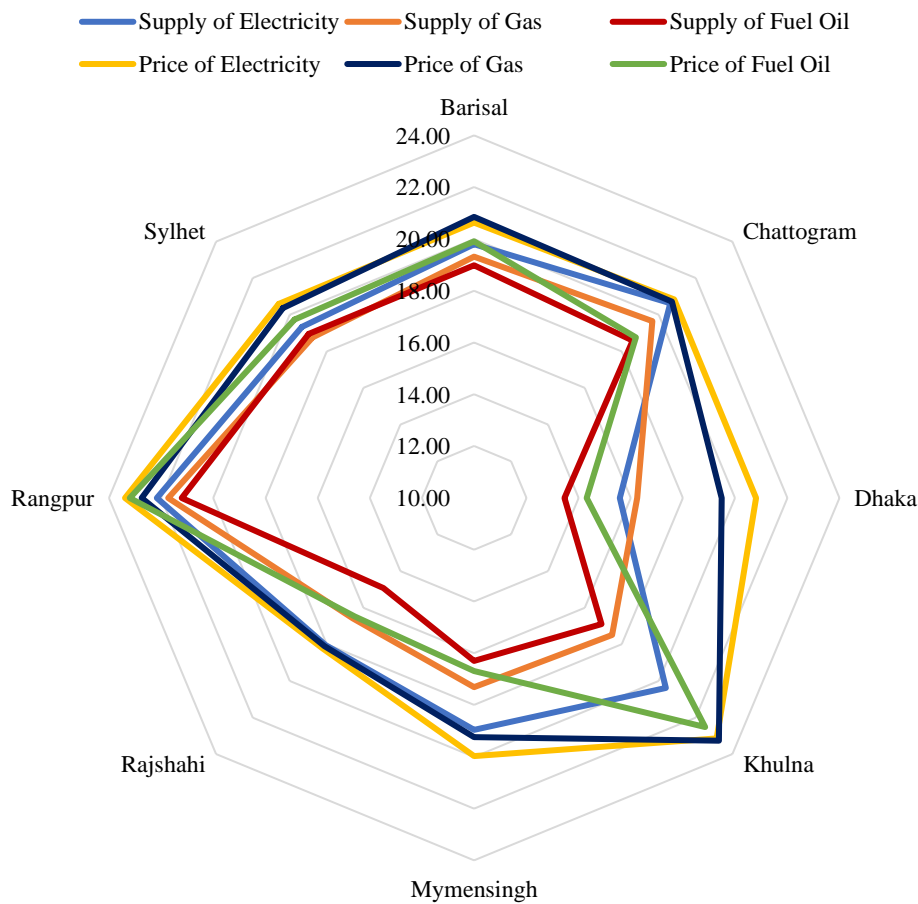

**Source:** Authors' Illustration.

It is evident from the analysis that there is more stress in Chattogram, Khulna, and Barisal households compared to that in Dhaka households. Rangpur and Khulna have the maximum levels of price stress of electricity, and Rajshahi households show the minimum level. In the case of supply of gas, Dhaka and Rajshahi show minimum levels of stress, either due to less severe issues, adequate distributors of LPG, or improved coping mechanisms, and Rangpur Division shows maximum levels of stress, reflecting severe supply issues or a lack of appropriate infrastructure. Gas price stress overrides each of the other indicated supply issues with each division having heavy stress over gas price levels, reflecting a general issue with economic effects. Rajshahi is the least stressed in the gas price narrative, reflecting improved affordability or insusceptibility to gas price effects, and Rangpur and Khulna are the worst hit, reflecting potential economic inequality regardless of centrally controlled gas price levels. In addition to gas price stress, Rangpur is highly stressed by fuel oil price levels. In total, stress from fuel oil price is lower than that linked with electricity and gas.

**Figure 6: Average Perceived Stress Scale Across Divisions: Overall Stress**

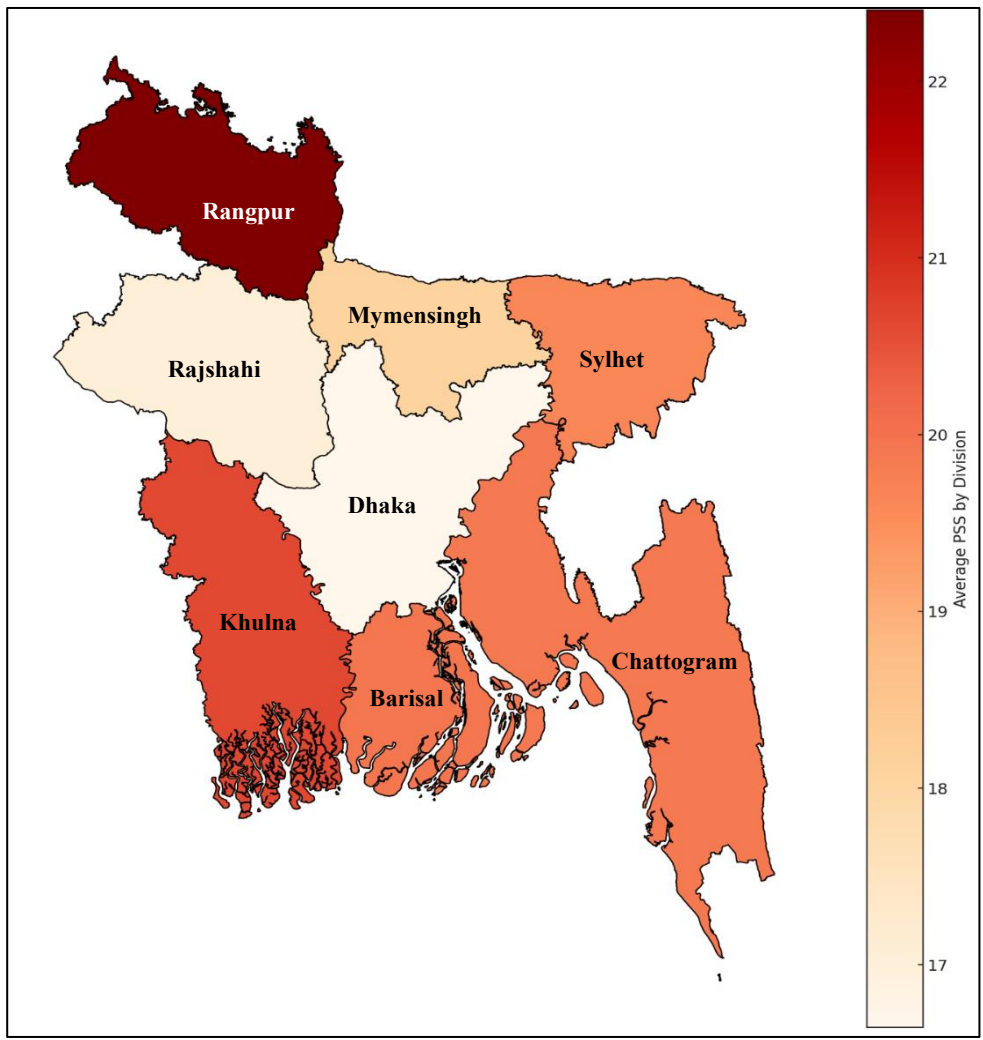

**Source**: Authors' Calculation.

Here, figure 6 highlights this regional disparity, signalling the North's critical need for support and highlighting the divisional priorities the government should establish to alleviate stress among the populace. In conclusion, Rangpur division is exposed to the maximum stresses in almost all the different scenarios, emphasising the demand to introduce immediate intervention. On the contrary, Dhaka shows the least stresses, implying superior quality supply, economic robustness, and resilience. This inequality indicates a centre bias in policy and operations that could be overlooking the regional demands.

### 4.3 Endogeneity Problem of Heat Index

It is not simple to draw inferences about the relationship between heat and stress due to the potentially high confounding effects from other factors, such as socio-economic status, infrastructure, and regional policies.

**Figure 7: Heat Index Across Divisions**

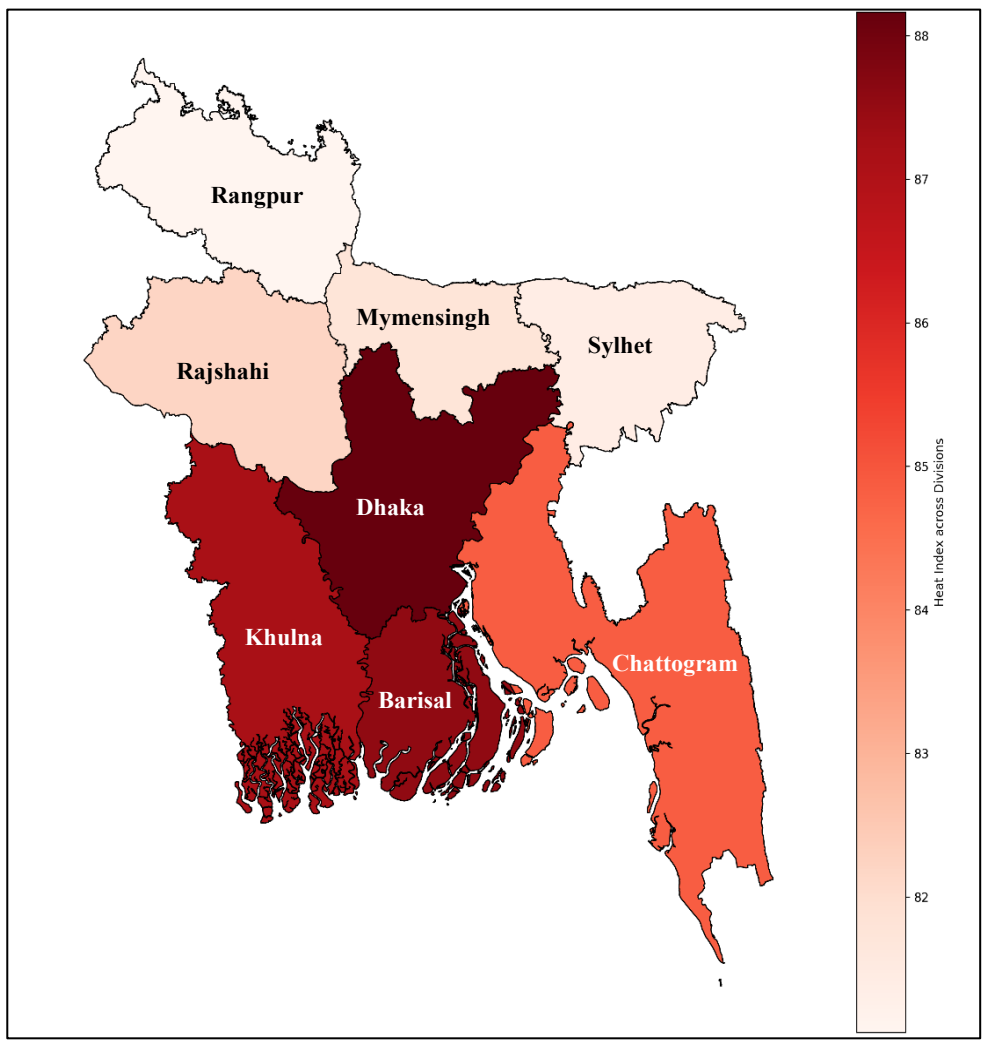

**Source**: Authors' Illustration.

As evident from Figure 6, stress levels rise from the centre (Dhaka) towards the outside. So, this implies that areas further from the capital have greater stress related to power and energy crisis. Figure 7, though, indicates an increase in the heat index when moving towards the centre from outside, with Dhaka having more heat than other areas.

The reversed relationship for heat and stress indicates the existence of confounding effects and reverse causality. In particular, with temperature expected to increase stress, data presented in Figures 6 and 7 indicate that where temperatures are higher, such as in Dhaka, stress is lower. This is because better infrastructure, more wealth, and greater adaptability in this area reduce the stress-inducing effects that may result from high temperature. In areas where temperature is lower but infrastructure less developed (like Rangpur or Khulna), stress is greater, suggesting that lower heat cannot always translate to lower stress. This interaction implies that stress is not independent of temperature, and other factors such as socio-economic status and regional development can distort the effect of temperature on stress. These confounding effects and this reverse causality where stress may have an effect on temperature to some degree introduce endogeneity bias, rendering an accurate estimation of actual causality difficult in a cross-sectional analysis.

This is a particular problem considering that our current analysis is conducted on cross-sectional data, under which it is more complicated to separate reverse causality and confounding effects. In case panel data were available, it would enable us to keep track over time on the same households, thus making it possible to examine how stress and heat develop over different intervals, considering seasonal changes in temperature as well as long-term socio-economic changes. This alternative method would also account for time-invariant characteristics at the level of a household such as wealth and infrastructure, which would otherwise act as confounders in a model based on a cross-section. However, panel data analysis is outside the scope of this analysis. So, considering that there is a problem with confounding as well as with reverse coefficients through a cross-sectional analysis, perhaps it would be better to exclude heat from our model based on a cross-section altogether.

### 4.4 Statistical Tests

Our analysis indicates the presence of heteroskedasticity, which can potentially affect the reliability. To address this issue, we have employed robust standard errors in our analysis. However, it is important to note that certain Likert variables representing environmental and political opinions, as shown in Appendix 2, do not follow a normal distribution.

### 4.5 Summary Results of the Political and Environmental Opinions

Table 2 presents the frequency distribution of political and environmental opinions.

**Table 2: Summary Results of Environmental and Political Opinions**

| Statement Variables | Fully Disagreed | Roughly Disagreed | Neutral | Roughly Agreed | Fully Agreed |
|---|---|---|---|---|---|
| Environmental pollution has adverse effects on weather and climate change (e_1) | 2.1% | 6.5% | 13.3% | 24.6% | 53.5% |

| | | | | | | |
|---|---|---|---|---|---|---|
| International organisations are mainly responsible for fighting climate change (e_2) | 3.5% | 8.8% | 20.9% | 36.9% | 29.9% |
| Government and relevant authorities are mainly responsible for fighting climate change (e_3) | 3.6% | 8.3% | 22.5% | 24.6% | 41% |
| Without nuclear energy, energy crisis will not resolve (e_4) | 3.9% | 9.6% | 30.8% | 26.5% | 29.2% |
| Without renewable energy, energy crisis will not resolve (e_5) | 3.4% | 15.1% | 33% | 22.1% | 26.4% |
| Bangladesh is ready for nuclear energy production (e_6) | 8.7% | 10.7% | 36.4% | 28.2% | 16% |
| The media is overexaggerating the crisis (e_7) | 5.6% | 13.1% | 42.5% | 22.1% | 16.7% |
| The respondent has no problem in establishing a nuclear power plant in the neighbourhood (e_8) | 9.3% | 10.5% | 41.7% | 23.5% | 15% |
| The household is prepared to reduce power and energy consumption to save environment and climate (e_9) | 4% | 12.1% | 27.4% | 28% | 28.5% |

**Source:** Authors' Calculation.

## 4.6 Result and Discussion from Multivariate OLS Model

The result of the regression analysis from multivariate OLS is tabulated in Table 3.

**Table 3: Summary Results of MOLS Regression Analysis**

| VARIABLES | (1) PSS: Elc. Supply | (2) PSS: Elc. Price | (3) PSS: Gas Supply | (4) PSS: Gas Price | (5) PSS: Fuel Supply | (6) PSS: Fuel Price | (7) Price: Overall |
|---|---|---|---|---|---|---|---|
| Urban (sub-Urban = 0) | 2.67*** | 0.91** | 5.17*** | 2.57*** | 3.03*** | 2.22*** | 2.81*** |
| | (0.36) | (0.41) | (0.40) | (0.41) | (0.33) | (0.37) | (0.27) |
| Division (Base: Dhaka) | | | | | | | |
|   Barisal | 5.08*** | 0.77 | 3.91*** | 1.87** | 5.9*** | 5.03*** | 3.67*** |
| | (0.73) | (0.81) | (0.70) | (0.83) | (0.76) | (0.72) | (0.52) |
|   Chattogram | 4.38*** | 0.003 | 2.37*** | 0.60 | 4.22*** | 3.59*** | 2.52*** |
| | (0.5) | (0.59) | (0.51) | (0.54) | (0.4) | (0.46) | (0.34) |
|   Khulna | 5.19*** | 3.13*** | 1.84*** | 3.49*** | 2.86*** | 7.48*** | 3.96*** |
| | (0.9) | (0.98) | (0.69) | (0.97) | (0.61) | (0.96) | (0.67) |
|   Mymensingh | 1.72*** | -1.14* | -1.0 | -1.64** | 0.47 | 0.08 | -0.26 |
| | (0.57) | (0.61) | (0.62) | (0.66) | (0.5) | (0.59) | (0.45) |
|   Rajshahi | 1.02 | -3.47*** | -1.18 | -2.86*** | -0.66 | -0.66 | -1.30** |
| | (0.69) | (0.81) | (0.75) | (0.71) | (0.68) | (0.76) | (0.56) |
|   Rangpur | 7.33*** | 3.11*** | 6.36*** | 3.12*** | 6.84*** | 7.0*** | 5.75*** |
| | (0.64) | (0.78) | (0.59) | (0.66) | (0.58) | (0.72) | (0.48) |
|   Sylhet | 4.95*** | 0.86 | 4.7*** | 2.17** | 6.54*** | 6.01*** | 4.18*** |
| | (0.76) | (0.90) | (0.75) | (0.84) | (0.71) | (0.81) | (0.63) |
| Sex | -1.16 | -1.24 | -2.07** | -1.21 | -1.22 | -1.87* | -1.5 |
| | (1.16) | (1.37) | (0.96) | (1.35) | (0.86) | (1.02) | (0.93) |
| Age | -0.025 | -0.02 | -0.04** | -0.01 | -0.06*** | -0.05*** | -0.03*** |
| | (0.02) | (0.02) | (0.02) | (0.02) | (0.014) | (0.02) | (0.01) |
| Years of Education | -0.08** | -0.01 | -0.16*** | -0.04 | -0.08** | 0.05 | -0.06** |
| | (0.03) | (0.04) | (0.04) | (0.04) | (0.03) | (0.04) | (0.03) |
| Household Size | 0.09 | 0.32** | -0.13 | 0.01 | -0.01 | -0.04 | 0.11 |

|  | (0.14) | (0.15) | (0.19) | (0.19) | (0.11) | (0.14) | (0.12) |
|---|---|---|---|---|---|---|---|
| Number of Students | 0.07 | -0.32 | -0.09 | -0.10 | 0.15 | -0.13 | -0.07 |
|  | (0.22) | (0.25) | (0.21) | (0.23) | (0.17) | (0.22) | (0.16) |
| Income Source (Base: Agri) | | | | | | | |
| Service | -0.94** | -1.23** | -1.05** | -1.2** | -2.25*** | -2.11*** | -1.45*** |
|  | (0.42) | (0.51) | (0.44) | (0.50) | (0.4) | (0.44) | (0.34) |
| Industry | 0.14 | -0.03 | 0.70 | 0.12 | -0.82 | -1.14* | -0.16 |
|  | (0.62) | (0.70) | (0.68) | (0.65) | (0.62) | (0.67) | (0.48) |
| Income Group (Base: Below BDT 10000) | | | | | | | |
| BDT 10000 to 24999 | -0.66 | 1.63 | -2.09* | 0.46 | -1.33 | 0.38 | -0.27 |
|  | (1.39) | (1.47) | (1.16) | (1.44) | (0.95) | (1.32) | (1.09) |
| BDT 25000 to 39999 | -0.16 | 1.88 | -1.63 | 0.81 | -0.87 | 0.78 | 0.16 |
|  | (1.41) | (1.48) | (1.19) | (1.47) | (0.97) | (1.34) | (1.11) |
| BDT 40000 to 79999 | -1.81 | 0.33 | -1.39 | -0.54 | -1.72 | -0.57 | -0.95 |
|  | (1.52) | (1.66) | (1.36) | (1.60) | (1.08) | (1.45) | (1.18) |
| Above 80000 | -1.95 | 2.80 | -4.95*** | -2.19 | -1.73 | -0.18 | -1.36 |
|  | (2.44) | (2.85) | (1.74) | (2.64) | (1.74) | (2.05) | (1.92) |
| Availability of Personal Vehicle (1 = Yes) | 0.82** | 0.22 | 1.31*** | 0.86** | 2.92*** | 3.06*** | 1.36*** |
|  | (0.40) | (0.45) | (0.40) | (0.42) | (0.40) | (0.49) | (0.31) |
| Consciousness Variables | | | | | | | |
| e_1 | -0.62*** | -0.41** | -0.77*** | -0.35** | -0.55*** | -0.47** | -0.53*** |
|  | (0.16) | (0.18) | (0.17) | (0.18) | (0.17) | (0.18) | (0.13) |
| e_2 | -0.22 | 0.68*** | -0.11 | 0.39* | -0.38** | -0.07 | 0.05 |
|  | (0.19) | (0.23) | (0.19) | (0.21) | (0.17) | (0.2) | (0.16) |
| e_3 | 0.02 | 0.17 | -0.43** | -0.25 | -0.35** | -0.55*** | -0.24* |
|  | (0.17) | (0.19) | (0.18) | (0.18) | (0.16) | (0.17) | (0.13) |
| e_4 | 0.15 | 0.76*** | 0.02 | -0.37 | 0.13 | -0.21 | 0.08 |
|  | (0.2) | (0.26) | (0.21) | (0.24) | (0.19) | (0.23) | (0.16) |
| e_5 | 1.12*** | 0.07 | 0.83*** | 0.86*** | 0.43*** | 1.08*** | 0.73*** |
|  | (0.19) | (0.23) | (0.18) | (0.20) | (0.16) | (0.21) | (0.14) |
| e_6 | 0.52*** | -0.32 | 0.93*** | 0.48** | 1.23*** | 0.79*** | 0.61*** |
|  | (0.18) | (0.22) | (0.19) | (0.19) | (0.17) | (0.20) | (0.14) |
| e_7 | 0.77*** | 0.68*** | 1.04*** | 0.79*** | 1.13*** | 1.0*** | 0.92*** |
|  | (0.20) | (0.22) | (0.19) | (0.20) | (0.17) | (0.19) | (0.15) |
| e_8 | -0.60*** | -0.43** | -0.19 | -0.36* | -0.39** | -0.29* | -0.38*** |
|  | (0.19) | (0.21) | (0.19) | (0.19) | (0.16) | (0.17) | (0.14) |
| e_9 | -0.62*** | -0.84*** | -0.88*** | -0.55*** | -0.78*** | -0.46*** | -0.69*** |
|  | (0.18) | (0.21) | (0.19) | (0.19) | (0.16) | (0.17) | (0.13) |
| Electricity Consumption pc | 0.003*** | 0.002 | | | | | |
|  | (0.001) | (0.001) | | | | | |
| Gas Consumption pc | | | -0.003 | 0.001 | | | |
|  | | | (0.002) | (0.002) | | | |
| Oil Consumption pc | | | | | -0.002 | -0.002 | |
|  | | | | | (0.001) | (0.002) | |
| Power and Energy Consumption pc | | | | | | | 0.001 |
|  | | | | | | | (0.001) |
| Observations | 1,000 | 1,000 | 1,000 | 1,000 | 1,000 | 1,000 | 1,000 |
| R-squared | 0.309 | 0.145 | 0.400 | 0.196 | 0.481 | 0.397 | 0.404 |

**Note**: The parentheses indicate robust errors in parentheses (*** p<0.01, ** p<0.05, * p<0.1). The constant terms are dropped from the model although the terms are included in regression models.
**Source:** Authors' Calculation.

Table 3 shows that, in comparison to sub-urban households, urban households are consistently associated with higher levels of stress across a range of power and energy scenarios. When compared to the baseline division - Dhaka, the analysis of stress levels across various power and energy scenarios in Bangladesh shows consistent regional disparities. Rajshahi is the sole exception when it comes to gas supply and price scenarios. The findings support our previous analysis.

Sex of the household head is not significantly or consistently associated with stress levels, possibly due to lower sample size of female-led households. It would be wiser to not comment any conclusive remark regarding the dimension of sex of the household head. Conversely, when controlling for other factors, the age of the household head appears to be negatively associated with stress levels, suggesting that older individuals may experience lower stress in response to energy supply challenges. This might be explained by improved coping mechanisms or increased resilience of the older people (Aldwin, Yancura, & Lee, 2021; Kurth, Igarashi, & Aldwin, 2024). In particular, because they are more likely to have other working family members, households with older household heads may have access to more resources, which could improve their financial security and enable them to have greater access to energy. This result may also reflect accumulated life experience, adaptability, and the development of well-established coping mechanisms over time (Aldwin, Yancura, & Lee, 2021). However, it is important to note that no significant association is observed between age and stress attributed to electricity supply, prices and gas prices (specification 1, 2 and 4 respectively). Additionally, older heads of households, after their retirements, are more likely than younger ones to spend more time at home (Ornstein, et al., 2020; Lawton, Moss, & Fulcomer, 1987), which could expose them to the more immediate effects of energy shortages (Yagita & Iwafune, 2021). The younger household heads are likely to be involved in employment and so, spend more time outside of home. It may be wise for policymakers to concentrate interventions on areas primarily occupied by households with younger heads, who may have a harder time managing stress related to energy, as the stress level attributed to the overall scenario of power and energy crisis has an inverse relationship with age (specification 7). However, it's important to recognise that older individuals may face health challenges that are closely tied to their energy consumption. These challenges can make it even harder for them to cope with energy shortages, as their daily lives often depend on consistent access to energy (Yagita & Iwafune, 2021; Chathuranga, Rajapaksha, Sajjad, & Siriwardana, 2024).

Years of education of household heads show a significant negative relationship with stress levels in various power and energy supply scenarios, holding other variables constant. It implies the potential importance of awareness and knowledge in effectively managing energy-related stress. Additionally, higher educational attainment might be linked to improved access to resources, analytical ability, or information that can mitigate the adverse impact of power and energy challenges. However, the variable does not show any statistical significance in any of the price scenario cases. It emphasises the widespread stress within the households due to the price regardless of their awareness or understanding.

Household size and the presence of students within household do not markedly have any association with stress levels. This aspect of the analysis suggests that while household size might affect others by the lifestyle of a household, the composition of the household does not distinctly sway perceptions or experiences of stress concerning power and energy issues. However, household size has a positive significant relationship with stress level associated with electricity price (specification 2).

Holding all other things constant, households' primary income sources have significant relationship with their stress levels regarding power and energy, with service sector employment linked to lower stress compared to agriculture or industry. This pattern reflects differences in the stability structure of income generation across sectors. Households engaged in service-related employment may experience relatively stable income flows, and greater flexibility in work arrangements. It also leads them to less direct exposure to production disruptions caused by energy shortages (Schettkat & Yocarini, 2006; Rani, Wang, Awad, & Zhao, 2023; Diamond, 1962). For industries and agriculture, consistent and reliable energy is often crucial for keeping production going smoothly (Djanibekov & Gaur, 2018). The stress caused by fluctuations in energy supply is heightened by the nature of work in these sectors, which usually involves greater physical exposure to environmental conditions and a heavy dependence on external inputs like fuel and electricity. These differences between sectors help explain why stress levels related to power and energy disruptions are often closely tied to the type of income a household relies on.

With a baseline income of less than BDT 10,000, income levels do not clearly show a pattern of significant associations with stress levels related to power and energy scenarios. On the other hand, higher-income households, show a noticeable decrease in stress related to gas supply, holding all other variables constant, especially those who are earning over BDT 80,000. It is possibly due to the fact that these higher-income households often reside in affluent areas with better infrastructure and resource allocation, ensuring more consistent energy supplies and possibly prioritising them in supply distribution, thereby reducing their susceptibility to general supply disruptions. An intriguing observation reveals that stress levels linked to the price of electricity, gas, and fuel oil remain consistent across all income brackets, indicating a widespread and uniform concern over power and energy expenses among households, regardless of their economic standing.

Vehicle ownership within households correlates with increased stress, reflecting the additional energy demands and heightened awareness of energy costs and supply issues such as availability, reliance on consistent energy supplies for transportation needs. The presence of vehicles in a household may also reflect a lifestyle with greater dependence on energy resources, thereby making these households more sensitive to disruptions or changes in the energy sector and policies must be designed by keeping this factor in account as well.

Our analysis of environmental and political opinion variables ('e_1' through 'e_9') leads to a complex relationship between personal beliefs, and stress levels attributed to energy-related situations. Households that recognise the link between environmental pollution and climate change ('e_1') tend to experience lower stress levels. This is likely because of their increased awareness of the environmental impact of energy and power crises. The increased awareness helps them better cope with these challenges. Individuals who attribute primary responsibility for climate action to the government ('e_3') or international organisations ('e_2') exhibit varying degrees of stress, which may be a reflection of how they view local versus global agency in tackling environmental and energy issues. Families that feel that international organisations should handle climate change are more stressed about electricity prices, probably because they think that the uncertainties lied in electricity pricing are mostly due to international factors. Families assigning primary responsibility for climate action to the government exhibit lower stress levels in fuel oil pricing and supply scenarios, as well as gas supply issues. This pattern suggests that expectations of governmental efficacy in environmental management can significantly shape stress responses to power and energy challenges. Hence, enhancing government transparency, responsibility, and practicality in policy formulation and execution becomes imperative.

Beliefs in the indispensability of nuclear ('e_4') and renewable energy ('e_5') for resolving the energy crisis similarly influence stress levels, highlighting the role of individual views in ensuring energy sustainability and security in stress experiences. Households, viewing nuclear energy as key to solving the energy crisis, report higher stress regarding electricity pricing, underscoring their anticipation of pricing stability through increased electricity supply. Households prioritising renewable energy may experience increased stress in almost every scenario. This could be due to concerns about the current energy mix's sustainability and the urgency of transitioning to greener energy sources. Households believing Bangladesh is ready for nuclear energy ('e_6') face higher stress across most energy and power crisis scenarios, suggesting an urgent desire for transitioning to stable, alternative energy sources and reflecting their perception of nuclear energy as a feasible and safe option.

Scepticism towards media portrayal of the energy crisis ('e_7') and willingness to accept nuclear power plants nearby ('e_8') or to reduce energy and power consumption for environmental reasons ('e_9') also correlate with stress levels, indicating how personal proximity to energy solutions, media perceptions, and environmental consciousness shape stress responses. Our analysis shows that scepticism towards media reports often increases stress, likely due to feelings of being misinformed about energy issues. Conversely, households open to nearby nuclear facilities generally experience less stress in various crisis scenarios, indicating a broader acceptance of nuclear energy as a feasible solution. Additionally, households willing to cut energy consumption for environmental reasons report lower stress levels, suggesting that a strong environmental consciousness contributes to a more resilient attitude towards energy crises. Our outcomes further demonstrate that higher environmental consciousness correlates with lower stress levels across all scenarios, underscoring the value of promoting environmental education at the household level to alleviate stress and enhance energy conservation in Bangladesh's power sector, which is consistent with previous literature in Bangladesh (Moazzem & Quaiyyum, 2024).

## 4.7 Results and Discussions from Simultaneous-Quantile Regressions

The results for the quantile regressions in the case of assessing stress level subject to overall power and energy crisis is given in Table 4. Despite the differences captured across quantiles, there are notable similarities between the MOLS and SQR results in this case.

**Table 4: Results from Quantile Regression: PSS – Overall Price and Energy Crisis Scenario**

| VARIABLES | (1) 15th Quantile | (2) 30th Quantile | (3) 50th Quantile | (4) 70th Quantile | (5) 85th Quantile |
|---|---|---|---|---|---|
| Urban (sub-Urban = 0) | 3.280*** | 2.730*** | 1.684*** | 1.093*** | 0.993*** |
|  | (0.409) | (0.367) | (0.323) | (0.228) | (0.247) |
| Division (Base: Dhaka) |  |  |  |  |  |
|   Barisal | 5.595*** | 3.515*** | 1.833*** | 1.153*** | 0.662 |
|  | (0.838) | (0.723) | (0.537) | (0.433) | (0.467) |
|   Chattogram | 4.546*** | 2.653*** | 1.447*** | 0.897*** | 0.637** |
|  | (0.639) | (0.460) | (0.298) | (0.263) | (0.286) |
|   Khulna | 1.910** | 1.648* | 2.244*** | 3.261*** | 6.802*** |
|  | (0.899) | (0.842) | (0.659) | (0.633) | (1.206) |
|   Mymensingh | 0.0984 | -0.872 | -1.056** | -0.993** | -0.604 |
|  | (0.871) | (0.831) | (0.499) | (0.427) | (0.401) |
|   Rajshahi | -2.179** | -3.190*** | 0.507 | 1.580*** | 1.459*** |
|  | (0.993) | (0.887) | (1.108) | (0.426) | (0.412) |
|   Rangpur | 6.798*** | 5.623*** | 4.559*** | 4.753*** | 5.055*** |
|  | (0.863) | (0.737) | (0.572) | (0.515) | (0.511) |
|   Sylhet | 5.587*** | 3.607*** | 2.816*** | 2.616*** | 2.169*** |
|  | (0.937) | (0.700) | (0.505) | (0.415) | (0.567) |
| Sex | -2.074 | -0.957 | -0.225 | -0.0309 | -0.120 |
|  | (2.370) | (1.338) | (0.847) | (0.513) | (0.543) |
| Age | -0.0226 | -0.0331** | -0.0273** | -0.00629 | -0.0122 |
|  | (0.0223) | (0.0163) | (0.0130) | (0.00895) | (0.0106) |
| Years of Education | -0.124*** | -0.106*** | -0.0564* | -0.0128 | -0.000290 |
|  | (0.0409) | (0.0357) | (0.0298) | (0.0224) | (0.0235) |
| Household Size | 0.0592 | -0.0702 | 0.0330 | 0.0272 | 0.0396 |
|  | (0.163) | (0.146) | (0.104) | (0.0885) | (0.0845) |
| Number of Students | -0.120 | 0.0612 | 0.0920 | -0.0639 | 0.00346 |
|  | (0.226) | (0.203) | (0.143) | (0.141) | (0.152) |
| Income Source (Base: Agri) |  |  |  |  |  |
|   Service | -1.459*** | -1.383*** | -0.997*** | -0.290 | -0.345 |
|  | (0.509) | (0.431) | (0.342) | (0.239) | (0.252) |
|   Industry | -0.661 | -0.670 | -0.135 | 0.211 | 0.0395 |
|  | (0.738) | (0.628) | (0.559) | (0.321) | (0.366) |
| Income Group (Base: Below BDT 10000) |  |  |  |  |  |
|   BDT 10000 to 24999 | -0.933 | -0.517 | -0.0716 | 0.236 | 0.756 |
|  | (1.696) | (1.139) | (0.874) | (0.816) | (1.618) |
|   BDT 25000 to 39999 | -0.355 | -0.0984 | 0.257 | 0.286 | 0.875 |
|  | (1.736) | (1.202) | (0.963) | (0.853) | (1.656) |
|   BDT 40000 to 79999 | -1.044 | -0.788 | -0.322 | 0.223 | 0.830 |
|  | (1.822) | (1.305) | (1.060) | (0.887) | (1.617) |
|   Above 80000 | -0.543 | 0.166 | -0.862 | -0.112 | 0.400 |
|  | (3.783) | (3.199) | (1.823) | (1.538) | (1.799) |
| Availability of Personal Vehicle (1 = Yes) | 0.948** | 1.116*** | 0.714** | 0.381 | 0.261 |
|  | (0.445) | (0.428) | (0.362) | (0.300) | (0.272) |
| Consciousness Variables |  |  |  |  |  |
|   e_1 | -0.856*** | -0.714*** | -0.297** | -0.371*** | -0.227** |
|  | (0.209) | (0.176) | (0.131) | (0.125) | (0.114) |
|   e_2 | -0.186 | -0.234 | -0.0415 | 0.0207 | -0.0622 |
|  | (0.216) | (0.168) | (0.147) | (0.149) | (0.128) |
|   e_3 | 0.0136 | -0.0191 | -0.313*** | -0.253** | -0.242** |
|  | (0.183) | (0.201) | (0.119) | (0.113) | (0.108) |
| e_4 | -0.00872 | 0.0515 | -0.131 | 0.0891 | 0.122 |
|  | (0.184) | (0.185) | (0.151) | (0.118) | (0.114) |
| e_5 | -0.0514 | 0.434** | 0.608*** | 0.668*** | 0.706*** |
|  | (0.209) | (0.178) | (0.139) | (0.127) | (0.121) |
| e_6 | 0.588*** | 0.968*** | 0.921*** | 0.513*** | 0.398*** |
|  | (0.189) | (0.189) | (0.150) | (0.134) | (0.118) |
| e_7 | 1.212*** | 1.188*** | 0.669*** | 0.262** | 0.0429 |
|  | (0.235) | (0.211) | (0.140) | (0.114) | (0.114) |
| e_8 | -0.362* | -0.370** | -0.349** | -0.178* | -0.288*** |
|  | (0.210) | (0.173) | (0.143) | (0.0985) | (0.107) |
| e_9 | -0.851*** | -0.563*** | -0.165 | -0.104 | 0.105 |
|  | (0.204) | (0.194) | (0.129) | (0.101) | (0.114) |
| Power and Energy Consumption pc | 0.000668 | 0.000449 | 0.000431 | 0.000107 | -0.000244 |
|  | (0.000855) | (0.000734) | (0.000519) | (0.000370) | (0.000401) |
| Observations | 1,000 | 1,000 | 1,000 | 1,000 | 1,000 |

**Note**: The parentheses indicate bootstrapped errors in parentheses (*** p<0.01, ** p<0.05, * p<0.1). The constant terms are dropped from the model although the terms are included in regression models.
**Source:** Authors' Calculation.

Table 4 shows that urban households consistently experience higher stress across all quantiles, with coefficients decreasing from 3.280 at the 15th quantile to 0.993 at the 85th quantile. This indicates that while urban households feel a considerable amount of stress from the overall energy and price crises, the intensity of this stress decreases as the perceived stress level increases, possibly due to adaptive coping mechanisms or greater access to resources in urban areas.

Regional variations, similar to those observed in previous models, remain significant, with Barisal and Chattogram showing relatively high stress at the lower quantiles (Barisal: 5.595 at the 15th quantile, Chattogram: 4.546 at the 15th quantile), but this impact weakens as stress levels rise. For example, Barisal's stress impact significantly declines across quantiles, reaching 0.662 at the 85th quantile, and Chattogram's effect similarly drops from 4.546 at the 15th quantile to 0.637 at the 85th quantile. These results suggest that while these divisions face notable stress at lower levels, their ability to cope with increasing stress is reflected in the diminishing coefficients at higher quantiles.

Khulna division stands out as a key exception, with the stress coefficients steadily increasing as the quantiles rise (coefficient: 1.910 at the 15th quantile to 6.802 at the 85th quantile), indicating that Khulna households experience growing stress levels even as stress increases. This may be reflective of a unique vulnerability in Khulna, where the cumulative effect of energy and price crises intensifies at higher levels of stress.

Age continues to show a consistent negative association with stress, with older household heads experiencing lower stress across all quantiles. The negative coefficient ranges from -0.0226 at the 15th quantile to -0.0122 at the 85th quantile, suggesting that older individuals are better equipped to handle the impacts of energy crises. The effect, however, weakens at higher stress levels, which could indicate that while older individuals may experience less stress under normal circumstances, their coping capacity might be overstretched as stress escalates.

Education also plays a crucial role in reducing stress, with higher education levels associated with lower stress at lower to moderate quantiles. The coefficients range from -0.124 at the 15th quantile to -0.000290 at the 85th quantile, reinforcing the notion that more educated individuals have better access to resources and coping mechanisms, helping them manage stress from the energy crisis. The diminishing effect at the 85th quantile suggests that education may have limited impact on alleviating stress when it becomes extreme.

The income source variable shows consistent results in line with prior analyses, with service sector employment linked to lower stress levels compared to agriculture, especially at lower quantiles. The negative coefficients for service sector employment across the 15th, 30th, and 50th quantiles (-1.459, -1.383, and -0.997, respectively) indicate that households in the service sector experience less stress due to the stability of their income, contrasting with the more vulnerable agricultural sector.

The income group variable also exhibits varied patterns across quantiles. Higher-income groups, particularly those earning above BDT 80,000, show no significant stress reduction at the higher quantiles, with coefficients oscillating between -0.543 and 0.400. This suggests that wealthier households may be less affected by the energy crisis at lower stress levels but do not experience considerable stress alleviation when the stress becomes more extreme.

Vehicle ownership continues to correlate with higher stress at lower quantiles, with a significant positive relationship at the 15th and 30th quantiles (coefficients: 0.948 and 1.116, respectively), reinforcing the MOLS finding that households dependent on energy-intensive assets like vehicles are more vulnerable to energy supply disruptions.

Environmental consciousness variables show similar trends to those in previous scenarios. Belief in the impact of environmental pollution ($e\_1$) significantly reduces stress at lower quantiles (coefficients: -0.856 at the 15th quantile to -0.227 at the 85th quantile), indicating that environmental awareness helps households manage stress. Conversely, belief in renewable energy ($e\_5$) shows a positive relationship with stress, especially at higher quantiles (coefficients: 0.434 at the 30th quantile to 0.706 at the 85th quantile), suggesting that households prioritising renewable energy may feel heightened stress due to concerns over the urgency of the energy crisis. Scepticism towards media portrayals of the energy crisis ($e\_7$) continues to increase stress at lower quantiles (coefficient: 1.212 at the 15th quantile), but the effect diminishes at higher stress levels, supporting the earlier findings that media scepticism exacerbates stress at lower levels but becomes less impactful when stress escalates. Similarly, willingness to reduce energy consumption for environmental reasons ($e\_9$) reduces stress at lower quantiles but shows diminishing returns at higher stress levels, aligning with the overall finding that environmental consciousness can alleviate moderate stress but is less effective when stress becomes extreme.

In this part, we integrate the insights from the scenario-specific quantile regression (SQR) results (as seen in Tables G.1 to G.6 from Appendix G) with the overall PSS quantile regression findings presented in Table 4, underscoring the necessity of incorporating scenario-specific approaches for a more nuanced understanding of stress levels in response to varying energy crises. Although both approaches share broad trends, the scenario-specific models reveal distinct patterns that enhance the robustness and depth of our findings, ultimately supporting the case for their inclusion.

One of the key differences that emerges is the varying impact of urban versus sub-urban households across different energy scenarios. While the overall PSS analysis (Table 4) identifies consistently higher stress levels for urban households, the scenario-specific regressions highlight how this urban stress premium behaves differently in specific contexts. For instance, in the Gas Price scenario (Table G.4), the impact on urban households is most pronounced at lower stress levels but gradually diminishes at higher quantiles, suggesting that urban households experience acute stress at moderate levels, but are somewhat insulated from extreme stress. Similarly, urban households show a decrease in stress intensity in the Electricity Supply and Fuel Price scenarios. These findings underscore the nuanced nature of urban household stress, which can vary depending on the specific energy crisis.

Moreover, the scenario-specific regressions provide a more detailed look at regional disparities. While the overall PSS model offers general regional patterns, the scenario-specific models reveal complexities that are otherwise obscured. For example, Barisal in the Gas Supply scenario (Table G.3) shows a significant positive effect at lower quantiles (5.633 at the 15th quantile), but this effect weakens as stress levels rise, even reversing to a negative association at the 85th quantile. This regional variation is mirrored across different scenarios, such as the Fuel Supply and Gas Price regressions, which show how different regions react differently depending on the specific nature of the crisis. These differences suggest that regional stress factors are highly contextual and deserve more focused attention in policy design.

A notable insight from the scenario-specific approach is the concept of the 'middle-income squeeze', which becomes particularly apparent in the Fuel Price scenario (Table G.6). Here, households in the middle-income bracket (BDT 25,000–79,999) exhibit increasing stress at higher quantiles, reflecting how the rising cost of fuel strains these households, who are caught between the lower-income groups' need to limit consumption and the financial resilience of higher-income groups. This finding highlights the vulnerability of middle-income households to energy price volatility, an insight that is not fully captured by the overall PSS analysis. By identifying this 'squeeze', the scenario-specific approach emphasises the importance of recognising the financial pressures faced by this group, which are exacerbated by both rising energy costs and lifestyle expectations tied to energy consumption.

The scenario-specific models also provide a more nuanced understanding of how environmental and political attitudes influence stress levels. For instance, belief in renewable energy ($e\_5$) shows a significant positive relationship with stress at higher quantiles in the Gas Price scenario (Table G.4), suggesting that households prioritising renewable energy may feel more stressed due to the perceived urgency of addressing energy crises. This contrasts with the more general environmental consciousness variables in the overall PSS model, where the effects are generally negative at lower quantiles but diminish as stress levels rise. These findings point to the complex interplay between environmental attitudes and energy stress, highlighting that the perception of urgency or dissatisfaction with the pace of energy transitions can contribute to heightened stress at higher levels of distress.

Finally, the scenario-specific quantile regressions also reveal the limitations of the overall PSS model, especially in terms of income and vehicle ownership. The Fuel Price scenario specifically highlights the rising stress among middle-income households, which is not as evident in the PSS model. Additionally, vehicle ownership, which consistently correlates with higher stress in the scenario-specific regressions (especially in the Gas and Fuel Supply scenarios), shows a more nuanced impact that is valuable in understanding how energy reliance on assets like vehicles exacerbates stress at different levels.

In conclusion, while the overall PSS quantile regression provides a broad view of how stress varies across quantiles, the scenario-specific regressions offer deeper insights into the dynamics of energy-related stress across different contexts. The scenario-specific approach highlights regional differences, identifies the middle-income squeeze phenomenon, and reveals the more complex interactions between household characteristics, environmental attitudes, and stress levels. These insights collectively demonstrate why it is crucial to adopt scenario-specific models alongside the overall PSS approach, offering a more granular and comprehensive understanding of the factors influencing stress during energy crises.

### 4.8 Results and Discussions from Random Forest Model

The table 5 presents the $R^2$ values and MSE (mean squared error) associated with the fitted values obtained from the Random Forest model:

**Table 5: $R^2$ Values and MSE from Random Forest Models**

| Dependent Variables | $R^2$ Values | MSE |
|---|---|---|
| PSS: Electricity Supply | 0.47 | 23.25 |
| PSS: Electricity Price | 0.423 | 24.14 |
| PSS: Gas Supply | 0.60 | 17.85 |
| PSS: Gas Supply | 0.33 | 28.67 |
| PSS: Fuel Oil Supply | 0.63 | 13.26 |
| PSS: Fuel Oil Price | 0.62 | 18.06 |
| PSS | 0.67 | 8.79 |

**Source:** Authors' Calculation.

The analysis of Random Forest models applied to stress levels across various power and energy crisis scenarios reveals a high effectiveness in explaining the variance in stress, with $R^2$ values indicating a substantial explanation of the variance and generally moderate to low MSE values highlighting the models' accuracy. Overall, these models exhibit strong predictive power, effectively capturing the underlying factors influencing perceived stress in different energy and power crisis contexts. Despite their robust performance, a notable portion of variance in stress levels remains unexplained, suggesting room for further model refinement and investigation into additional explanatory variables.

**Table 6: Top 10 Key Features Influencing Stress Scores in Various Scenarios (From Most Important to Less Important)[7]**

| Rank | Electricity Supply | Electricity Price | Gas Supply | Gas Price | Fuel Oil Supply | Fuel Oil Price | PSS |
|---|---|---|---|---|---|---|---|
| 1 | Division (0.176) | Division (0.116) | Division (0.16) | Division (0.11) | Division (0.2) | Division (0.22) | Division (0.19) |
| 2 | Electricity Cons. (0.072) | Electricity Con. (0.104) | Urban (0.11) | Ready for Nuclear (0.086) | Ready for Nuclear (0.11) | Resp. of Int. Org. (0.12) | Ready for Nuclear (0.084) |
| 3 | Scepticism towards media (0.067) | Age (0.089) | Govt. Resp. (0.1) | Age (0.084) | Scepticism towards media (0.09) | Vehicle Availability (0.088) | Resp. of Int. Org. (0.072) |
| 4 | Ready for Nuclear (0.0663) | Ready for Nuclear (0.08) | Resp. of Int. Org. (0.09) | Gas Cons. (0.08) | Resp. of Int. Org. (0.083) | Ready for Nuclear (0.083) | Scepticism towards media (0.07) |
| 5 | Willingness to Accept Nuclear (0.0662) | Years of Education (0.077) | Scepticism towards media (0.08) | Years of Education (0.07) | Willingness to Accept Nuclear (0.057) | Scepticism towards media (0.08) | Support for Nuclear (0.06) |
| 6 | Prepared for Cons. Red. (0.058) | Resp. of Int. Org. (0.053) | Env. Consc. (0.056) | Support for Nuclear (0.059) | Prepared for Cons. Red. (0.055) | Age (0.048) | Age (0.059) |
| 7 | Age (0.0653) | Scepticism towards media (0.05) | Gas Cons. (0.053) | Support for RE (0.056) | Age (0.046) | Support for RE (0.046) | Power and Energy Cons. (0.057) |
| 8 | Years of Education (0.062) | Willingness to Accept Nuclear (0.049) | Years of Education (0.044) | Willingness to Accept Nuclear (0.053) | Vehicle Availability (0.043) | Support for Nuclear (0.042) | Prepared for Cons. Red. (0.056) |
| 9 | Support for RE (0.056) | Support for Nuclear (0.048) | Age (0.042) | Prepared for Cons. Red. (0.052) | Support for Nuclear (0.04) | Years of Education (0.041) | Willingness to Accept Nuclear (0.052) |
| 10 | Resp. of Int. Org. (0.054) | Prepared for Cons. Red. (0.0479) | Prepared for Cons. Red. (0.04) | Resp. of Int. Org. (0.05) | Years of Education (0.039) | Willingness to Accept Nuclear (0.034) | Years of Education (0.048) |

**Source:** Authors' Calculation.

---

[7] The importance score of each feature is presented in the parenthesis.

The analysis underscores the significant role of regional factors, especially the division variable, in influencing stress levels related to energy supply and pricing. The division consistently emerges as one of the top-ranking features across all scenarios, indicating that regional differences play a pivotal role in shaping how households perceive and respond to challenges in energy availability and costs. This suggests that localised factors, such as regional policies, infrastructure, and resource availability, are crucial in determining energy stress levels.

Perception-based features, particularly those related to nuclear energy, and scepticism towards media also stand out as key influencers. The readiness for nuclear energy and willingness to accept nuclear solutions rank highly across different energy scenarios, particularly in electricity and fuel oil pricing contexts. This highlights the strong connection between personal beliefs about nuclear energy and stress levels associated with energy supply and pricing. Individuals who are more open to nuclear energy seem to experience less stress, suggesting that perception and acceptance of energy solutions significantly affect how households respond to energy crises. On the other hand, scepticism towards media consistently appears as an important factor, suggesting that individuals who are more sceptical of media sources may experience heightened stress, possibly due to distrust in information related to energy issues.

The analysis also reveals that socioeconomic factors, such as education and, significantly shape stress perceptions. Higher educational levels are correlated with a better understanding of energy policies, which may reduce stress levels associated with energy supply and pricing. Age and urban residency further contribute to stress in specific contexts. Age emerges as an influential factor, particularly in the context of fuel oil pricing, where older individuals are likely more affected by pricing changes. Similarly, urban residency plays a major role, especially in scenarios related to gas supply, where urban households face distinct challenges in accessing and affording gas. The analysis suggests that urban households experience more pronounced stress in gas supply, which may be due to higher demand and infrastructure limitations in urban areas.

Finally, environmental and political factors, such as support for renewable energy and the responsibility of international organisations, are key players in shaping stress perceptions. Support for renewable energy stands out in the gas pricing scenario, reflecting a broader societal shift towards sustainable energy sources and how this affects public sentiment and stress levels. Similarly, the responsibility of international organisations is closely tied to stress perceptions, especially when it comes to energy pricing, indicating that how international entities are perceived in managing global energy crises impacts individual stress levels.

In summary, the findings demonstrate a complex interplay of demographic, socioeconomic, and perception-based factors in shaping energy-related stress. Demographics, such as age and urban residency, provide insight into the vulnerability of certain groups, while socioeconomic status and educational attainment reflect how knowledge and economic stability influence stress. Perceptions of energy policies, especially regarding nuclear energy and renewable resources, significantly affect stress levels, particularly in pricing scenarios. These insights highlight the need for targeted policies and communication strategies that consider these multidimensional factors to reduce stress across various segments of society. The consistency of these findings across different models strengthens the robustness of the analysis and underscores the importance of addressing these variables in energy policy and planning.

### 4.9 Results and Discussion from Ordered Probit Model

In Appendix H, Table H.1, we present the ordered Probit model results. The primary interest of our study is on the marginal effects of variables on the probability of transitioning between stress ranges.

In Appendix H, Tables ranging from H.2 to H.7, we presented the marginal effects of various factors on scenario-specific stress levels. In table 7, we present the marginal effects on stress level associated with overall power and energy crisis.

**Table 7: Marginal Effects from the Ordered Probit Models (Scenario: Overall Stress Level)**

| Variables | Prob (pss_rng = 1) | | Prob (pss_rng = 2) | | Prob (pss_rng = 3) | |
|---|---|---|---|---|---|---|
| | Mg. Eff. | (P-Val.) | Mg. Eff. | (P-Val.) | Mg. Eff. | (P-Val.) |
| Urban | -0.150 | 0.00 | 0.088 | 0.00 | 0.062 | 0.00 |
| Division (Base: Dhaka) | | | | | | |
|    Barisal | -0.158 | 0.00 | 0.095 | 0.00 | 0.063 | 0.00 |
|    Chattogram | -0.131 | 0.00 | 0.098 | 0.00 | 0.032 | 0.00 |
|    Khulna | -0.173 | 0.00 | 0.07 | 0.008 | 0.103 | 0.001 |
|    Mymensingh | -0.073 | 0.016 | 0.064 | 0.012 | 0.010 | 0.091 |
|    Rajshahi | 0.286 | 0.00 | -0.279 | 0.00 | -0.01 | 0.001 |
|    Rangpur | -0.184 | 0.00 | 0.013 | 0.703 | 0.171 | 0.00 |
|    Sylhet | -0.122 | 0.00 | 0.095 | 0.00 | 0.027 | 0.062 |
| Sex | 0.053 | 0.177 | -0.031 | 0.178 | -0.022 | 0.181 |
| Age | 0.001 | 0.186 | -0.001 | 0.196 | 0.00 | 0.179 |
| Years of Education | 0.00 | 0.932 | 0.00 | 0.932 | 0.00 | 0.932 |
| Household Size | -0.004 | 0.599 | 0.002 | 0.60 | 0.001 | 0.60 |
| Number of Students | 0.003 | 0.714 | -0.002 | 0.713 | -0.001 | 0.714 |
| Income Source (Base: Agri) | | | | | | |
|    Service | 0.058 | 0.001 | -0.034 | 0.003 | -0.024 | 0.001 |
|    Industry | 0.014 | 0.589 | -0.008 | 0.591 | -0.006 | 0.587 |
| Income Group (Base: Below BDT 10000) | | | | | | |
|    BDT 10000 to 24999 | 0.021 | 0.692 | -0.013 | 0.692 | -0.009 | 0.683 |
|    BDT 25000 to 39999 | 0.001 | 0.988 | -0.001 | 0.988 | 0.00 | 0.988 |
|    BDT 40000 to 79999 | 0.051 | 0.381 | -0.03 | 0.38 | -0.021 | 0.384 |
| Above 80000 | 0.122 | 0.196 | -0.072 | 0.196 | -0.050 | 0.201 |
| Availability of Personal Vehicle (1 = Yes) | -0.04 | 0.022 | 0.023 | 0.025 | 0.017 | 0.023 |
| Consciousness Variables | | | | | | |
| e_1 | 0.030 | 0.00 | -0.018 | 0.00 | -0.039 | 0.00 |
| e_2 | 0.006 | 0.40 | -0.004 | 0.404 | -0.003 | 0.397 |
| e_3 | 0.007 | 0.307 | -0.004 | 0.307 | -0.003 | 0.311 |
| e_4 | 0.000 | 0.992 | 0.00 | 0.992 | 0.00 | 0.992 |
| e_5 | -0.049 | 0.00 | 0.029 | 0.00 | 0.02 | 0.00 |
| e_6 | -0.015 | 0.013 | 0.009 | 0.017 | 0.006 | 0.014 |
| e_7 | -0.044 | 0.00 | 0.026 | 0.00 | 0.018 | 0.00 |
| e_8 | 0.015 | 0.02 | -0.009 | 0.023 | -0.006 | 0.021 |
| e_9 | 0.034 | 0.00 | -0.02 | 0.00 | -0.014 | 0.00 |
| Power and Energy Consumption pc | 0.00 | 0.74 | 0.00 | 0.74 | 0.00 | 0.736 |

**Source:** Authors' Calculation.

The results show that urban residency consistently plays a crucial role in determining stress levels, with urban households being more likely to experience higher stress ranges compared to suburban areas. This is particularly evident in scenarios related to gas supply and fuel oil supply, where urban residents show a higher probability of transitioning into moderate to high stress ranges. This suggests that urban households are more vulnerable to energy crises, likely due to the concentration of demand and infrastructure challenges in cities.

Regional disparities further complicate the picture. For instance, households in Rangpur are more likely to experience high stress related to gas supply, as reflected in a strong marginal effect of 0.22 for the third stress range in gas supply scenario. Similarly, regions such as Khulna and Sylhet also exhibit significant variations in stress, indicating the importance of local infrastructure and energy availability in shaping household stress responses.

Income source is another critical factor influencing stress transitions. Households with service-based incomes are generally more resilient to energy crises, as seen in scenarios like fuel oil pricing, where these households tend to stay in lower stress ranges. Conversely, agriculture-based households show a higher likelihood of experiencing moderate to high stress, particularly due to the financial vulnerabilities they face in coping with rising energy prices. This suggests that the nature of one's livelihood plays a significant role in determining how stress is distributed across different energy scenarios.

In terms of income levels, the analysis shows that higher-income households are generally better able to absorb the impact of energy price fluctuations, particularly for gas pricing (pss_gaspr), where households earning above BDT 80,000 are less likely to experience high stress. However, income levels appear to have less influence in the context of electricity pricing, where other factors, such as environmental and political opinions, seem to have a more profound effect.

Consciousness variables, particularly regarding environmental beliefs and political opinions, emerge as highly influential factors across all energy scenarios. The concern about environmental pollution (e_1) correlates with lower stress levels in the context of electricity supply, as households that are more environmentally conscious show a greater willingness to adapt to energy conservation. Similarly, a readiness to reduce energy consumption for climate protection (e_9) also aligns with lower stress regarding electricity supply and fuel oil pricing scenarios. On the other hand, scepticism towards media portrayals of the energy crisis (e_7) plays an opposite role, being associated with higher stress levels, particularly in scenarios related to gas and electricity pricing. This indicates that perceptions of media exaggeration around the energy crisis might amplify the stress felt by households, especially in the face of rising energy prices.

In addition, views on the necessity of nuclear energy (e_6) have a notable influence, with those in favour of nuclear energy demonstrating a higher likelihood of lower stress levels for fuel oil pricing and electricity pricing. The belief in the role of international organisations in managing energy crises (e_5) also correlates with a reduced probability of high stress in certain scenarios, especially related to fuel oil supply. These findings underscore the strong link between household beliefs and attitudes toward energy policy, the environment, and media portrayals, which significantly shape stress levels in energy crises.

Moreover, the ownership of a personal vehicle significantly increases the probability of higher stress levels, particularly in fuel oil pricing scenarios, as vehicle-owning households are directly impacted by fluctuations in fuel prices. This underscores the importance of transportation costs as a significant stressor for households with vehicles.

Finally, while age shows a slight increase in stress for older individuals in the electricity pricing scenario, education levels have minimal impact on stress across all energy scenarios. This suggests that financial stability, socioeconomic status, and regional factors are more significant in shaping household stress levels than education alone.

## 5.0 Policy Discussions

The integration of the PSS-10 into policy dimensions represents a shift in the way policymakers understand and address societal issues, particularly those arising in the context of power and energy crises. Historically, energy policy has predominantly focused on technical, economic, and infrastructural considerations, such as the generation capacity of power plants, fuel supply chains, energy market regulation, and pricing models (MOPEMR, 2023; Taghizadeh-Hesary & Zhang, 2023; Desprairies, 1983). These are essential elements for ensuring reliable energy access, but they fail to capture the full spectrum of human experiences during energy shortages or fluctuations in energy prices.

In traditional policy approaches, the focus has been on meeting quantitative targets, such as ensuring X gigawatts of energy supply, maintaining fuel reserves, or reducing energy costs by a specific percentage, without considering the psychological burden these challenges impose on households (Taghizadeh-Hesary & Zhang, 2023; Prontera, 2020). This shift to the new approach aligns with an evolving political economy perspective that calls for a more holistic understanding of policy outcomes. Traditionally, policies have been evaluated based on their economic outcomes (e.g., GDP growth, poverty reduction, energy access) (Taghizadeh-Hesary & Zhang, 2023; Tebbe, Mailloux, & Nemet, 2024). However, psychological well-being has become an increasingly important dimension in the discourse on public policy. The PSS-10 offers a human-centred approach that acknowledges the individual and collective psychological costs of energy policies, which is especially relevant in the current political economy where governments are increasingly focused on improving the quality of life rather than merely economic metrics. This approach, instead of assuming that households respond uniformly to energy disruptions, enables a nuanced understanding of how different groups (urban vs. rural, age groups, income classes) perceive the stress of energy supply and pricing issues.

By utilising the PSS-10 into the policy process, energy policies can move beyond the conventional metrics of supply and demand, and begin addressing how these crises affect the mental health and well-being of households. For instance, energy crises are often accompanied by energy poverty, which can lead to feelings of helplessness and anxiety, particularly when energy costs rise rapidly or when energy access is intermittent (Stojilovska, Thomson, & Mejía-Montero, 2023; Grazini, 2024). These factors contribute to a growing sense of vulnerability among households, especially in urban areas where energy consumption is often higher, and in low-income communities that lack the financial flexibility to absorb price shocks (Grazini, 2024; Halkos & Gkampoura, 2021; Chevalier & Ouédraogo, 2009). In this study, we have shown how the PSS-10 can be used as a diagnostic tool to understand how these psychological factors intersect with physical and economic factors, such as income levels, housing types, and geographic location, helping policymakers identify the most stressed communities and households that are most at risk of psychological distress.

The inclusion of the PSS-10 into energy policy also brings human-centred insights into political and economic decision-making. For example, policies like energy price subsidies, fuel price caps, or alternative energy solutions could be tailored not only to reduce economic hardships but also to alleviate psychological distress. Policymakers can use the PSS-10 to assess the psychological resilience of different segments of the population and develop interventions that support both economic recovery and mental well-being. For instance, communities experiencing high levels of stress might benefit from psychosocial support programmes alongside energy-related financial relief programmes. Our findings also reinforce that the environmentally conscious and more educated households tend to feel less stress attributed to the crisis situations, implying that transparency and well-designed education programmes can help the population understand the structural and stoic problems which would require a long-term to be addressed properly. A policy intervention that includes mental health resources for communities facing power outages could enhance policy effectiveness, ensuring that the population feels emotionally supported in addition to financially assisted.

Incorporating the PSS-10 also provides an opportunity for policymakers to better understand public perceptions of energy crises. Understanding how households perceive energy issues, whether they feel empowered or helpless in response to energy scarcity, can help design policies that not only address technical and economic factors but also boost public confidence in government actions. This could be particularly useful in addressing public scepticism about government efforts to resolve energy crises. For instance, households that exhibit high stress levels may feel that energy solutions are not being implemented effectively, creating a feedback loop of frustration and distrust. By using the PSS-10 to monitor these perceptions, policymakers could refine communication strategies and policy interventions to ensure that they address the psychological needs of the population as well as their material needs. From a political economy perspective, integrating the PSS-10 in energy policy could have profound implications for policy legitimacy and public trust. Policies that account for the emotional and psychological toll of crises are likely to resonate more with citizens, particularly in democratic systems where public opinion is a critical determinant of policy success. Politicians and policymakers who acknowledge the stress caused by energy issues and incorporate strategies to reduce it would enhance their political capital, positioning themselves as responsive and responsible actors in times of crisis.

From an economic standpoint, the PSS-10 integrates seamlessly into a comprehensive welfare approach. By considering the psychological effects of energy disruptions, policies can be designed not only to mitigate economic losses (e.g., energy subsidies, price caps) but also to alleviate stress and uncertainty within households, which may have long-term implications for social welfare, health, and even productivity. For instance, the introduction of policies focused on mental health support during energy crises (such as counselling or stress-relief programmes for households) could prove highly effective in supporting the resilience of citizens. In the same vein, understanding the stress impact of rising energy prices could shape targeted subsidies or price stabilisation measures to ensure that the most vulnerable households are not only financially supported but also mentally protected.

One example of the practical implementation of the PSS-10 could be in community-based energy solutions. Rather than focusing solely on the economic feasibility of decentralised energy systems (such as solar energy cooperatives), a policy informed by the PSS-10 could also consider psychological resilience. Communities with high levels of energy-related stress could benefit from tailored interventions that reduce anxiety and stress around energy access and supply. This could include community engagement activities to ensure that people feel they have agency and control over their energy consumption, or psychological interventions that reduce the emotional burden of an energy crisis. The PSS-10 can also serve as an essential tool in policy evaluation. Instead of assessing only economic metrics, policymakers can now use the scale to track psychological outcomes associated with different energy policies. For instance, energy pricing policies could be assessed not just on their ability to meet financial goals, but also on how well they mitigate household stress and improve perceived security.

Moreover, mental health experts and psychologists could be incorporated into energy policy-making processes, ensuring that energy strategies do not only solve logistical problems but also promote the mental well-being of citizens because of its high dependency towards the lifestyle which require intensive use of power and energy. This could be particularly relevant for low-income households or those in regions with frequent power outages, where the mental strain of living through repeated crises may lead to long-term psychological effects, further exacerbating economic difficulties.

The PSS-10 can also serve as an essential tool in policy evaluation. Instead of assessing only economic metrics, policymakers can now use the scale to track psychological outcomes associated with different energy policies. For instance, energy pricing policies could be assessed not just on their ability to meet financial goals, but also on how well they mitigate household stress and improve perceived security. The evaluation of energy access policies, therefore, would take on a more multidimensional nature, accounting for both economic and psychological dimensions of well-being.

This study's findings highlight specific factors, such as age, income source, vehicle ownership, education, and environmental perceptions, that directly influence how stress manifests across different populations. These findings provide the evidence base necessary to propose specific policies targeted at vulnerable groups. For example, households with service sector incomes experience lower stress during energy crises, suggesting that policy interventions could target agricultural households, who show higher stress levels, by offering more flexible energy pricing or subsidies. The findings also point out that higher-income households are better able to weather energy price fluctuations, specifically for gas pricing. However, the study shows that socioeconomic resilience does not provide complete immunity to stress, especially for households depending on fuel for transportation. Thus, policies aimed at higher-income households could be designed to buffer their stress in terms of transportation costs, potentially through subsidies or alternative fuel incentives.

The study reveals significant regional disparities, particularly with Khulna, Barisal, and Chattogram showing marked differences in stress levels. These areas, with different energy access patterns, would benefit from localised energy support measures tailored to regional needs, including enhanced infrastructure development and targeted stress-relief programmes that focus on mental health support during energy crises. For instance, Khulna, with rising stress levels even at higher quantiles, could benefit from energy security policies that address both infrastructure stability and the mental resilience of the population.

The study consistently shows that households with lower incomes, especially those in agriculture-dependent sectors, are more likely to experience higher stress. Policies aimed at income support, financial subsidies, and alternative energy solutions (such as clean cookstoves or solar-powered irrigation) could significantly alleviate energy-related stress for these households. Conditional cash transfer programmes tied to energy efficiency could incentivise better energy management and reduce stress, particularly in rural areas.

One notable finding from this study is the significant relationship between stress levels in power and energy scenarios and the degree of reliance on each respective energy source. Households with a higher dependency on electricity, gas, and fuel oil tend to exhibit greater stress in response to disruptions or price increases in those energy sources. The issue is compounded by the growing demand for energy, which is increasing alongside urbanisation, industrialisation, and changing consumption patterns. As reliance on energy continues to rise, particularly in urban and middle-income households, the psychological burden associated with energy supply and pricing will likely intensify. If current policy frameworks fail to account for this rising vulnerability, by promoting energy diversification, improving efficiency, and enhancing resilience strategies for households with high energy dependence, future energy crises may exacerbate stress levels. Thus, policymakers must move beyond traditional infrastructure and supply-side solutions, adopting a more holistic, anticipatory approach that incorporates the evolving psychosocial dynamics of energy consumption.

## 6.0 Conclusions

While this study presents valuable insights into the psychological aspects of energy crises, it has some limitations that should be addressed in future research. Sample size and diversity could be expanded to include more households across different income groups, ethnic backgrounds, and urban-rural divides. Future studies could also explore longitudinal data to examine the long-term effects of energy crises on stress levels and mental health. Additionally, integrating energy consumption behaviours into the PSS-10 model would allow for a more detailed understanding of how household decisions around energy use contribute to or alleviate stress.

This study advances the understanding of the complex relationship between energy crises and household stress, presenting a robust framework for integrating psychological dimensions into energy policy. By utilising the PSS-10, we not only quantify the stress induced by disruptions in energy supply and pricing but also reveal a multi-layered response to such crises, one that is deeply influenced by demographics, socioeconomic status, regional factors, and environmental perceptions. Our methodology establishes the statistical reliability and contextual validity of the parameters that dictate stress within power and energy situations. An important point of replicability should be that the methodology goes beyond the boundary of Bangladesh as long as the native language translated questionnaires are reliable and valid within non-clinical setting for a particular country. The contextual validity can be drawn from the statistically significant variables from our econometric and machine learning models which solidifies our case by providing pragmatic and tangible case of how stress level could be associated with real-life crises scenarios.

The findings demonstrate that stress levels are not just a function of economic or infrastructural vulnerabilities but are significantly shaped by individual beliefs, age, income sources, and education levels. Urban households, low-income groups, and those with lower environmental awareness are particularly vulnerable, suggesting that targeted policy interventions are essential. These interventions should not only address the technical aspects of energy crises but also incorporate psychosocial support and behavioural incentives to mitigate stress.

From a policy perspective, integrating the psychological dimension into energy planning could radically transform the way governments approach energy security. The insights from this study urge policymakers to consider how mental resilience, behavioural adaptation, and perceptions of energy solutions influence the public's ability to manage the stresses associated with energy shortages and price volatility. While the traditional focus on infrastructure and market stability remains crucial, this study advocates for a holistic approach that integrates psychological well-being into the design of socioeconomic and environmental policies.

However, the study also highlights critical gaps in current energy policy, primarily the lack of focus on mental health and social resilience in times of energy crises. Region-specific solutions, financial resilience strategies, and greater public engagement on energy transitions, particularly regarding nuclear and renewable energy, are necessary to address these gaps. The psychosocial factors identified in this research should now be incorporated into energy policy frameworks, providing a more comprehensive and human-centric approach to energy security.

This research serves as a call to action for future studies and policy initiatives to not only focus on the technical dimensions of energy but to also account for the psychological resilience of populations in the face of mounting energy challenges. The integration of stress metrics into policy decision-making will enable a more responsive, inclusive, and effective energy policy, one that acknowledges the intricate dynamics of human behaviour and psychological well-being in energy crises.

# Appendix

## Appendix A: Sample Distribution

**Table A.1: Sample Distribution**

| Division Name | Sub-district Name[8] | Number of Samples |
|---|---|---|
| Barishal | Talhati (23.9) | 13 |
|  | Banaripara (21.7) | 13 |
|  | Bhola Sadar (15) | 14 |
|  | Mathbaria (28.6) | 14 |
| **Total** |  | **54** |
| Chattogram | Halishahar (8.7) | 37 |
|  | Chandpur Sadar (21.6) | 37 |
|  | Bijoynagar (12.4) | 37 |
|  | Muradnagar (26.2) | 37 |
|  | Companiganj (23.1) | 36 |
| **Total** |  | **184** |
| Dhaka | Khilgaon (18.6) | 40 |
|  | Kapasia (13.6) | 41 |
|  | Madaripur Sadar (3.2) | 40 |
|  | Roypura (15.9) | 41 |
|  | Ghatail (20.5) | 41 |
|  | Adabor (28.6) | 40 |
|  | Belabo (13.2) | 41 |
| **Total** |  | **284** |
| Khulna | Kachua (22.5) | 28 |
|  | Kalia (16.5) | 28 |
|  | Khulna Sadar (28.6) | 27 |
|  | Kaliganj (14) | 27 |
| **Total** |  | **110** |
| Mymensingh | Mymensingh Sadar (27.4) | 19 |
|  | Bhaluka (15.5) | 18 |
|  | Nakla (35.3) | 18 |
|  | Jamalpur Sadar (49.3) | 19 |
| **Total** |  | **74** |
| Rajshahi | Kahaloo (17.1) | 33 |
|  | Kalai (21.5) | 32 |
|  | Matihar (17.4) | 33 |
|  | Sirajganj Sadar | 32 |
| **Total** |  | **130** |
| Rangpur | Dinajpur Sadar (64.1) | 28 |
|  | Kaliganj (36.9) | 27 |
|  | Debiganj (17.8) | 28 |
|  | Taraganj (38.1) | 27 |
| **Total** |  | **110** |
| Sylhet | Jagannathpur (22.6) | 14 |
|  | Nabiganj (16.9) | 13 |
|  | Barlekha (10.4) | 14 |
|  | Fenchuganj (19) | 13 |
| **Total** |  | **54** |
| **Total Sample** |  | **1000** |

**Source:** Authors' Calculation.

---

[8] In parentheses, headcount ratio of upper poverty level is illustrated from HIES-2016.

# Appendix B: Variable Description

**Table B.1: Explanation of the variables**

| Variable | Explanation |
|---|---|
| pss_elcs | PSS for the 'Supply of Electricity' scenario. |
| pss_elcp | PSS for the 'Price of Electricity' scenario. |
| pss_gassp | PSS for the 'Supply of Gas' scenario. |
| pss_gaspr | PSS for the 'Price of Gas' scenario. |
| pss_fuelsp | PSS for the 'Supply of Fuel Oil' scenario. |
| pss_fuelpr | PSS for the 'Price of Fuel Oil' scenario. |
| pss | PSS for the overall power and energy scenario. |
| pss_elcs_rng | PSS for the 'Supply of Electricity' scenario, expressed as various stress range. |
| pss_elcp_rng | PSS for the 'Price of Electricity' scenario, expressed as various stress range. |
| pss_gassp_rng | PSS for the 'Supply of Gas' scenario, expressed as various stress range. |
| pss_gaspr_rng | PSS for the 'Price of Gas' scenario, expressed as various stress range. |
| pss_fuelsp_rng | PSS for the 'Supply of Fuel Oil' scenario, expressed as various stress range. |
| pss_fuelpr_rng | PSS for the 'Price of Fuel Oil' scenario, expressed as various stress range. |
| pss_rng | PSS for the overall power and energy scenario, expressed as various stress range. |
| urban | This variable indicates whether the households surveyed fall in urban neighbourhoods or sub-urban neighbourhoods. By the precondition of gas and electricity availability in the households, the rural households are not included. If urban = 0, it means sub-urban, and if urban= 1, otherwise. |
| division_code_bar | Division code: Barisal. If division_code_bar = 1, it indicates Barisal, or otherwise. |
| division_code_ctg | Division code: Chattogram. If division_code_ctg = 1, it indicates Chattogram, or otherwise. |
| division_code_khl | Division code: Khulna. If division_code_khl = 1, it indicates Khulna, or otherwise. |
| division_code_mym | Division code: Mymensingh. If division_code_mym = 1, it indicates Mymensingh, or otherwise. |
| division_code_raj | Division code: Rajshahi. If division_code_raj = 1, it indicates Rajshahi, or otherwise. |
| division_code_rang | Division code: Rangpur. If division_code_rang = 1, it indicates Rangpur, or otherwise. |
| division_code_syl | Division code: Sylhet. If division_code_syl = 1, it indicates Sylhet, or otherwise. |
| a_5 | Sex of the household head. If a_5 = 0, then the household head is male. If a_5 = 1, the household head is female. |
| a_6 | Age of the household head |
| a_7 | Years of education of a household head |
| a_8 | Number of members in a household |
| a_9 | Number of students in a household |
| a_10_agri | Household's primary source of income: Agriculture. This is the base category of households' source of primary income. |
| a_10_ind | Household's primary source of income: Industry. If a_10_ind = 1, then the household's primary source of income is industry sector, or, if a_10_ind = 0, otherwise. |
| a_10_serv | Household's primary source of income: Service |
| a_11_10k | Household's monthly income: Below BDT 10,000 |
| a_11_25k | Household's monthly income: BDT 10,000 to BDT 24,999. If a_11_25k = 1, then household's monthly income falls in the range of BDT 10,000 to BDT 24,999. If a_11_25k = 0, otherwise. |
| a_11_40k | Household's monthly income: BDT 25,000 to BDT 39,999. If a_11_40k = 1, then household's monthly income falls in the range of BDT 25,000 to BDT 39,999. If a_11_40k = 0, otherwise. |
| a_11_80k | Household's monthly income: BDT 40,000 to BDT 79,999. If a_11_80k = 1, then household's monthly income falls in the range of BDT 40,000 to BDT 79,999. If a_11_80k = 0, otherwise. |
| a_11_a80k | Household's monthly income: above BDT 80,000. If a_11_a80k = 1, then household's monthly income falls in the category of households with income above BDT 80,000. If a_11_a80k, otherwise. |
| elc_cons_pc | Per capita electricity consumption within a household for the previous month. We used electricity bill as a metric of consumption. |
| gas_cons_pc | Per capita gas consumption within a household for the previous month. We divide total expenditure of a household on gas by the household size for the previous month of the survey. |
| oil_cons_pc | Per capita oil consumption within a household for the previous month. The same formula of per capita gas consumption has been used. |
| peo_cons_pc | Per capita power and energy consumption within a household. We add the three components: per capita electricity, gas and fuel oil consumption to derive this variable. |
| a_12 | Availability of personal/office vehicles |
| e_1 | Statement: Environmental pollution has adverse effects on weather and climate change. |
| e_2 | Statement: International organizations are mainly responsible for fighting climate change. |
| e_3 | Statement: Government and relevant authorities are mainly responsible for fighting climate change. |
| e_4 | Statement: Without nuclear energy, energy crisis will not resolve. |
| e_5 | Statement: Without renewable energy, energy crisis will not resolve. |
| e_6 | Statement: Bangladesh is ready for nuclear energy production. |
| e_7 | Statement: The media is overexaggerating the crisis. |
| e_8 | Statement: The respondent has no problem in establishing a nuclear power plant in the neighbourhood. |
| e_9 | Statement: The household is prepared to reduce power and energy consumption to save environment and climate. |

**Source:** Authors' Calculation.

## Appendix C: Shapiro-Wilk W test for normal data
### Table C.1: Shapiro-Wilk W test for normal data

| Variable | Observation | W | V | z | Prob>z |
|---|---|---|---|---|---|
| e_1 | 1,000 | 0.9688 | 19.676 | 7.378 | 0.000*** |
| e_2 | 1,000 | 0.98526 | 9.295 | 5.521 | 0.000*** |
| e_3 | 1,000 | 0.98881 | 7.055 | 4.838 | 0.000*** |
| e_4 | 1,000 | 0.99618 | 2.411 | 2.18 | 0.014*** |
| e_5 | 1,000 | 0.99656 | 2.172 | 1.921 | 0.027*** |
| e_6 | 1,000 | 0.99493 | 3.201 | 2.881 | 0.002*** |
| e_7 | 1,000 | 0.9984 | 1.007 | 0.017 | 0.500 |
| e_8 | 1,000 | 0.9966 | 2.146 | 1.891 | 0.029*** |
| e_9 | 1,000 | 0.99567 | 2.732 | 2.489 | 0.006*** |

*** *Variables that are not normally distributed*

**Source:** Authors' Calculation.

# Appendix D: Internal Consistency Test of Variables Associated with Environmental and Political Opinions

**Table D.1: Internal Consistency Test of E-values**

| | |
|---|---|
| Average interitem covariance: | 0.273498 |
| Number of items in the scale: | 9 |
| Scale reliability coefficient: | 0.7225 |

**Source:** Authors' Calculation.

# Appendix E: Internal Consistency Test of PSS-10 Values Associated with Power and Energy Crisis Scenarios

**Table E.1: Internal Consistency Test for PSS-10 values**

| Variable Name | Average interitem covariance | Number of items in the scale | Scale reliability coefficient |
|---|---|---|---|
| Electricity Supply | 0.51428 | 10 | 0.8368 |
| Electricity Price | 0.608461 | 10 | 0.8677 |
| Gas Supply | 0.843214 | 10 | 0.894 |
| Gas Price | 0.60014 | 10 | 0.8627 |
| Fuel Oil Supply | 1.101739 | 10 | 0.9252 |
| Fuel Oil Price | 1.24061 | 10 | 0.9248 |
| PSS | 22.15277 | 6 | 0.877 |

**Source:** Authors' Calculation.

# Appendix F: Econometric Test

**Table F.1: Breusch-Pagan test for heteroskedasticity**

| Variables: fitted values of PSS: Elc. Supply | | Variables: fitted values of PSS: Fuel Oil Supply | |
|---|---|---|---|
| chi2(1)   = | 103.39 | chi2(1)   = | 111.46 |
| Prob > chi2 = | 0.0000 | Prob > chi2 = | 0.0000 |
| Variables: fitted values of PSS: Elc. Price | | Variables: fitted values of PSS: Fuel Oil Price | |
| chi2(1)   = | 14.61 | chi2(1)   = | 39.16 |
| Prob > chi2 = | 0.0001 | Prob > chi2 = | 0.0000 |
| Variables: fitted values of PSS: Gas Supply | | Variables: fitted values of PSS: Overall | |
| chi2(1)   = | 82.79 | chi2(1)   = | 135.00 |
| Prob > chi2 = | 0.0000 | Prob > chi2 = | 0.0000 |
| Variables: fitted values of PSS: Gas Price | | | |
| chi2(1)   = | 58.79 | | |
| Prob > chi2 = | 0.0000 | | |

**Source:** Authors' Calculation

**Appendix G: Scenario-Specific Results and Discussions from Simultaneous-Quantile Regression Models**

The results for the quantile regressions in the case of assessing stress level subject to electricity supply is given in Table G.1. Despite the differences captured across quantiles, there are notable similarities between the MOLS and SQR results in this case.

**Table G.1: Results from Quantile Regression: PSS – Electricity Supply Scenario**

| VARIABLES | (1) 15th Quantile | (2) 30th Quantile | (3) 50th Quantile | (4) 70th Quantile | (5) 85th Quantile |
|---|---|---|---|---|---|
| Urban (sub-Urban = 0) | 3.320*** | 3.004*** | 2.028*** | 1.435*** | 1.083** |
|  | (0.637) | (0.557) | (0.405) | (0.381) | (0.532) |
| Division (Base: Dhaka) | | | | | |
|   Barisal | 8.173*** | 5.539*** | 3.761*** | 2.798*** | 0.995 |
|  | (1.305) | (1.023) | (0.792) | (0.705) | (1.207) |
|   Chattogram | 8.141*** | 4.601*** | 3.011*** | 2.322*** | 1.083* |
|  | (0.955) | (0.807) | (0.442) | (0.425) | (0.573) |
|   Khulna | 3.712*** | 1.842 | 2.943*** | 4.071*** | 7.242*** |
|  | (1.309) | (1.227) | (0.956) | (0.947) | (2.193) |
|   Mymensingh | 4.262*** | 1.498* | 0.542 | 0.498 | -0.0198 |
|  | (1.338) | (0.835) | (0.547) | (0.650) | (0.862) |
|   Rajshahi | 3.630*** | 0.652 | 1.353* | 2.005*** | 1.558* |
|  | (1.302) | (1.147) | (0.740) | (0.715) | (0.897) |
|   Rangpur | 10.35*** | 7.532*** | 6.531*** | 6.106*** | 5.436*** |
|  | (1.278) | (1.032) | (0.724) | (0.713) | (1.068) |
|   Sylhet | 8.133*** | 4.709*** | 3.306*** | 2.801*** | 1.222 |
|  | (1.178) | (0.971) | (0.701) | (0.698) | (1.166) |
| Sex | -3.198 | -2.081 | -0.759 | 0.747 | 0.676 |
|  | (2.120) | (2.076) | (1.317) | (1.193) | (1.085) |
| Age | -0.0548** | -0.0407* | -0.0227 | -0.0173 | -0.00929 |
|  | (0.0264) | (0.0214) | (0.0155) | (0.0147) | (0.0231) |
| Years of Education | -0.132** | -0.183*** | -0.124*** | -0.0478 | -0.00912 |
|  | (0.0551) | (0.0465) | (0.0301) | (0.0365) | (0.0456) |
| Household Size | 0.00319 | 0.0545 | -0.0187 | 0.0164 | 0.120 |
|  | (0.207) | (0.179) | (0.124) | (0.114) | (0.150) |
| Number of Students | 0.0479 | 0.354 | 0.283 | 0.109 | -0.109 |
|  | (0.290) | (0.282) | (0.208) | (0.183) | (0.229) |
| Income Source (Base: Agri) | | | | | |
|   Service | -2.274*** | -1.123** | -0.722* | -0.150 | -0.141 |
|  | (0.732) | (0.564) | (0.391) | (0.384) | (0.528) |
|   Industry | -1.660 | -0.0411 | 0.553 | 0.219 | -0.343 |
|  | (1.111) | (1.059) | (0.529) | (0.494) | (0.816) |
| Income Group (Base: Below BDT 10000) | | | | | |
|   BDT 10000 to 24999 | -0.0988 | -0.758 | 0.219 | 0.900 | -0.795 |
|  | (1.364) | (1.421) | (0.935) | (1.505) | (3.070) |
|   BDT 25000 to 39999 | 0.0610 | -0.447 | 1.024 | 1.450 | 0.565 |
|  | (1.455) | (1.511) | (1.000) | (1.556) | (3.130) |
|   BDT 40000 to 79999 | -0.686 | -1.029 | 0.928 | 1.443 | -0.210 |
|  | (1.756) | (1.847) | (1.088) | (1.554) | (3.160) |
|   Above 80000 | -2.226 | 0.763 | 0.188 | 1.564 | 0.191 |
|  | (3.725) | (3.738) | (2.228) | (2.336) | (3.700) |
| Availability of Personal Vehicle (1 = Yes) | 0.930 | 0.848* | -0.0339 | -0.202 | -0.222 |
|  | (0.615) | (0.489) | (0.365) | (0.381) | (0.529) |
| Consciousness Variables | | | | | |
|   e_1 | -0.907*** | -0.551** | -0.439*** | -0.315* | -0.286 |
|  | (0.269) | (0.214) | (0.168) | (0.179) | (0.248) |
|   e_2 | -0.672** | -0.539** | -0.155 | -0.0345 | 0.0723 |
|  | (0.304) | (0.233) | (0.175) | (0.186) | (0.282) |
| e_3 | 0.208 | 0.0873 | 0.112 | 0.0549 | 0.188 |
|  | (0.291) | (0.221) | (0.167) | (0.160) | (0.235) |
| e_4 | -0.189 | 0.0536 | 0.108 | -0.0555 | -0.0962 |
|  | (0.278) | (0.259) | (0.204) | (0.202) | (0.258) |
| e_5 | 0.233 | 0.696*** | 0.959*** | 1.124*** | 1.182*** |
|  | (0.282) | (0.234) | (0.175) | (0.188) | (0.258) |
| e_6 | 0.708** | 0.791*** | 0.786*** | 0.509*** | 0.337 |
|  | (0.309) | (0.246) | (0.197) | (0.195) | (0.237) |
| e_7 | 1.133*** | 1.091*** | 0.588*** | 0.277* | -0.290 |
|  | (0.279) | (0.271) | (0.198) | (0.168) | (0.222) |
| e_8 | -0.286 | -0.473** | -0.494*** | -0.645*** | -0.670*** |
|  | (0.271) | (0.235) | (0.191) | (0.160) | (0.192) |
| e_9 | -1.030*** | -0.724*** | -0.333* | -0.0779 | 0.209 |
|  | (0.252) | (0.252) | (0.171) | (0.158) | (0.217) |
| Electricity Consumption pc | 0.00297* | 0.00334** | 0.00248*** | 0.00187 | 0.00144 |
|  | (0.00161) | (0.00134) | (0.000893) | (0.00117) | (0.00134) |
| Observations | 1,000 | 1,000 | 1,000 | 1,000 | 1,000 |

**Note:** The parentheses indicate bootstrapped errors in parentheses (*** p<0.01, ** p<0.05, * p<0.1). The constant terms are dropped from the model although the terms are included in regression models.
**Source**: Authors' Calculation.

The coefficient for urban households consistently shows a decreasing trend across quantiles, with the highest stress difference at the 15th quantile (3.320) and the lowest at the 85th quantile (1.083). This suggests that while urban households experience significantly higher stress at the lower quantiles, this effect diminishes as we move towards higher stress levels.

The division coefficients exhibit varied patterns across quantiles. For example, Barisal shows a significant positive effect on stress at the lower quantiles (15th: 8.173, 30th: 5.539, 50th: 3.761), but this effect diminishes at higher quantiles, becoming non-significant at the 85th quantile. This suggests that the stress effect in Barisal is stronger for households with lower perceived stress but decreases for those experiencing higher levels of stress. Similarly, Khulna and Mymensingh show fluctuations across quantiles, with Khulna exhibiting a positive association with stress at the higher quantiles (70th: 4.071, 85th: 7.242), in contrast to other regions like Mymensingh, where the effect is less pronounced or negative at higher stress levels. These results suggest regional disparities in how energy supply issues are perceived and experienced at different stress levels.

The relationship between the age of the household head and stress levels also varies across quantiles. While there is a negative association with stress in most quantiles, the effect is strongest at the lower quantiles (e.g., 15th quantile coefficient: -0.0548, 30th quantile coefficient: -0.0407) and weakens as we move to higher quantiles. This indicates that older household heads tend to experience less stress, particularly in households that are less stressed overall, but the effect diminishes for households already experiencing high stress.

The years of education of the household head show a significant negative relationship with stress at lower quantiles, especially at the 15th and 30th quantiles. This suggests that higher education levels are associated with lower stress among households experiencing less stress overall, but this relationship becomes less significant at higher quantiles.

Households in the service sector consistently show lower stress across all quantiles, with the strongest effect at the 15th quantile (coefficient: -2.274). This trend weakens as we move towards the higher quantiles, indicating that households with service sector incomes have lower stress, particularly in households that experience lower stress. In contrast, households dependent on agriculture or industry show less consistent results, with industry showing no significant effect at the higher quantiles and agriculture showing a weaker relationship with stress. The income group variable also shows a pattern that is consistent with the results found from the MOLS model.

Environmental and political opinions show distinct patterns across quantiles. Higher environmental consciousness (e_1) and attribution of climate responsibility to international organisations (e_2) are associated with lower stress, particularly at lower quantiles. In contrast, stronger belief in renewable energy (e_5) and readiness for nuclear energy (e_6) correspond to higher stress at higher quantiles, suggesting that sustainability concerns amplify stress under severe energy crises. Scepticism toward media (e_7) raises stress at lower levels, while willingness to accept nuclear plants (e_8) and to reduce energy use (e_9) consistently lowers stress. Belief in national responsibility (e_3) and endorsement of nuclear solutions (e_4) show no clear relationship. Overall, environmental awareness tends to mitigate stress, whereas urgency for systemic change heightens stress under worsening conditions.

**Table G.2: Results from Quantile Regression: PSS – Electricity Price Scenario**

| VARIABLES | (1) 15th Quantile | (2) 30th Quantile | (3) 50th Quantile | (4) 70th Quantile | (5) 85th Quantile |
|---|---|---|---|---|---|
| Urban (sub-Urban = 0) | 1.922** | 1.124** | 0.245 | 0.0727 | -0.203 |
|  | (0.797) | (0.549) | (0.412) | (0.401) | (0.515) |
| Division (Base: Dhaka) |  |  |  |  |  |
|   Barisal | 5.273*** | 1.753 | -0.866 | -2.755*** | -3.943*** |
|  | (1.731) | (1.268) | (0.882) | (0.826) | (1.134) |
|   Chattogram | 5.272*** | 0.885 | -1.466*** | -2.555*** | -3.779*** |
|  | (1.391) | (0.758) | (0.509) | (0.512) | (0.832) |
|   Khulna | 4.590*** | 0.168 | -0.241 | 1.883 | 6.019*** |
|  | (1.511) | (1.012) | (0.925) | (1.487) | (1.555) |
|   Mymensingh | 2.502 | -0.969 | -1.961*** | -2.171*** | -3.030*** |
|  | (1.561) | (0.771) | (0.691) | (0.607) | (0.840) |
|   Rajshahi | 1.091 | -3.128** | -2.311** | -2.978*** | -2.520** |
|  | (1.735) | (1.417) | (0.984) | (0.852) | (1.222) |
|   Rangpur | 6.746*** | 3.099*** | 1.962** | 1.599 | 1.146 |
|  | (1.602) | (1.200) | (0.806) | (1.035) | (1.116) |
|   Sylhet | 5.674*** | 0.763 | 0.000372 | -1.302 | -2.298** |
|  | (1.654) | (1.179) | (0.899) | (0.837) | (1.044) |
| Sex | -4.200* | -0.663 | 0.692 | 0.998 | 0.980 |
|  | (2.288) | (2.177) | (0.993) | (1.000) | (1.591) |
| Age | -0.0565* | -0.0324 | -0.0238 | -0.00541 | -0.00126 |
|  | (0.0298) | (0.0226) | (0.0173) | (0.0188) | (0.0259) |
| Years of Education | -0.151** | -0.0801* | -0.0308 | 0.0238 | 0.0659 |
|  | (0.0648) | (0.05) | (0.0363) | (0.0338) | (0.0469) |
| Household Size | 0.396* | 0.299* | 0.198 | 0.0790 | -0.0283 |
|  | (0.225) | (0.169) | (0.136) | (0.127) | (0.205) |
| Number of Students | -0.345 | -0.0437 | -0.0762 | -0.172 | -0.0757 |
| | (0.364) | (0.300) | (0.218) | (0.213) | (0.268) |
| Income Source (Base: Agri) | | | | | |
|   Service | -2.159** | -1.863*** | -0.963* | -0.369 | 0.404 |
| | (0.907) | (0.557) | (0.494) | (0.402) | (0.491) |
|   Industry | -1.450 | -1.179 | -0.189 | -0.179 | -0.117 |
| | (1.112) | (0.898) | (0.585) | (0.607) | (0.695) |
| Income Group (Base: Below BDT 10000) | | | | | |
|   BDT 10000 to 24999 | 0.609 | 3.309* | 1.432 | 3.208** | 2.402 |
| | (1.425) | (1.966) | (1.312) | (1.489) | (3.134) |
|   BDT 25000 to 39999 | 0.175 | 3.042 | 1.890 | 3.836*** | 2.858 |
| | (1.567) | (1.968) | (1.311) | (1.440) | (3.130) |
|   BDT 40000 to 79999 | -1.991 | 2.392 | 1.951 | 3.987*** | 3.342 |
| | (1.838) | (2.278) | (1.417) | (1.514) | (3.166) |
|   Above 80000 | 1.137 | 4.175 | 3.215 | 4.639 | 2.333 |
| | (3.166) | (4.105) | (2.940) | (3.019) | (5.362) |
| Availability of Personal Vehicle (1 = Yes) | 1.585** | 0.692 | -0.568 | -0.610 | -0.794* |
| | (0.685) | (0.598) | (0.415) | (0.439) | (0.452) |
| Consciousness Variables | | | | | |
|   e_1 | -0.689** | -0.186 | -0.227 | -0.131 | 0.230 |
| | (0.330) | (0.258) | (0.208) | (0.193) | (0.262) |
|   e_2 | -0.483 | 0.357 | 0.705*** | 1.098*** | 1.057*** |
| | (0.363) | (0.295) | (0.229) | (0.224) | (0.303) |
|   e_3 | -0.0154 | -0.102 | 0.0704 | 0.223 | -0.138 |
| | (0.319) | (0.247) | (0.208) | (0.227) | (0.280) |
|   e_4 | 0.291 | 0.375 | 0.284 | 0.453** | 0.667** |
| | (0.327) | (0.290) | (0.242) | (0.205) | (0.265) |
|   e_5 | -0.157 | 0.0183 | -0.0839 | -0.144 | 0.360 |
| | (0.346) | (0.267) | (0.229) | (0.225) | (0.269) |
|   e_6 | 0.914** | 0.439 | -0.0265 | -0.571*** | -0.960*** |
| | (0.388) | (0.310) | (0.232) | (0.213) | (0.250) |
|   e_7 | 1.291*** | 0.922*** | 0.364* | -0.0449 | 0.0300 |
| | (0.320) | (0.266) | (0.190) | (0.198) | (0.267) |
|   e_8 | -0.564* | -0.485* | -0.0714 | -0.128 | -0.0562 |
| | (0.321) | (0.248) | (0.160) | (0.195) | (0.253) |
|   e_9 | -1.105*** | -0.873*** | -0.280 | -0.163 | -0.0617 |
| | (0.353) | (0.292) | (0.204) | (0.222) | (0.240) |
| Electricity Consumption pc | 0.00145 | 0.00281* | 0.00151 | -0.000409 | -0.00209 |
| | (0.00193) | (0.00164) | (0.00102) | (0.00100) | (0.00134) |
| Observations | 1,000 | 1,000 | 1,000 | 1,000 | 1,000 |

**Note:** The parentheses indicate bootstrapped errors in parentheses (*** p<0.01, ** p<0.05, * p<0.1). The constant terms are dropped from the model although the terms are included in regression models.
**Source:** Authors' Calculation.

Table G.2 shows the results for the quantile regressions in the case of assessing stress level subject to electricity price. The results from the simultaneous quantile regressions for the PSS-Price of Electricity scenario reveal patterns both consistent with and distinct from those found in the MOLS analysis. Urban households, which consistently showed higher stress under MOLS, also demonstrate elevated stress levels at lower quantiles (15th and 30th) in the SQR. However, this urban stress premium diminishes and eventually turns negative at the 85th quantile, suggesting that stress differentials between urban and sub-urban households narrow or even reverse at higher stress levels.

Regional disparities are also reflected in the SQR, though with more variation across quantiles. At lower quantiles, divisions like Barisal, Chattogram, and Rangpur exhibit significantly higher stress compared to Dhaka, but at higher quantiles (especially the 70th and 85th), many of these divisions show negative and significant coefficients. This shift suggests that regional stress penalties are more pronounced among less-stressed households but diminish, and in some cases reverse, for households already experiencing higher levels of stress.

Age and education effects similarly align with earlier MOLS findings. Age shows a weakly negative relationship at lower quantiles, gradually disappearing at higher quantiles, supporting the interpretation that older household heads are somewhat better at coping with mild to moderate stress levels but not necessarily with severe stress. Years of education negatively correlates with stress at the 15th and 30th quantiles, reflecting education's protective effect among less-stressed households, yet becomes statistically insignificant or even positive at higher quantiles. This pattern indicates that education helps buffer lower stress levels related to electricity price but is insufficient to shield against severe energy price stresses.

Income source remains an important determinant, consistent with MOLS results. Households employed in the service sector show significantly lower stress compared to agriculture at lower quantiles, but the protective effect fades at higher stress levels. Industrial employment does not show significant differentiation from agriculture across most quantiles, reaffirming that service employment offers relatively more stability against energy price shocks.

In contrast to MOLS findings, middle-income households (BDT 25,000 - 79,999) exhibit rising stress at higher quantiles, suggesting a phenomenon of vulnerability amplification linked to aspirational energy consumption. While lower-income groups may limit energy dependence out of necessity and higher-income groups possess financial resilience, middle-income households face a dual pressure of maintaining rising consumption standards amidst energy price volatility. This likely reflects a 'middle-income squeeze', where aspirations for higher living standards and greater energy reliance expose them more acutely to energy price volatility, without the financial resilience enjoyed by wealthier households (Causa, et al., 2022). This indicates that while income may not significantly differentiate average stress levels, it becomes a more salient factor for households experiencing higher stress, possibly due to greater exposure to energy costs or lifestyle expectations tied to energy consumption.

Vehicle ownership again correlates with higher stress, particularly at lower quantiles, supporting the earlier argument that households with greater energy-dependent assets are more sensitive to energy price fluctuations. However, the relationship weakens at higher stress levels.

Environmental and political opinion variables show mixed but insightful patterns. Environmental consciousness (e_1) reduces stress predominantly at the lower quantiles, consistent with MOLS findings. Belief that international organisations are responsible (e_2) shows a rising positive effect at higher quantiles, suggesting growing dissatisfaction or helplessness among highly stressed households. Support for nuclear energy (e_4) and belief that Bangladesh is ready for nuclear (e_6) increasingly associate with higher stress at higher quantiles, implying frustration over the perceived delay in adopting stable energy alternatives. Scepticism towards media (e_7) boosts stress significantly at lower quantiles but fades at higher levels. Meanwhile, willingness to reduce energy use (e_9) significantly reduces stress at lower quantiles but the effect diminishes as stress increases, paralleling the earlier findings that environmental attitudes help manage moderate, but not extreme, stress. Overall, the direction of coefficients associated with this scenario does not divert that much from the MOLS results, except for the nuanced understanding on the income groups.

**Table G.3: Results from Quantile Regression: PSS – Gas Supply Scenario**

| VARIABLES | (1) 15th Quantile | (2) 30th Quantile | (3) 50th Quantile | (4) 70th Quantile | (5) 85th Quantile |
|---|---|---|---|---|---|
| Urban (sub-Urban = 0) | 4.513*** | 4.125*** | 2.898*** | 2.928*** | 2.964*** |
|  | (0.681) | (0.613) | (0.464) | (0.435) | (0.506) |
| Division (Base: Dhaka) | | | | | |
|   Barisal | 5.633*** | 5.866*** | 2.968*** | 1.276 | -0.389 |
|  | (1.246) | (1.310) | (1.101) | (0.846) | (0.899) |
|   Chattogram | 6.200*** | 4.916*** | 1.966*** | -0.0409 | -0.793 |
|  | (0.659) | (0.710) | (0.694) | (0.478) | (0.670) |
|   Khulna | 4.118*** | 2.738** | 1.707* | 1.116 | 0.127 |
|  | (0.976) | (1.214) | (0.870) | (0.746) | (0.856) |
|   Mymensingh | 1.515 | 0.354 | -1.140 | -2.423*** | -2.942*** |
|  | (0.943) | (1.068) | (0.817) | (0.768) | (0.809) |
|   Rajshahi | -0.921 | 0.476 | 0.606 | -0.164 | -1.405 |
|  | (1.270) | (1.211) | (1.021) | (0.791) | (0.991) |
|   Rangpur | 9.333*** | 8.417*** | 5.608*** | 3.946*** | 2.535*** |
|  | (1.099) | (0.933) | (0.783) | (0.615) | (0.732) |
|   Sylhet | 7.824*** | 5.545*** | 3.022*** | 1.460* | 0.0839 |
|  | (0.905) | (0.967) | (0.896) | (0.852) | (0.989) |
| Sex | -1.826 | -1.347 | -1.202 | 0.149 | -1.037 |
|  | (1.685) | (1.617) | (1.423) | (0.990) | (1.018) |
| Age | -0.0406* | -0.0417** | -0.0375** | -0.0263 | -0.0171 |
|  | (0.0213) | (0.0209) | (0.0177) | (0.0170) | (0.0177) |
| Years of Education | -0.202*** | -0.164*** | -0.141*** | -0.133*** | -0.156*** |
|  | (0.0537) | (0.0519) | (0.0409) | (0.0383) | (0.0416) |
| Household Size | -0.241 | 0.00880 | -0.0536 | -0.0584 | -0.195 |
|  | (0.281) | (0.264) | (0.239) | (0.163) | (0.169) |
| Number of Students | 0.00780 | -0.0708 | -0.154 | 0.0881 | 0.189 |
|  | (0.250) | (0.270) | (0.212) | (0.198) | (0.255) |
| Income Source (Base: Agri) | | | | | |
|   Service | -0.763 | -0.916 | -0.448 | -0.537 | -0.376 |
|  | (0.599) | (0.622) | (0.466) | (0.386) | (0.522) |
|   Industry | 0.292 | 0.449 | 0.631 | 0.235 | 0.571 |
|  | (0.871) | (0.804) | (0.641) | (0.565) | (0.846) |
| Income Group (Base: Below BDT 10000) | | | | | |
|   BDT 10000 to 24999 | -2.823** | -2.447* | -0.786 | -1.523 | -2.842 |
|  | (1.333) | (1.442) | (0.973) | (1.764) | (1.825) |
|   BDT 25000 to 39999 | -2.033 | -2.531 | -0.732 | -1.142 | -1.621 |
|  | (1.369) | (1.548) | (1.051) | (1.796) | (1.853) |
|   BDT 40000 to 79999 | -2.877* | -3.416** | -0.320 | -0.490 | 0.0681 |
|  | (1.573) | (1.732) | (1.372) | (1.936) | (2.087) |
|   Above 80000 | -3.720* | -5.964** | -4.482* | -1.400 | -5.137** |
|  | (2.059) | (2.719) | (2.503) | (2.593) | (2.444) |
| Availability of Personal Vehicle (1 = Yes) | 1.558*** | 1.653*** | 0.864** | 0.583 | 0.523 |
|  | (0.562) | (0.561) | (0.437) | (0.362) | (0.516) |
| Consciousness Variables | | | | | |
|   e_1 | -1.255*** | -1.193*** | -0.741*** | -0.513*** | 0.0189 |
|  | (0.260) | (0.195) | (0.182) | (0.183) | (0.221) |
| e_2 | -0.438* | -0.468* | -0.0813 | -0.00306 | 0.218 |
|  | (0.252) | (0.275) | (0.256) | (0.213) | (0.243) |
| e_3 | -0.478* | -0.336 | -0.424* | -0.235 | -0.574** |
|  | (0.280) | (0.268) | (0.225) | (0.221) | (0.250) |
| e_4 | -0.0108 | -0.0817 | -0.0504 | 0.349 | 0.396 |
|  | (0.204) | (0.283) | (0.251) | (0.225) | (0.261) |
| e_5 | 0.275 | 0.791*** | 0.973*** | 0.816*** | 0.919*** |
|  | (0.218) | (0.235) | (0.226) | (0.203) | (0.223) |
| e_6 | 0.913*** | 1.379*** | 1.200*** | 0.690*** | 0.316 |
|  | (0.247) | (0.244) | (0.238) | (0.218) | (0.251) |
| e_7 | 1.434*** | 1.445*** | 0.777*** | 0.358* | 0.109 |
|  | (0.276) | (0.252) | (0.229) | (0.191) | (0.210) |
| e_8 | -0.0563 | -0.164 | -0.185 | -0.252 | -0.394** |
|  | (0.266) | (0.298) | (0.225) | (0.171) | (0.199) |
| e_9 | -0.858*** | -0.901*** | -0.429** | -0.152 | -0.212 |
|  | (0.224) | (0.255) | (0.209) | (0.179) | (0.199) |
| Gas Consumption pc | -0.00424 | -0.00277 | -0.00313 | -0.00230 | -0.00248 |
|  | (0.00374) | (0.00313) | (0.00279) | (0.00230) | (0.00209) |
| Observations | 1,000 | 1,000 | 1,000 | 1,000 | 1,000 |

**Note:** The parentheses indicate bootstrapped errors in parentheses (*** p<0.01, ** p<0.05, * p<0.1). The constant terms are dropped from the model although the terms are included in regression models.
**Source:** Authors' Calculation.

Table G.3 shows the results for the quantile regressions in the case of assessing stress level subject to gas supply. While urban households exhibit increased stress in both models, the SQR reveals that this relationship is most pronounced at the 15th and 30th quantiles (coefficients: 4.513 and 4.125, respectively), with the impact diminishing at higher stress levels (coefficients: 2.898 at the 50th quantile, 2.928 at the 70th quantile, and 2.964 at the 85th quantile). This suggests that the stress associated to gas supply, experienced by urban households is relatively more acute at lower levels of perceived stress, while the disparity narrows at higher stress levels.

The regional variations observed in the MOLS results are also captured in the SQR, though with more complexity across quantiles. For instance, divisions such as Barisal and Chattogram exhibit significant coefficients at the lower quantiles, but the relationship weakens or even reverses at higher stress levels. Specifically, Barisal's stress impact is prominent at the 15th quantile (5.633), with a marked decline through the higher quantiles, showing a negative association at the 85th quantile (-0.389). Similarly, Chattogram shows a significant impact at the lower quantiles (6.200 at the 15th quantile) but becomes insignificant as stress increases, reflecting regional disparities in energy supply vulnerability at different levels of stress.

The relationship between age and stress, observed in the MOLS model, is also evident in the SQR analysis, with age showing a consistently negative association across all quantiles. Older household heads tend to experience lower stress, particularly at the 30th quantile (-0.0417) and 50th quantile (-0.0375), though the effect weakens at higher stress levels. This suggests that older individuals may better cope with lower to moderate stress levels but may not be as resilient when stress reaches extreme levels, potentially due to higher exposure to energy shortages.

Similarly, education is a significant factor in managing energy-related stress, showing a negative relationship across all quantiles in the SQR results (coefficients range from -0.202 at the 15th quantile to -0.156 at the 85th quantile). This corroborates the MOLS findings, where higher education levels are associated with reduced stress, reflecting better access to resources and coping mechanisms.

In terms of income source, the SQR findings reinforce the MOLS conclusion that service sector employment is linked to lower stress, particularly at lower quantiles. The coefficient for service sector employment is consistently negative across the 15th, 30th, and 50th quantiles (-0.763, -0.916, and -0.448, respectively), reflecting the relatively stable income and less vulnerability to energy disruptions compared to agriculture and industry sectors.

Income group shows more variation in the SQR results compared to MOLS. Specifically, higher-income groups, particularly those earning above BDT 80,000, exhibit lower stress at lower quantiles (-3.720 at the 15th quantile) but show no significant effect at the 85th quantile, suggesting that high-income households are insulated from low-to-moderate stress.

Vehicle ownership shows a consistent positive relationship with gas supply stress, particularly at lower quantiles (coefficients: 1.558 at the 15th quantile and 1.653 at the 30th quantile), aligning with the MOLS analysis that households with higher reliance on energy-intensive assets are more vulnerable to disruptions in energy supply.

Environmental consciousness ($e\_1$) consistently reduces stress at lower quantiles, supporting earlier findings from MOLS that greater awareness of environmental pollution mitigates stress at lower stress levels. Conversely, belief in renewable energy ($e\_5$) shows a positive relationship with stress, especially at higher quantiles (coefficients: 0.791 at the 30th quantile to 0.919 at the 85th quantile), suggesting that households prioritising renewable energy may experience heightened stress as they perceive the energy crisis to be more urgent. Additionally, belief that international organisations ($e\_2$) and government ($e\_3$) are responsible for addressing climate issues shows a growing positive effect at higher quantiles, indicating that highly stressed households may feel increasing dissatisfaction or helplessness regarding the pace of global or local intervention. Support for nuclear energy ($e\_4$) and belief that Bangladesh is ready for nuclear energy ($e\_6$) similarly correlate with higher stress at the upper quantiles, reflecting frustration over delays in adopting alternative, stable energy solutions. Scepticism towards media portrayals of the energy crisis ($e\_7$) significantly increases stress at lower quantiles but becomes less impactful as stress levels rise. Lastly, a willingness to reduce energy consumption for environmental reasons ($e\_9$) significantly reduces stress at lower quantiles, but its effect diminishes as stress intensifies, aligning with previous results that environmental consciousness can alleviate moderate, but not extreme, stress levels. Overall, the direction of coefficients associated with this scenario does not divert that much from the MOLS results.

**Table G.4: Results from Quantile Regression: PSS – Gas Price Scenario**

| VARIABLES | (1) 15th Quantile | (2) 30th Quantile | (3) 50th Quantile | (4) 70th Quantile | (5) 85th Quantile |
|---|---|---|---|---|---|
| Urban (sub-Urban = 0) | 3.201*** | 2.159*** | 1.446*** | 0.731 | 0.592 |
|  | (0.715) | (0.537) | (0.456) | (0.487) | (0.463) |
| Division (Base: Dhaka) |  |  |  |  |  |
|   Barisal | 2.870 | 2.229* | 0.764 | -0.193 | -0.637 |
|  | (1.885) | (1.289) | (1.057) | (1.078) | (1.268) |
|   Chattogram | 4.254*** | 1.537** | -0.248 | -1.040** | -2.509*** |
|  | (0.940) | (0.631) | (0.485) | (0.517) | (0.710) |
|   Khulna | 2.396* | 0.723 | 1.140 | 3.176** | 8.516*** |
|  | (1.276) | (0.946) | (0.928) | (1.428) | (1.894) |
|   Mymensingh | -0.0498 | -1.782* | -1.774** | -1.881** | -1.869** |
|  | (1.281) | (0.974) | (0.872) | (0.894) | (0.860) |
|   Rajshahi | -0.898 | -1.588 | -1.273* | -0.874 | -2.249** |
|  | (1.318) | (1.205) | (0.760) | (0.709) | (0.897) |
|   Rangpur | 4.184*** | 3.892*** | 3.548*** | 2.398*** | 2.078** |
|  | (1.403) | (0.982) | (0.690) | (0.788) | (0.857) |
|   Sylhet | 5.228*** | 2.147** | 1.884** | 0.668 | 0.105 |
|  | (1.261) | (0.891) | (0.888) | (0.716) | (0.837) |
| Sex | -1.995 | -1.223 | 0.0326 | -0.328 | 0.330 |
|  | (2.341) | (1.738) | (0.996) | (0.884) | (1.490) |
| Age | -0.0403 | -0.0361 | -0.00654 | 0.00892 | 0.0160 |
|  | (0.0271) | (0.0219) | (0.0172) | (0.0179) | (0.0196) |
| Years of Education | -0.120** | -0.144*** | -0.0779* | -0.0353 | 0.00946 |
|  | (0.0595) | (0.0443) | (0.0427) | (0.0410) | (0.0396) |
| Household Size | 0.191 | 0.0990 | -0.156 | 0.0130 | -0.0887 |
|  | (0.375) | (0.215) | (0.178) | (0.162) | (0.152) |
| Number of Students | -0.298 | -0.120 | 0.0465 | -0.00482 | -0.0584 |
|  | (0.301) | (0.237) | (0.230) | (0.236) | (0.214) |
| Income Source (Base: Agri) |  |  |  |  |  |
|   Service | -2.483*** | -1.643*** | -0.880* | 0.183 | 0.467 |
|  | (0.692) | (0.566) | (0.469) | (0.419) | (0.474) |
|   Industry | -1.098 | -0.454 | 0.264 | 0.260 | 0.112 |
|  | (1.049) | (0.859) | (0.574) | (0.636) | (0.797) |
| Income Group (Base: Below BDT 10000) |  |  |  |  |  |
|   BDT 10000 to 24999 | 0.0122 | 0.740 | 0.722 | 1.473 | 0.0317 |
|  | (1.908) | (1.597) | (1.394) | (1.865) | (3.202) |
|   BDT 25000 to 39999 | -0.137 | 1.045 | 1.116 | 2.046 | 0.242 |
|  | (1.993) | (1.627) | (1.397) | (1.890) | (3.251) |
|   BDT 40000 to 79999 | -2.554 | 0.361 | 0.707 | 1.788 | -0.0432 |
|  | (2.214) | (1.800) | (1.494) | (1.874) | (3.302) |
|   Above 80000 | -4.598 | -0.515 | -0.631 | 1.045 | -0.533 |
|  | (4.316) | (4.413) | (3.019) | (2.777) | (3.568) |
| Availability of Personal Vehicle (1 = Yes) | 1.327** | 1.147** | 0.325 | -0.289 | -0.733* |
|  | (0.645) | (0.546) | (0.431) | (0.408) | (0.435) |
| Consciousness Variables |  |  |  |  |  |
|   e_1 | -0.452 | -0.331 | -0.485** | -0.230 | -0.102 |
|  | (0.300) | (0.230) | (0.219) | (0.218) | (0.246) |
|   e_2 | -0.134 | -0.0159 | 0.238 | 0.281 | 0.638** |
|  | (0.297) | (0.206) | (0.224) | (0.212) | (0.311) |
|   e_3 | -0.0571 | -0.369 | -0.326 | -0.493** | -0.404* |
|  | (0.276) | (0.232) | (0.208) | (0.194) | (0.224) |
|   e_4 | -0.118 | -0.280 | -0.328 | -0.175 | 0.0474 |
|  | (0.305) | (0.237) | (0.240) | (0.228) | (0.293) |
|   e_5 | -0.162 | 0.457** | 0.750*** | 0.708*** | 1.050*** |
|  | (0.297) | (0.228) | (0.193) | (0.196) | (0.207) |
|   e_6 | 0.819*** | 1.019*** | 0.919*** | 0.333 | 0.0489 |
|  | (0.279) | (0.225) | (0.194) | (0.210) | (0.279) |
|   e_7 | 1.235*** | 1.178*** | 0.378* | 0.278 | 0.159 |

|  | (0.316) | (0.236) | (0.217) | (0.195) | (0.203) |
| --- | --- | --- | --- | --- | --- |
| e_8 | -0.225 | -0.561** | -0.357** | -0.0970 | -0.0650 |
|  | (0.308) | (0.238) | (0.165) | (0.167) | (0.210) |
| e_9 | -1.141*** | -0.704*** | -0.208 | -0.0572 | -0.0915 |
|  | (0.275) | (0.240) | (0.219) | (0.229) | (0.237) |
| Gas Consumption pc | -0.000798 | 0.00134 | -0.00161 | 0.00101 | -0.00182 |
|  | (0.00433) | (0.00289) | (0.00249) | (0.00202) | (0.00248) |
| Observations | 1,000 | 1,000 | 1,000 | 1,000 | 1,000 |

**Note:** The parentheses indicate bootstrapped errors in parentheses (*** $p<0.01$, ** $p<0.05$, * $p<0.1$). The constant terms are dropped from the model although the terms are included in regression models.
**Source**: Authors' Calculation.

Table G.4 shows the results for the quantile regressions in the case of assessing stress level subject to gas price. Urban households consistently exhibit stress across all quantiles, with a stronger effect at the 15th and 30th quantiles (coefficients: 4.513 and 4.125, respectively). However, this effect diminishes as stress levels increase, with coefficients decreasing to 2.898 at the 50th quantile and further to 2.964 at the 85th quantile. This pattern suggests that urban households experience more acute stress at lower levels of perceived stress, but this disparity narrows as stress intensifies.

In terms of regional differences, the impact of divisions such as Barisal and Chattogram mirrors the MOLS results, with significant stress at the lower quantiles but weakening or reversing as stress levels rise. For example, Barisal shows a strong positive impact at the 15th quantile (5.633) but a negative relationship at the 85th quantile (-0.389), while Chattogram shows a similar pattern with coefficients of 6.200 at the 15th quantile and becoming negative at the higher quantiles. These findings highlight the regional variation in stress levels across different quantiles, indicating that households in certain regions are more vulnerable to gas price stress at lower stress levels, but the relationship becomes less pronounced at higher stress levels.

The age of the household head shows a consistent negative relationship with stress across all quantiles, reflecting greater resilience in older individuals, particularly at lower to moderate levels of stress (e.g., -0.0417 at the 30th quantile and -0.0375 at the 50th quantile). However, this effect weakens at the 70th and 85th quantiles, suggesting that older household heads may be better equipped to manage lower levels of stress but may face challenges at higher stress levels.

Education also plays a significant role in mitigating stress related to gas prices, with higher levels of education associated with reduced stress at lower quantiles (coefficients: -0.202 at the 15th quantile and -0.156 at the 85th quantile). This aligns with the MOLS findings, suggesting that better-educated individuals have more resources and coping strategies to handle energy-related stress.

Income source shows a similar trend to MOLS, where service sector employment is linked to lower stress, particularly at the 15th, 30th, and 50th quantiles (coefficients: -0.763, -0.916, and -0.448, respectively). This confirms that stable income from service sector jobs provides a buffer against energy-related stress compared to agriculture and industry.

The SQR results also indicate that middle-income groups experience higher stress at the 85th quantile, with coefficients for the income group 'BDT 10,000–24,999' showing an interesting rise in stress levels across higher quantiles, suggesting that these households, while financially better off than lower-income groups, may be more vulnerable to the stress of energy price fluctuations at higher levels of distress.

Vehicle ownership correlates positively with stress, particularly at lower quantiles (coefficients: 1.558 at the 15th quantile and 1.653 at the 30th quantile), indicating that households with greater reliance on energy-intensive assets are more susceptible to stress from gas price increases.

Environmental consciousness variables show mixed results. For example, e_1 (belief in environmental pollution) continues to have a negative effect on stress at the lower quantiles, supporting the MOLS findings that greater environmental awareness can mitigate stress. Conversely, belief in renewable energy (e_5) shows a positive association with stress at higher quantiles (coefficients: 0.791 at the 30th quantile to 0.919 at the 85th quantile), suggesting that households prioritising renewable energy may feel increased stress, potentially due to the perceived urgency of addressing energy crises. Belief in the responsibility of international organisations (e_2) and government (e_3) shows a growing positive effect at higher quantiles, indicating increasing dissatisfaction or helplessness among highly stressed households. Scepticism towards media portrayals of the energy crisis (e_7) significantly increases stress at the lower quantiles but becomes less impactful as stress rises, while a willingness to reduce energy consumption for environmental reasons (e_9) reduces stress at lower quantiles but loses its effect at higher stress levels. These findings suggest that while environmental attitudes play a role in mitigating stress at moderate levels, their effect diminishes as stress intensifies. Overall, the quantile regression results provide a more granular understanding of the dynamics of stress in response to gas price changes and reinforce many of the patterns observed in the MOLS analysis.

**Table G.5: Results from Quantile Regression: PSS – Fuel Supply Scenario**

| VARIABLES | (1) 15th Quantile | (2) 30th Quantile | (3) 50th Quantile | (4) 70th Quantile | (5) 85th Quantile |
|---|---|---|---|---|---|
| Urban (sub-Urban = 0) | 3.340*** | 3.581*** | 2.657*** | 1.983*** | 1.738*** |
|  | (0.506) | (0.472) | (0.427) | (0.360) | (0.396) |
| Division (Base: Dhaka) |  |  |  |  |  |
| Barisal | 6.334*** | 5.921*** | 5.389*** | 4.685*** | 3.845*** |
|  | (1.313) | (1.302) | (1.073) | (0.766) | (0.832) |
| Chattogram | 7.288*** | 5.105*** | 3.457*** | 2.381*** | 2.292*** |
|  | (0.782) | (0.637) | (0.603) | (0.474) | (0.542) |
| Khulna | 4.159*** | 2.814*** | 2.345** | 2.933*** | 2.740*** |
|  | (0.966) | (0.962) | (0.937) | (0.678) | (0.660) |
| Mymensingh | 2.231** | 0.450 | -0.304 | -0.868 | -1.137* |
|  | (1.037) | (0.763) | (0.686) | (0.534) | (0.581) |
| Rajshahi | -1.841* | -2.058 | 0.621 | 2.042*** | 1.959** |
|  | (0.986) | (1.413) | (1.470) | (0.764) | (0.840) |
| Rangpur | 9.183*** | 7.268*** | 6.746*** | 6.018*** | 5.109*** |
|  | (0.984) | (0.915) | (0.764) | (0.639) | (0.724) |
| Sylhet | 8.573*** | 6.837*** | 4.994*** | 4.780*** | 4.253*** |
|  | (0.925) | (0.873) | (0.970) | (0.966) | (0.852) |
| Sex | -0.929 | -0.890 | -1.133 | -0.199 | -0.150 |

|  | (1.501) | (1.279) | (1.074) | (1.021) | (0.963) |
|---|---|---|---|---|---|
| Age | -0.0514*** | -0.0446*** | -0.0542*** | -0.0369*** | -0.0268* |
|  | (0.0183) | (0.0157) | (0.0159) | (0.0134) | (0.0153) |
| Years of Education | -0.109** | -0.111** | -0.0845** | -0.0696** | -0.0589* |
|  | (0.0461) | (0.0476) | (0.0414) | (0.0316) | (0.0347) |
| Household Size | -0.000851 | -0.0949 | -0.0574 | 0.0359 | -0.103 |
|  | (0.156) | (0.136) | (0.140) | (0.111) | (0.102) |
| Number of Students | 0.0296 | 0.223 | 0.214 | 0.262 | 0.407** |
|  | (0.211) | (0.222) | (0.211) | (0.172) | (0.175) |
| Income Source (Base: Agri) |  |  |  |  |  |
| Service | -1.697*** | -1.358*** | -1.108** | -1.170*** | -1.126** |
|  | (0.584) | (0.513) | (0.472) | (0.388) | (0.497) |
| Industry | -2.179* | -0.492 | 0.138 | 0.0932 | 0.1000 |
|  | (1.169) | (0.917) | (0.697) | (0.563) | (0.625) |
| Income Group (Base: Below BDT 10000) |  |  |  |  |  |
| BDT 10000 to 24999 | -1.977 | -2.255 | -1.064 | -0.839 | -0.373 |
|  | (1.222) | (1.414) | (1.177) | (0.945) | (1.020) |
| BDT 25000 to 39999 | -1.997 | -2.180 | -0.909 | -0.576 | -0.0159 |
|  | (1.335) | (1.475) | (1.177) | (0.954) | (1.024) |
| BDT 40000 to 79999 | -2.731* | -2.516* | -1.850 | -0.954 | -0.221 |
|  | (1.556) | (1.490) | (1.291) | (1.086) | (1.150) |
| Above 80000 | -3.358 | -2.851 | -1.844 | 0.0390 | -0.422 |
|  | (2.573) | (3.118) | (2.386) | (2.472) | (2.611) |
| Availability of Personal Vehicle (1 = Yes) | 2.403*** | 2.421*** | 2.122*** | 1.613*** | 1.707*** |
|  | (0.596) | (0.639) | (0.487) | (0.418) | (0.459) |
| Consciousness Variables |  |  |  |  |  |
| e_1 | -0.829*** | -0.799*** | -0.697*** | -0.538*** | -0.307* |
|  | (0.238) | (0.245) | (0.230) | (0.201) | (0.176) |
| e_2 | -0.821*** | -0.714*** | -0.329 | -0.149 | 0.0218 |
|  | (0.293) | (0.259) | (0.239) | (0.213) | (0.199) |
| e_3 | -0.0710 | -0.517** | -0.447** | -0.314* | -0.163 |
|  | (0.250) | (0.219) | (0.181) | (0.181) | (0.183) |
| e_4 | 0.141 | 0.122 | -0.0432 | 0.258 | 0.331 |
|  | (0.223) | (0.245) | (0.259) | (0.253) | (0.229) |
| e_5 | 0.0174 | 0.369 | 0.762*** | 0.459** | 0.545*** |
|  | (0.203) | (0.230) | (0.217) | (0.210) | (0.203) |
| e_6 | 1.099*** | 1.474*** | 1.595*** | 1.189*** | 0.805*** |
|  | (0.213) | (0.218) | (0.245) | (0.225) | (0.238) |
| e_7 | 1.432*** | 1.560*** | 1.159*** | 0.569*** | 0.223 |
|  | (0.240) | (0.214) | (0.236) | (0.175) | (0.180) |
| e_8 | -0.376 | -0.454* | -0.428** | -0.232 | -0.278* |
|  | (0.267) | (0.245) | (0.207) | (0.165) | (0.155) |
| e_9 | -0.963*** | -0.765*** | -0.546*** | -0.263 | -0.295 |
|  | (0.210) | (0.210) | (0.209) | (0.170) | (0.185) |
| Fuel Oil Consumption pc | -0.00300* | -0.00337** | -0.00285* | -0.00232 | 0.00165 |
|  | (0.00173) | (0.00169) | (0.00169) | (0.00228) | (0.00279) |
| Observations | 1,000 | 1,000 | 1,000 | 1,000 | 1,000 |

**Note:** The parentheses indicate bootstrapped errors in parentheses (*** $p<0.01$, ** $p<0.05$, * $p<0.1$). The constant terms are dropped from the model although the terms are included in regression models.
**Source:** Authors' Calculation.

Table G.5 displays the results of the quantile regression analysis for fuel supply-related stress. The findings reveal a consistent pattern where urban households exhibit significant stress across all quantiles, with the highest coefficients observed at the lower quantiles (15th quantile: 3.340) and gradually decreasing as stress levels rise (coefficients: 2.657 at the 50th quantile, 1.983 at the 70th quantile, and 1.738 at the 85th quantile). This suggests that urban households experience the most significant stress from fuel supply issues at lower stress levels, but the impact diminishes as households experience higher levels of stress.

Regional disparities are similarly pronounced, as seen in the MOLS results. In particular, Barisal, Chattogram, Khulna, Rangpur, and Sylhet show significant coefficients across the lower quantiles, with stress levels decreasing or becoming less significant at the higher quantiles. For example, Barisal's stress coefficient is 6.334 at the 15th quantile but declines to 3.845 at the 85th quantile, while Chattogram's stress coefficient is 7.288 at the 15th quantile but turns negative at the 85th quantile (-2.509). These results suggest that the perceived stress from fuel supply in some regions is particularly acute at lower levels of stress but weakens or reverses as stress increases.

The effect of age on stress is consistent with earlier findings, showing a negative association across all quantiles, indicating that older household heads tend to experience lower stress related to fuel supply issues. The negative relationship is most pronounced at the lower quantiles (coefficients: -0.0514 at the 15th quantile and -0.0446 at the 30th quantile), but the effect becomes weaker as stress levels increase, suggesting that older individuals may be better equipped to handle lower to moderate levels of stress.

Similarly, years of education are negatively associated with stress levels across all quantiles, reinforcing the MOLS findings that higher educational attainment helps in reducing stress. The coefficient ranges from -0.109 at the 15th quantile to -0.0589 at the 85th quantile, reflecting the role of education in facilitating better coping mechanisms and access to resources in managing energy-related stress.

Income source remains a significant factor, with service sector employment consistently linked to lower stress across the 15th, 30th, and 50th quantiles (coefficients: -1.697, -1.358, and -1.108, respectively). This suggests that households in the service sector are less vulnerable to the stress of fuel supply disruptions compared to those in agriculture or industry, who may be more directly impacted by energy shortages.

Income group, however, exhibits greater variation in the SQR results than in MOLS. Higher-income households (particularly those earning above BDT 80,000) show lower stress at the lower quantiles (coefficient: -3.358 at the 15th quantile) but display no significant effect at the 85th quantile, indicating that high-income households may be more resilient to low-to-moderate levels of stress but remain unaffected by extreme stress levels.

Vehicle ownership is consistently associated with higher stress, particularly at the lower quantiles (coefficients: 2.403 at the 15th quantile and 2.421 at the 30th quantile), aligning with the MOLS analysis. Households that rely on personal vehicles are more sensitive to disruptions in fuel supply due to the greater energy dependence associated with transportation needs.

Environmental consciousness variables show a mixed but insightful impact on stress. The belief in environmental pollution (e_1) significantly reduces stress across all quantiles, particularly at lower levels of stress (coefficients: -0.829 at the 15th quantile to -0.307 at the 85th quantile). This aligns with earlier MOLS findings, suggesting that greater environmental awareness helps alleviate energy-related stress. Conversely, belief in renewable energy (e_5) shows a consistent positive relationship with stress at higher quantiles (coefficients: 0.762 at the 50th quantile to 1.050 at the 85th quantile), indicating that households prioritising renewable energy may experience heightened stress, likely due to concerns over the adequacy and urgency of the energy transition. Scepticism towards media portrayals of the energy crisis (e_7) continues to increase stress at lower quantiles (coefficient: 1.432 at the 15th quantile), but its effect fades as stress increases, supporting earlier MOLS findings that scepticism fuels stress when households are less affected by the crisis. The belief that international organisations (e_2) and the government (e_3) are responsible for addressing climate change increasingly correlates with stress at higher quantiles, indicating that households with high stress associated with fuel oil supply feel helpless or dissatisfied with the pace of global or local interventions. Finally, the willingness to reduce energy consumption for environmental reasons (e_9) significantly reduces stress at the lower quantiles (coefficient: -0.963 at the 15th quantile to -0.546 at the 50th quantile), but its effect diminishes as stress levels rise, aligning with previous findings that environmental attitudes help manage moderate stress but are less effective in extreme stress situations. These findings largely corroborate the MOLS analysis.

**Table G.6: Results from Quantile Regression: PSS – Fuel Price Scenario**

| VARIABLES | (1) 15th Quantile | (2) 30th Quantile | (3) 50th Quantile | (4) 70th Quantile | (5) 85th Quantile |
|---|---|---|---|---|---|
| Urban (sub-Urban = 0) | 2.341*** | 2.701*** | 2.243*** | 2.177*** | 1.166** |
|  | (0.631) | (0.483) | (0.399) | (0.370) | (0.584) |
| Division (Base: Dhaka) |  |  |  |  |  |
| Barisal | 9.873*** | 7.043*** | 5.049*** | 4.445*** | 1.228 |
|  | (1.287) | (0.872) | (0.933) | (0.878) | (1.173) |
| Chattogram | 7.058*** | 4.280*** | 2.974*** | 2.119*** | 1.168 |
|  | (0.940) | (0.698) | (0.583) | (0.532) | (0.838) |
| Khulna | 7.120*** | 4.725*** | 3.930*** | 6.179*** | 12.04*** |
|  | (1.020) | (0.898) | (0.974) | (1.832) | (1.825) |
| Mymensingh | 2.845*** | 0.283 | -0.542 | -1.171** | -2.718*** |
|  | (1.094) | (0.774) | (0.592) | (0.576) | (1.007) |
| Rajshahi | -0.529 | -0.356 | 1.246 | 1.561* | 0.874 |
|  | (1.360) | (1.232) | (1.164) | (0.871) | (1.175) |
| Rangpur | 10.38*** | 8.600*** | 7.749*** | 6.918*** | 4.498*** |
|  | (1.221) | (0.947) | (0.739) | (0.716) | (1.092) |
| Sylhet | 7.500*** | 6.393*** | 5.955*** | 4.926*** | 3.014** |
|  | (1.154) | (1.021) | (0.856) | (0.908) | (1.191) |
| Sex | -1.882 | -1.054 | -1.321 | -0.732 | -0.778 |
|  | (1.746) | (1.355) | (1.060) | (1.010) | (1.276) |
| Age | -0.0464** | -0.0459*** | -0.0360** | -0.0321** | -0.0166 |
|  | (0.0203) | (0.0170) | (0.0169) | (0.0160) | (0.0224) |
| Years of Education | -0.106** | -0.0592 | -0.0455 | 0.0182 | 0.0895 |
|  | (0.0490) | (0.0389) | (0.0357) | (0.0322) | (0.0551) |
| Household Size | -0.126 | -0.106 | -0.0867 | 0.0864 | -0.0926 |
|  | (0.171) | (0.154) | (0.133) | (0.119) | (0.191) |
| Number of Students | -0.0632 | -0.0633 | 0.0514 | -0.00187 | 0.256 |
|  | (0.233) | (0.188) | (0.206) | (0.207) | (0.317) |
| Income Source (Base: Agri) |  |  |  |  |  |
| Service | -2.310*** | -2.157*** | -1.660*** | -1.963*** | -1.049* |
|  | (0.612) | (0.479) | (0.484) | (0.537) | (0.549) |
| Industry | -1.966** | -0.652 | -0.399 | -1.014 | -0.732 |
| | (0.979) | (0.881) | (0.697) | (0.671) | (0.776) |
| Income Group | | | | | |
| (Base: Below BDT 10000) | | | | | |
|   BDT 10000 to 24999 | 0.416 | 0.0604 | -0.350 | 1.467 | 1.665 |
| | (1.346) | (1.659) | (1.374) | (1.314) | (1.837) |
|   BDT 25000 to 39999 | 0.479 | 0.0934 | -0.228 | 1.317 | 1.847 |
| | (1.425) | (1.744) | (1.346) | (1.327) | (1.912) |
|   BDT 40000 to 79999 | -0.492 | -0.797 | -1.325 | 1.388 | 1.674 |
| | (1.581) | (1.844) | (1.440) | (1.448) | (1.968) |
|   Above 80000 | 0.575 | 0.0487 | -1.519 | 2.290 | 2.269 |
| | (2.673) | (2.957) | (2.332) | (2.200) | (2.492) |
| Availability of Personal Vehicle | 1.794*** | 1.814*** | 2.041*** | 1.898*** | 1.904*** |
| (1 = Yes) | (0.603) | (0.520) | (0.474) | (0.449) | (0.607) |
| Consciousness Variables | | | | | |
|   e_1 | -0.554* | -0.791*** | -0.436** | -0.383* | -0.567* |
| | (0.293) | (0.254) | (0.209) | (0.212) | (0.290) |
|   e_2 | -0.784*** | -0.522* | -0.233 | 0.0541 | 0.326 |
| | (0.287) | (0.271) | (0.215) | (0.246) | (0.294) |
|   e_3 | -0.649** | -0.537** | -0.719*** | -0.477** | -0.652*** |
| | (0.292) | (0.228) | (0.198) | (0.198) | (0.243) |
|   e_4 | -0.300 | -0.406* | -0.451** | -0.277 | -0.0765 |
| | (0.230) | (0.227) | (0.220) | (0.260) | (0.290) |
|   e_5 | 0.330 | 0.784*** | 0.961*** | 1.013*** | 1.226*** |
| | (0.213) | (0.234) | (0.233) | (0.230) | (0.309) |
|   e_6 | 0.997*** | 1.260*** | 1.442*** | 0.861*** | 0.0531 |
| | (0.214) | (0.201) | (0.255) | (0.274) | (0.356) |
|   e_7 | 1.640*** | 1.442*** | 1.170*** | 0.613*** | 0.382** |
| | (0.261) | (0.226) | (0.244) | (0.179) | (0.186) |
|   e_8 | -0.190 | -0.236 | -0.344* | -0.136 | 0.200 |
| | (0.268) | (0.215) | (0.180) | (0.181) | (0.214) |
|   e_9 | -0.736*** | -0.549*** | -0.438** | -0.0767 | 0.179 |
| | (0.221) | (0.207) | (0.191) | (0.186) | (0.242) |
| Fuel Oil Consumption pc | -0.00162 | -0.00249* | -0.00276 | -0.000427 | -0.000943 |
| | (0.00210) | (0.00147) | (0.00242) | (0.00298) | (0.00433) |
| Observations | 1,000 | 1,000 | 1,000 | 1,000 | 1,000 |

Note: The parentheses indicate bootstrapped errors in parentheses (*** p<0.01, ** p<0.05, * p<0.1). The constant terms are dropped from the model although the terms are included in regression models.
Source: Authors' Calculation.

Table G.6 presents the quantile regression results for fuel price-related stress. Urban households experience consistent stress across all quantiles, with coefficients decreasing from 2.341 at the 15th quantile to 1.166 at the 85th quantile. This suggests that the perceived stress from fuel price increases is more pronounced at lower stress levels, but becomes less significant as stress intensifies, possibly due to coping mechanisms or other factors reducing the impact of fuel price hikes at higher stress levels.

Regional disparities are evident across divisions, mirroring the trends seen in the MOLS analysis. For instance, Barisal and Chattogram show high stress at the lower quantiles (Barisal: 9.873 at the 15th quantile, Chattogram: 7.058 at the 15th quantile), with a notable decline at higher quantiles, reflecting how fuel price stress is felt more acutely in certain regions at lower stress levels. Barisal's stress impact decreases significantly, reaching a negative coefficient (-0.389) at the 85th quantile. Similarly, Chattogram's stress coefficient drops from 7.058 at the 15th quantile to -2.509 at the 85th quantile, indicating a reduction in fuel price-related stress at higher stress levels.

The effect of age on stress is consistent with prior findings, showing a negative relationship with stress across all quantiles, especially at the lower levels (coefficients: -0.0464 at the 15th quantile and -0.0459 at the 30th quantile). This suggests that older individuals tend to experience less stress related to fuel prices, likely due to greater financial stability or established coping mechanisms. However, the negative effect diminishes as stress levels rise, potentially reflecting increased exposure to the impact of price fluctuations at higher stress levels.

Education also plays a crucial role, with higher levels of education associated with lower stress across all quantiles. The coefficient ranges from -0.106 at the 15th quantile to -0.0589 at the 85th quantile, supporting the notion that more educated individuals are better equipped to manage stress related to fuel prices, possibly due to better access to information and resources.

Income source consistently shows that households engaged in service sector employment experience lower stress than those in agriculture or industry. This trend is most pronounced at the lower quantiles, with significant negative coefficients for service sector employment across the 15th, 30th, and 50th quantiles (-2.310, -2.157, and -1.660, respectively), reflecting the relative stability of income in the service sector compared to agriculture and industry, which are more vulnerable to energy disruptions.

The income group variable shows more variation in the SQR results than in the MOLS analysis. Households with incomes above BDT 80,000 show lower stress at the lower quantiles (coefficient: 0.575 at the 15th quantile) but experience no significant effect at higher stress levels (coefficient: -0.533 at the 85th quantile), suggesting that high-income households are less susceptible to the effects of fuel price increases at lower stress levels but are less resilient when stress reaches higher levels.

Vehicle ownership remains significantly correlated with stress across all quantiles, with the highest coefficients observed at the lower quantiles (coefficients: 1.794 at the 15th quantile and 1.814 at the 30th quantile). This reinforces the MOLS finding that households with higher energy dependence, especially those reliant on personal vehicles, are more vulnerable to fuel price-related stress.

Environmental consciousness variables exhibit varying effects across quantiles, similar to the results in previous scenarios. Belief in environmental pollution's impact (e_1) significantly reduces stress at the lower quantiles (coefficients: -0.554 at the 15th quantile and -0.567 at the 85th quantile), suggesting that households with higher environmental awareness may be more resilient to fuel price stress. However, belief in renewable energy (e_5) continues to show a positive relationship with stress, particularly at higher quantiles (coefficients: 0.784 at the 30th quantile to 1.226 at the 85th quantile), reflecting that households prioritising renewable energy may feel more stressed as they perceive the energy crisis to be more urgent. Scepticism towards media portrayals of the energy crisis (e_7) increases stress at lower quantiles (coefficient: 1.432 at the 15th quantile) but loses significance at higher quantiles, supporting previous findings that scepticism exacerbates stress at lower levels but is less impactful when stress intensifies. On the other hand, willingness to reduce energy consumption for environmental reasons (e_9) significantly reduces stress at lower quantiles (coefficient: -0.736 at the 15th quantile to -0.438 at the 50th quantile), but its effect diminishes at higher stress levels, indicating that environmental consciousness helps alleviate moderate stress but is less effective in extreme situations. In conclusion, the results from the quantile regression reinforce the findings from MOLS

**Appendix H: Results of Ordered Probit Model in Details**
**Table H.1: Regression Results from Ordered Probit Model**

| VARIABLES | (1) PSS: Elc. Supply | (2) PSS: Elc. Price | (3) PSS: Gas Supply | (4) PSS: Gas Price | (5) PSS: Fuel Supply | (6) PSS: Fuel Price | (7) Price: Overall |
|---|---|---|---|---|---|---|---|
| Urban (sub-Urban = 0) | 0.72*** | 0.24*** | 1.42*** | 0.71*** | 0.83*** | 0.49*** | 1.15*** |
|  | (0.11) | (0.09) | (0.15) | (0.10) | (0.14) | (0.10) | (0.13) |
| Division (Base: Dhaka) | | | | | | | |
|   Barisal | 1.07*** | 0.10 | 0.96*** | 0.64*** | 1.28*** | 0.58*** | 1.30*** |
|  | (0.20) | (0.14) | (0.17) | (0.20) | (0.37) | (0.18) | (0.17) |
|   Chattogram | 0.81*** | -0.01 | 0.37** | -0.03 | 1.25*** | 0.70*** | 0.92*** |
|  | (0.13) | (0.11) | (0.14) | (0.12) | (0.21) | (0.11) | (0.12) |
|   Khulna | 1.09*** | 0.53*** | -0.11 | 0.42** | 0.39* | 1.05*** | 1.64*** |
|  | (0.22) | (0.19) | (0.19) | (0.21) | (0.23) | (0.22) | (0.25) |
|   Mymensingh | 0.62*** | -0.21* | 0.001 | -0.21 | 0.82*** | 0.42*** | 0.43** |
|  | (0.15) | (0.12) | (0.21) | (0.15) | (0.22) | (0.14) | (0.20) |
|   Rajshahi | -0.43** | -1.04*** | -0.81*** | -1.08*** | -1.18*** | -0.96*** | -1.14*** |
|  | (0.18) | (0.17) | (0.21) | (0.17) | (0.21) | (0.18) | (0.18) |
|   Rangpur | 1.75*** | 0.83*** | 1.32*** | 0.78*** | 2.33*** | 0.96*** | 2.05*** |
|  | (0.19) | (0.18) | (0.19) | (0.19) | (0.33) | (0.23) | (0.22) |
|   Sylhet | 0.87*** | -0.18 | 0.61*** | 0.08 | 0.99*** | 0.42** | 0.83*** |
|  | (0.21) | (0.18) | (0.22) | (0.18) | (0.29) | (0.20) | (0.26) |
| Sex | -0.34 | -0.41* | -0.62*** | -0.39 | -0.28 | -0.40* | -0.40 |
|  | (0.27) | (0.24) | (0.19) | (0.27) | (0.33) | (0.22) | (0.30) |
| Age | -0.01** | -0.01 | -0.01** | -0.004 | -0.02*** | -0.01* | -0.008 |
|  | (0.01) | (0.004) | (0.004) | (0.004) | (0.01) | (0.01) | (0.006) |
| Years of Education | -0.012 | -0.01 | -0.04*** | 0.003 | -0.04*** | 0.013 | 0.001 |
|  | (0.01) | (0.01) | (0.01) | (0.01) | (0.01) | (0.01) | (0.01) |
| Household Size | 0.017 | 0.05 | -0.09* | -0.03 | -0.02 | -0.02 | 0.03 |
|  | (0.035) | (0.03) | (0.05) | (0.04) | (0.05) | (0.03) | (0.05) |
| Number of Students | -0.032 | -0.04 | 0.02 | 0.01 | 0.10 | 0.01 | -0.06 |
|  | (0.06) | (0.05) | (0.06) | (0.06) | (0.07) | (0.05) | (0.07) |
| Income Source (Base: Agri) | | | | | | | |
|   Service | -0.23* | -0.30*** | -0.33*** | -0.30** | -0.87*** | -0.38*** | -0.44*** |
|  | (0.12) | (0.11) | (0.12) | (0.12) | (0.19) | (0.12) | (0.14) |
|   Industry | -0.028 | 0.12 | 0.25 | 0.05 | -0.34 | -0.22 | -0.11 |
|  | (0.20) | (0.17) | (0.22) | (0.18) | (0.31) | (0.17) | (0.19) |
| Income Group (Base: Below BDT 10000) | | | | | | | |
|   BDT 10000 to 24999 | -0.45 | 0.17 | -0.71* | -0.16 | -0.63* | -0.03 | -0.16 |
|  | (0.39) | (0.32) | (0.41) | (0.35) | (0.35) | (0.29) | (0.42) |
|   BDT 25000 to 39999 | -0.24 | 0.22 | -0.48 | -0.03 | -0.22 | 0.11 | -0.01 |
|  | (0.40) | (0.33) | (0.42) | (0.36) | (0.34) | (0.30) | (0.42) |
|   BDT 40000 to 79999 | -0.65 | 0.04 | -0.49 | -0.41 | -0.55 | -0.37 | -0.39 |
|  | (0.41) | (0.35) | (0.46) | (0.38) | (0.38) | (0.32) | (0.44) |
|   Above 80000 | -0.96 | 0.54 | -1.38** | -0.68 | 0.30 | -0.56 | -0.93 |
|  | (0.65) | (0.81) | (0.62) | (0.54) | (0.78) | (0.50) | (0.72) |
| Availability of Personal Vehicle (1 = Yes) | 0.24** | 0.07 | 0.19 | 0.19* | 1.02*** | 0.80*** | 0.30** |
|  | (0.11) | (0.10) | (0.12) | (0.10) | (0.18) | (0.13) | (0.13) |
| Consciousness Variables | | | | | | | |
| e_1 | -0.17*** | -0.08** | -0.18*** | -0.09** | -0.42*** | -0.07 | -0.23*** |
|  | (0.05) | (0.04) | (0.06) | (0.04) | (0.10) | (0.05) | (0.05) |
| e_2 | -0.066 | 0.13*** | -0.02 | 0.08* | -0.21*** | -0.04 | -0.05 |
|  | (0.05) | (0.05) | (0.05) | (0.05) | (0.06) | (0.05) | (0.06) |
| e_3 | 0.002 | 0.08* | -0.05 | -0.01 | 0.01 | -0.06 | -0.05 |
|  | (0.047) | (0.04) | (0.06) | (0.05) | (0.06) | (0.05) | (0.05) |
| e_4 | 0.042 | 0.18*** | -0.08 | -0.08 | 0.05 | -0.02 | 0.001 |
|  | (0.050) | (0.05) | (0.06) | (0.06) | (0.06) | (0.06) | (0.06) |
| e_5 | 0.27*** | -0.02 | 0.15*** | 0.18*** | 0.10 | 0.20*** | 0.37*** |
|  | (0.05) | (0.05) | (0.05) | (0.05) | (0.06) | (0.06) | (0.06) |
| e_6 | 0.136*** | -0.14*** | 0.18*** | 0.06 | 0.41*** | 0.14*** | 0.19** |
|  | (0.05) | (0.05) | (0.05) | (0.05) | (0.06) | (0.05) | (0.05) |
| e_7 | 0.127** | 0.11** | 0.19*** | 0.21*** | 0.29*** | 0.19*** | 0.34*** |
|  | (0.05) | (0.05) | (0.06) | (0.05) | (0.08) | (0.05) | (0.05) |
| e_8 | -0.21*** | -0.104** | -0.03 | -0.06 | -0.15*** | -0.10** | -0.11** |
|  | (0.05) | (0.04) | (0.05) | (0.04) | (0.06) | (0.04) | (0.05) |
| e_9 | -0.097** | -0.16*** | -0.23*** | -0.15*** | -0.27*** | -0.12*** | -0.26*** |
|  | (0.049) | (0.04) | (0.052) | (0.05) | (0.06) | (0.04) | (0.05) |
| Electricity Consumption pc | 0.0004 | 0.0002 |  |  |  |  |  |
|  | (0.0003) | (0.0003) |  |  |  |  |  |
| Gas Consumption pc |  |  | -0.001 | -0.001 |  |  |  |
|  |  |  | (0.001) | (0.001) |  |  |  |
| Oil Consumption pc |  |  |  |  | -0.001 | -0.001* |  |
|  |  |  |  |  | (0.001) | (0.0005) |  |
| Power and Energy Consumption pc |  |  |  |  |  |  | 7.50e-05 |
|  |  |  |  |  |  |  | (0.0002) |
| Observations | 1,000 | 1,000 | 1,000 | 1,000 | 1,000 | 1,000 | 1,000 |
| Pseudo R-squared | 0.251 | 0.14 | 0.32 | 0.174 | 0.48 | 0.397 | 0.37 |

Robust Standard errors in parentheses
*** p<0.01, ** p<0.05, * p<0.1

Source: Authors' Calculation.

**Table H.2: Marginal Effects from the Ordered Probit Model (Scenario: Supply of Electricity)**

| Variables | Prob (pss_elcs_rng = 1) | | Prob (pss_elcs_rng = 2) | | Prob (pss_elcs_rng = 3) | |
|---|---|---|---|---|---|---|
|  | Mg. Eff. | (P-Val.) | Mg. Eff. | (P-Val.) | Mg. Eff. | (P-Val.) |
| Urban (sub-Urban = 0) | -0.13 | 0.00 | 0.06 | 0.00 | 0.06 | 0.00 |
| Division (Base: Dhaka) |  |  |  |  |  |  |
|     Barisal | -0.2 | 0.00 | 0.11 | 0.00 | 0.09 | 0.00 |
|     Chattogram | -0.16 | 0.00 | 0.11 | 0.00 | 0.05 | 0.00 |
|     Khulna | -0.19 | 0.00 | 0.11 | 0.00 | 0.09 | 0.00 |
|     Mymensingh | -0.13 | 0.00 | 0.1 | 0.00 | 0.03 | 0.00 |
|     Rajshahi | 0.13 | 0.02 | -0.12 | 0.02 | -0.01 | 0.02 |
|     Rangpur | -0.24 | 0.00 | 0.02 | 0.63 | 0.22 | 0.00 |
|     Sylhet | -0.17 | 0.00 | 0.11 | 0.00 | 0.06 | 0.00 |
| Sex | 0.06 | 0.21 | -0.03 | 0.22 | -0.03 | 0.21 |
| Age | 0.002 | 0.04 | -0.001 | 0.05 | -0.001 | 0.04 |
| Years of Education | 0.002 | 0.21 | -0.001 | 0.21 | -0.001 | 0.21 |
| Household Size | -0.003 | 0.62 | 0.001 | 0.62 | 0.002 | 0.63 |
| Number of Students | 0.01 | 0.56 | -0.003 | 0.56 | -0.003 | 0.56 |
| Income Source (Base: Agri) |  |  |  |  |  |  |
|     Service | 0.04 | 0.06 | -0.02 | 0.07 | -0.02 | 0.06 |
|     Industry | 0.01 | 0.9 | -0.002 | 0.9 | -0.002 | 0.87 |
| Income Group (Base: Below BDT 10000) |  |  |  |  |  |  |
| BDT 10000 to 24999 | 0.08 | 0.25 | -0.04 | 0.25 | -0.04 | 0.25 |
| BDT 25000 to 39999 | 0.04 | 0.54 | -0.02 | 0.54 | -0.02 | 0.54 |
| BDT 40000 to 79999 | 0.11 | 0.11 | -0.06 | 0.12 | -0.06 | 0.12 |
| Above 80000 | 0.17 | 0.14 | -0.08 | 0.14 | -0.09 | 0.14 |
| Availability of Personal Vehicle (1 = Yes) | -0.04 | 0.03 | 0.02 | 0.04 | 0.02 | 0.033 |
| Consciousness Variables | | | | | | |
| e_1 | 0.03 | 0.00 | -0.01 | 0.00 | -0.02 | 0.00 |
| e_2 | 0.01 | 0.18 | -0.01 | 0.20 | -0.01 | 0.20 |
| e_3 | 0.00 | 0.97 | 0.00 | 0.97 | 0.00 | 0.97 |
| e_4 | -0.01 | 0.40 | 0.004 | 0.40 | 0.004 | 0.40 |
| e_5 | -0.05 | 0.00 | 0.02 | 0.00 | 0.02 | 0.00 |
| e_6 | -0.02 | 0.00 | 0.012 | 0.01 | 0.01 | 0.00 |
| e_7 | -0.02 | 0.01 | 0.01 | 0.02 | 0.01 | 0.01 |
| e_8 | 0.04 | 0.00 | -0.02 | 0.00 | -0.02 | 0.00 |
| e_9 | 0.02 | 0.05 | -0.01 | 0.06 | -0.01 | 0.05 |
| Electricity Consumption pc | 0.00 | 0.16 | 0.00 | 0.16 | 0.00 | 0.15 |

**Source:** Authors' Calculation.

**Table H.3: Marginal Effects from the Ordered Probit Model (Scenario: Price of Electricity)**

| Variables | Prob (pss_elcp_rng = 1) | | Prob (pss_elcp_rng = 2) | | Prob (pss_elcp_rng = 3) | |
|---|---|---|---|---|---|---|
| | Mg. Eff. | (P-Val.) | Mg. Eff. | (P-Val.) | Mg. Eff. | (P-Val.) |
| Urban (sub-Urban = 0) | -0.038 | 0.01 | -0.005 | 0.20 | 0.042 | 0.01 |
| Division (Base: Dhaka) | | | | | | |
| Barisal | -0.014 | 0.47 | -0.004 | 0.511 | 0.018 | 0.471 |
| Chattogram | 0.001 | 0.939 | 0.000 | 0.941 | -0.001 | 0.939 |
| Khulna | -0.056 | 0.003 | -0.062 | 0.068 | 0.118 | 0.015 |
| Mymensingh | 0.036 | 0.090 | -0.004 | 0.552 | -0.032 | 0.089 |
| Rajshahi | 0.258 | 0.000 | -0.164 | 0.000 | -0.094 | 0.000 |
| Rangpur | -0.072 | 0.000 | -0.14 | 0.001 | 0.21 | 0.000 |
| Sylhet | 0.029 | 0.329 | -0.002 | 0.758 | -0.027 | 0.296 |
| Sex | 0.063 | 0.081 | 0.008 | 0.258 | -0.071 | 0.082 |
| Age | 0.001 | 0.158 | 0.000 | 0.293 | -0.001 | 0.156 |
| Years of Education | 0.001 | 0.570 | 0.000 | 0.583 | -0.001 | 0.568 |
| Household Size | -0.01 | 0.154 | -0.001 | 0.293 | 0.01 | 0.152 |
| Number of Students | 0.01 | 0.409 | 0.001 | 0.471 | -0.007 | 0.409 |
| Income Source (Base: Agri) | | | | | | |
| Service | 0.046 | 0.008 | 0.006 | 0.185 | -0.051 | 0.008 |
| Industry | -0.018 | 0.474 | -0.002 | 0.530 | 0.021 | 0.475 |
| Income Group (Base: Below BDT 10000) | | | | | | |
| BDT 10000 to 24999 | -0.026 | 0.596 | -0.003 | 0.617 | 0.029 | 0.596 |
| BDT 25000 to 39999 | -0.034 | 0.495 | -0.004 | 0.535 | 0.039 | 0.495 |
| BDT 40000 to 79999 | -0.006 | 0.916 | -0.001 | 0.916 | 0.007 | 0.916 |
| Above 80000 | -0.084 | 0.50 | -0.01 | 0.551 | 0.094 | 0.501 |
| Availability of Personal Vehicle (1 = Yes) | -0.01 | 0.506 | -0.001 | 0.531 | 0.011 | 0.505 |
| Consciousness Variables | | | | | | |
| e_1 | 0.012 | 0.051 | 0.002 | 0.181 | -0.014 | 0.044 |
| e_2 | -0.021 | 0.004 | -0.003 | 0.205 | 0.023 | 0.005 |
| e_3 | -0.013 | 0.060 | -0.002 | 0.23 | 0.014 | 0.058 |
| e_4 | -0.028 | 0.001 | -0.003 | 0.176 | 0.031 | 0.001 |
| e_5 | 0.002 | 0.743 | 0.000 | 0.744 | -0.002 | 0.743 |
| e_6 | 0.022 | 0.001 | 0.003 | 0.208 | -0.025 | 0.002 |
| e_7 | -0.017 | 0.018 | -0.002 | 0.168 | 0.019 | 0.015 |
| e_8 | 0.016 | 0.014 | 0.002 | 0.191 | -0.018 | 0.013 |
| e_9 | 0.024 | 0.000 | 0.003 | 0.138 | -0.027 | 0.000 |
| Electricity Consumption pc | 0.00 | 0.56 | 0.00 | 0.58 | 0.00 | 0.56 |

**Source**: Authors' Calculation.

**Table H.4: Marginal Effects from the Ordered Probit Model (Scenario: Supply of Gas)**

| Variables | Prob (pss_gassp_rng = 1) | | Prob (pss_gassp_rng = 2) | | Prob (pss_gassp_rng = 3) | |
|---|---|---|---|---|---|---|
| | Mg. Eff. | (P-Val.) | Mg. Eff. | (P-Val.) | Mg. Eff. | (P-Val.) |
| Urban | -0.23 | 0.00 | 0.14 | 0.00 | 0.095 | 0.00 |
| Division (Base: Dhaka) | | | | | | |
| Barisal | -0.13 | 0.00 | 0.04 | 0.008 | 0.090 | 0.00 |
| Chattogram | -0.064 | 0.011 | 0.040 | 0.019 | 0.024 | 0.01 |
| Khulna | 0.022 | 0.58 | -0.016 | 0.59 | -0.005 | 0.57 |
| Mymensingh | -0.00 | 0.996 | 0.00 | 0.996 | 0.00 | 0.996 |
| Rajshahi | 0.19 | 0.000 | -0.17 | 0.00 | -0.025 | 0.001 |
| Rangpur | -0.16 | 0.00 | 0.13 | 0.69 | 0.15 | 0.000 |
| Sylhet | -0.10 | 0.00 | 0.05 | 0.00 | 0.047 | 0.034 |
| Sex | 0.101 | 0.001 | -0.060 | 0.001 | -0.041 | 0.003 |
| Age | 0.002 | 0.017 | -0.001 | 0.021 | -0.001 | 0.018 |
| Years of Education | 0.006 | 0.002 | -0.004 | 0.002 | -0.002 | 0.002 |
| Household Size | 0.014 | 0.067 | -0.008 | 0.067 | -0.006 | 0.074 |
| Number of Students | -0.003 | 0.747 | 0.002 | 0.747 | 0.001 | 0.747 |
| Income Source (Base: Agri) | | | | | | |
| Service | 0.054 | 0.007 | -0.032 | 0.009 | -0.022 | 0.009 |
| Industry | -0.04 | 0.255 | 0.024 | 0.26 | 0.016 | 0.259 |
| Income Group (Base: Below BDT 10000) | | | | | | |
| BDT 10000 to 24999 | 0.115 | 0.084 | -0.070 | 0.086 | -0.048 | 0.090 |
| BDT 25000 to 39999 | 0.077 | 0.260 | -0.046 | 0.26 | -0.032 | 0.264 |
| BDT 40000 to 79999 | 0.08 | 0.29 | -0.047 | 0.293 | -0.032 | 0.292 |
| Above 80000 | 0.22 | 0.023 | -0.132 | 0.024 | -0.092 | 0.03 |
| Availability of Personal Vehicle (1 = Yes) | -0.03 | 0.12 | 0.018 | 0.13 | 0.012 | 0.12 |
| Consciousness Variables | | | | | | |
| e_1 | 0.030 | 0.002 | -0.018 | 0.003 | -0.012 | 0.002 |
| e_2 | 0.003 | 0.698 | -0.002 | 0.70 | -0.001 | 0.70 |
| e_3 | 0.008 | 0.39 | -0.005 | 0.40 | -0.003 | 0.40 |
| e_4 | 0.012 | 0.19 | -0.007 | 0.196 | -0.005 | 0.19 |
| e_5 | -0.024 | 0.003 | 0.014 | 0.004 | 0.01 | 0.005 |
| e_6 | -0.029 | 0.00 | 0.017 | 0.00 | 0.012 | 0.00 |
| e_7 | -0.03 | 0.00 | 0.018 | 0.00 | 0.012 | 0.00 |
| e_8 | 0.005 | 0.53 | -0.003 | 0.53 | -0.002 | 0.53 |
| e_9 | 0.037 | 0.00 | -0.022 | 0.00 | -0.015 | 0.00 |
| Gas Consumption pc | 0.00 | 0.13 | -0.00 | 0.13 | 0.00 | 1.42 |

**Source:** Authors' Calculation.

**Table H.5: Marginal Effects from the Ordered Probit Model (Scenario: Price of Gas)**

| Variables | Prob (pss_gaspr_rng = 1) | | Prob (pss_gaspr_rng = 2) | | Prob (pss_gaspr_rng = 3) | |
|---|---|---|---|---|---|---|
| | Mg. Eff. | (P-Val.) | Mg. Eff. | (P-Val.) | Mg. Eff. | (P-Val.) |
| Urban | -0.11 | 0.00 | 0.01 | 0.294 | 0.10 | 0.00 |
| Division (Base: Dhaka) | | | | | | |
| Barisal | -0.07 | 0.00 | -0.05 | 0.07 | 0.12 | 0.01 |
| Chattogram | 0.005 | 0.79 | -0.001 | 0.78 | -0.004 | 0.79 |
| Khulna | -0.05 | 0.03 | -0.02 | 0.304 | 0.07 | 0.074 |
| Mymensingh | 0.04 | 0.17 | -0.012 | 0.299 | -0.025 | 0.147 |
| Rajshahi | 0.264 | 0.00 | -0.191 | 0.00 | -0.073 | 0.00 |
| Rangpur | -0.075 | 0.00 | -0.081 | 0.01 | 0.16 | 0.00 |
| Sylhet | -0.011 | 0.662 | 0.001 | 0.807 | 0.011 | 0.67 |
| Sex | 0.059 | 0.156 | -0.005 | 0.393 | -0.054 | 0.156 |
| Age | 0.001 | 0.421 | -0.00 | 0.54 | -0.001 | 0.418 |
| Years of Education | -0.001 | 0.713 | 0.00 | 0.725 | 0.001 | 0.713 |
| Household Size | 0.005 | 0.434 | 0.00 | 0.518 | -0.004 | 0.436 |
| Number of Students | -0.001 | 0.896 | 0.00 | 0.897 | 0.001 | 0.896 |
| Income Source (Base: Agri) | | | | | | |
| Service | 0.046 | 0.014 | -0.004 | 0.326 | -0.042 | 0.014 |
| Industry | -0.007 | 0.798 | 0.001 | 0.801 | 0.006 | 0.799 |
| Income Group (Base: Below BDT 10000) | | | | | | |
| BDT 10000 to 24999 | 0.025 | 0.644 | -0.002 | 0.667 | -0.023 | 0.645 |
| BDT 25000 to 39999 | 0.005 | 0.923 | -0.000 | 0.923 | -0.005 | 0.923 |
| BDT 40000 to 79999 | 0.063 | 0.281 | -0.006 | 0.441 | -0.057 | 0.283 |
| Above 80000 | 0.103 | 0.212 | -0.009 | 0.402 | -0.094 | 0.215 |
| Availability of Personal Vehicle (1 = Yes) | -0.03 | 0.07 | 0.003 | 0.364 | 0.026 | 0.067 |
| Consciousness Variables | | | | | | |
| e_1 | 0.013 | 0.05 | -0.001 | 0.367 | -0.012 | 0.046 |
| e_2 | -0.012 | 0.092 | 0.001 | 0.32 | 0.011 | 0.101 |
| e_3 | 0.001 | 0.906 | -0.00 | 0.907 | -0.001 | 0.907 |
| e_4 | 0.012 | 0.141 | -0.001 | 0.383 | -0.011 | 0.142 |
| e_5 | -0.028 | 0.00 | 0.002 | 0.288 | 0.025 | 0.00 |
| e_6 | -0.009 | 0.186 | 0.001 | 0.423 | 0.008 | 0.181 |
| e_7 | -0.032 | 0.00 | 0.003 | 0.320 | 0.029 | 0.00 |
| e_8 | 0.01 | 0.155 | -0.001 | 0.391 | -0.01 | 0.155 |
| e_9 | 0.023 | 0.002 | -0.002 | 0.330 | -0.021 | 0.001 |
| Gas Consumption pc | 0.00 | 0.101 | 0.00 | 0.351 | 0.00 | 0.105 |

**Source:** Authors' Calculation.

**Table H.6: Marginal Effects from the Ordered Probit Model (Scenario: Supply of Fuel Oil)**

| Variables | Prob (pss_fuelsp_rng = 1) | | Prob (pss_fuelsp_rng = 2) | | Prob (pss_fuelsp_rng = 3) | |
|---|---|---|---|---|---|---|
| | Mg. Eff. | (P-Val.) | Mg. Eff. | (P-Val.) | Mg. Eff. | (P-Val.) |
| Urban | -0.113 | 0.00 | 0.095 | 0.00 | 0.018 | 0.001 |
| Division (Base: Dhaka) | | | | | | |
|     Barisal | -0.175 | 0.00 | 0.153 | 0.00 | 0.023 | 0.105 |
|     Chattogram | -0.173 | 0.00 | 0.151 | 0.00 | 0.022 | 0.003 |
|     Khulna | -0.069 | 0.076 | 0.066 | 0.073 | 0.003 | 0.233 |
|     Mymensingh | -0.129 | 0.00 | 0.119 | 0.00 | 0.01 | 0.046 |
|     Rajshahi | 0.268 | 0.00 | -0.266 | 0.00 | -0.002 | 0.064 |
|     Rangpur | -0.223 | 0.00 | 0.129 | 0.00 | 0.099 | 0.00 |
|     Sylhet | -0.15 | 0.00 | 0.134 | 0.00 | 0.013 | 0.096 |
| Sex | 0.041 | 0.364 | -0.035 | 0.369 | -0.007 | 0.349 |
| Age | 0.003 | 0.00 | -0.003 | 0.00 | -0.001 | 0.002 |
| Years of Education | 0.005 | 0.007 | -0.004 | 0.006 | -0.001 | 0.03 |
| Household Size | 0.002 | 0.724 | -0.002 | 0.724 | 0.00 | 0.724 |
| Number of Students | -0.014 | 0.153 | 0.011 | 0.154 | 0.002 | 0.173 |
| Income Source (Base: Agri) | | | | | | |
|     Service | 0.119 | 0.00 | -0.10 | 0.00 | -0.019 | 0.003 |
|     Industry | 0.046 | 0.266 | -0.040 | 0.267 | -0.008 | 0.275 |
| Income Group (Base: Below BDT 10000) | | | | | | |
|     BDT 10000 to 24999 | 0.086 | 0.072 | -0.072 | 0.077 | -0.014 | 0.073 |
|     BDT 25000 to 39999 | 0.030 | 0.521 | -0.025 | 0.523 | -0.005 | 0.517 |
|     BDT 40000 to 79999 | 0.076 | 0.140 | -0.063 | 0.146 | -0.012 | 0.135 |
|     Above 80000 | -0.041 | 0.704 | 0.034 | 0.704 | 0.007 | 0.707 |
| Availability of Personal Vehicle (1 = Yes) | -0.139 | 0.00 | 0.12 | 0.00 | 0.023 | 0.00 |
| Consciousness Variables | | | | | | |
|     e_1 | 0.058 | 0.00 | -0.048 | 0.00 | -0.009 | 0.002 |
|     e_2 | 0.028 | 0.001 | -0.024 | 0.001 | -0.004 | 0.006 |
|     e_3 | -0.001 | 0.883 | 0.001 | 0.883 | 0.000 | 0.884 |
|     e_4 | -0.007 | 0.379 | 0.006 | 0.376 | 0.001 | 0.401 |
|     e_5 | -0.014 | 0.101 | 0.011 | 0.10 | 0.002 | 0.132 |
|     e_6 | -0.057 | 0.00 | 0.048 | 0.00 | 0.009 | 0.00 |
|     e_7 | -0.04 | 0.00 | 0.034 | 0.00 | 0.007 | 0.001 |
|     e_8 | 0.021 | 0.009 | -0.018 | 0.008 | -0.004 | 0.037 |
|     e_9 | 0.037 | 0.00 | -0.031 | 0.00 | -0.006 | 0.001 |
| Oil Consumption pc | 0.00 | 0.248 | 0.00 | 0.25 | 0.00 | 0.258 |

**Source:** Authors' Calculation.

**Table H.7: Marginal Effects from the Ordered Probit Model (Scenario: Price of Fuel Oil)**

| Variables | Prob (pss_fuelpr_rng = 1) | | Prob (pss_fuelpr_rng = 2) | | Prob (pss_fuelpr_rng = 3) | |
|---|---|---|---|---|---|---|
| | Mg. Eff. | (P-Val.) | Mg. Eff. | (P-Val.) | Mg. Eff. | (P-Val.) |
| Urban | -0.092 | 0.00 | 0.041 | 0.00 | 0.051 | 0.00 |
| Division (Base: Dhaka) | | | | | | |
|     Barisal | -0.11 | 0.00 | 0.06 | 0.002 | 0.052 | 0.007 |
|     Chattogram | -0.128 | 0.00 | 0.06 | 0.002 | 0.07 | 0.00 |
|     Khulna | -0.163 | 0.00 | 0.036 | 0.182 | 0.13 | 0.00 |
|     Mymensingh | -0.086 | 0.002 | 0.053 | 0.006 | 0.033 | 0.004 |
|     Rajshahi | 0.29 | 0.00 | -0.26 | 0.00 | -0.023 | 0.00 |
|     Rangpur | -0.156 | 0.00 | 0.045 | 0.028 | 0.11 | 0.00 |
|     Sylhet | -0.086 | 0.023 | 0.053 | 0.016 | 0.033 | 0.075 |
| Sex | 0.075 | 0.069 | -0.033 | 0.073 | -0.041 | 0.072 |
| Age | 0.002 | 0.063 | -0.001 | 0.07 | -0.001 | 0.062 |
| Years of Education | -0.003 | 0.168 | 0.001 | 0.164 | 0.001 | 0.176 |
| Household Size | 0.003 | 0.653 | -0.001 | 0.654 | -0.002 | 0.652 |
| Number of Students | -0.001 | 0.883 | 0.001 | 0.883 | 0.001 | 0.883 |
| Income Source (Base: Agri) | | | | | | |
|     Service | 0.072 | 0.001 | -0.032 | 0.002 | -0.04 | 0.001 |
|     Industry | 0.042 | 0.189 | -0.02 | 0.196 | -0.023 | 0.189 |
| Income Group (Base: Below BDT 10000) | | | | | | |
|     BDT 10000 to 24999 | 0.006 | 0.916 | -0.003 | 0.916 | -0.003 | 0.916 |
|     BDT 25000 to 39999 | -0.021 | 0.707 | 0.01 | 0.706 | 0.012 | 0.707 |
|     BDT 40000 to 79999 | 0.069 | 0.241 | -0.031 | 0.245 | -0.039 | 0.242 |
|     Above 80000 | 0.104 | 0.261 | -0.046 | 0.264 | -0.058 | 0.263 |
| Availability of Personal Vehicle (1 = Yes) | -0.149 | 0.00 | 0.066 | 0.00 | 0.083 | 0.00 |
| Consciousness Variables | | | | | | |
|     e_1 | 0.013 | 0.129 | -0.006 | 0.141 | -0.007 | 0.125 |
|     e_2 | 0.008 | 0.356 | -0.004 | 0.363 | -0.005 | 0.353 |
|     e_3 | 0.011 | 0.20 | -0.005 | 0.204 | -0.006 | 0.202 |
|     e_4 | 0.004 | 0.672 | -0.002 | 0.673 | -0.002 | 0.672 |
|     e_5 | -0.037 | 0.00 | 0.017 | 0.00 | 0.021 | 0.00 |
|     e_6 | -0.026 | 0.003 | 0.012 | 0.006 | 0.015 | 0.003 |
|     e_7 | -0.035 | 0.00 | 0.016 | 0.00 | 0.020 | 0.00 |
|     e_8 | 0.018 | 0.021 | -0.01 | 0.25 | -0.01 | 0.023 |
|     e_9 | 0.023 | 0.005 | -0.01 | 0.008 | -0.013 | 0.004 |
| Oil Consumption pc | 0.00 | 0.092 | 0.00 | 0.097 | 0.00 | 0.094 |

**Source:** Authors' Calculation.